\documentclass[aps,preprint]{revtex4}
 \usepackage{amsmath,amssymb,amsthm,mathrsfs,amsfonts,dsfont} 
\usepackage{graphicx}
\usepackage{amsbsy}
\usepackage{bm}
\usepackage{algorithm}
\usepackage{algorithmicx}
\usepackage{algpseudocode}
\renewcommand{\algorithmicrequire}{ \textbf{Input:}}    
\renewcommand{\algorithmicensure}{ \textbf{Output:}}

\def\id{{\mathds{1}}}
\def\bx{\boldsymbol{\xi}}
\def\X{\mathbf{X}}
\def\Tr{\mathrm{Tr}}
\def\bhx{\hat{\boldsymbol{\xi}}}
\def\bJ{\mathbf{J}}
\def\bD{\mathbf{D}}
\def\cX{\boldsymbol{\Xi}}
\def\bz{\bar z}

\def\det{\operatorname{det}}
\def\hz{\hat{Z}}
\def\hy{\hat{Y}}
\def\hx{\hat{\xi}}
\def\da{\dagger}
\def\la{\left(}
\def\ra{\right)}
\def\lb{\left\langle}
\def\rb{\right\rangle}
\def\d{\mathrm{d}}
\def\dd{\mathrm{d}^2}
\def\pp{\Phi^{\prime}}
\def\D{\mathcal{D}[\cdots]}
\def\p{\prime}
\def\t{\mathrm{T}}
\def\i{\mathrm{i}}
\def\bL{\boldsymbol{\Lambda}}
\def\otp{\overline{\hat{t}^\p}}
\def\inta{\lb\exp\la-\i\bs{\hz}^\da(\bs{\hx}  \bs{\pp}) \bs{z}-\i\bs{\hy}^\da\bs{\hx} \bs{y}-\i\bs{z}^\da(\bs{\hx } \bs{\pp})^\da \bs{\hz}-\i\bs{y}^\da\bs{\hx}^\da \bs{\hy}\ra\rb_{\bs{\hx}}}
\newcommand{\bs}[1]{\boldsymbol{#1}}

\begin{document}

\title{Spectrum of non-Hermitian deep-Hebbian neural networks}
\author{Zijian Jiang$^{1}$}
\thanks{Equal contribution.}
\author{Ziming Chen$^{2}$}
\thanks{Equal contribution.}
\author{Tianqi Hou$^{3}$}
\author{Haiping Huang$^{1}$}
\email{huanghp7@mail.sysu.edu.cn}
\affiliation{$^{1}$PMI Lab, School of Physics,
Sun Yat-sen University, Guangzhou 510275, People's Republic of China}
\affiliation{$^{2}$Department of Electronic Engineering, Tsinghua University, Beijing 100084, People's Republic of China}
\affiliation{$^{3}$Theory Lab, Central Research Institute, 2012 Labs, Huawei Technologies Co., Ltd., Hong Kong Science Park, People's Republic of China}
\date{\today}

\begin{abstract}
Neural networks with recurrent asymmetric couplings are important to understand how episodic memories are encoded in the brain.
Here, we integrate the experimental observation of wide synaptic integration window into our model of sequence retrieval in the 
continuous time dynamics. The model with non-normal neuron-interactions is theoretically studied by deriving a random matrix theory of the Jacobian matrix in neural dynamics.
The spectra bears several distinct features, such as breaking rotational symmetry about the origin, and the emergence of nested voids within the spectrum boundary.
The spectral density is thus highly non-uniformly distributed in the complex plane. The random matrix theory also predicts a transition to chaos.
In particular, the edge of chaos provides computational benefits for the sequential retrieval of memories. Our work provides a systematic study of time-lagged
correlations with arbitrary time delays, and thus can inspire future studies of a broad class of memory models, and even big data analysis of biological time series.

\end{abstract}

 \maketitle

\section{Introduction}
Neural sequence specifies a temporally structured dynamics of neuronal population activity, which is ubiquitous in functions of many brain regions,
such as hippocampal replay~\cite{Wilson-1994,Wilson-2009,Lee-2002} and choice-selective neural dynamics~\cite{Harvey-2012}. Generation of neural sequence
requires general circuit-level mechanisms, which can be simulated by artificial neural networks composed of recurrent units~\cite{Rajan-2016}.
In essence, the synaptic coupling between each pair of neurons can be constructed by a temporally asymmetric Hebbian (TAH) rule, which has been extensively studied 
in the physics community~\cite{Haim-1986,Mezard-1988,ACC-1998} and recently the computational neuroscience community~\cite{Plos-2012,Brunel-2020}. This rule reflects
the asymmetric connectivity ubiquitous in many brain regions~\cite{Luo-2021}, bringing the network activity out of equilibrium and the patterns in each sequence can be reactivated
according to their intrinsic temporal order. Therefore, the TAH rule plays an important role in understanding how neural sequences are stored and retrieved, even episodic memory (memory traces of past experiences)~\cite{Emem-2002}.

As a limitation, the standard TAH rule incorporates only nearest-neighbor pairs of patterns in a sequence of uncorrelated patterns, whereas, the synaptic integration leading to the plasticity
could occur in a wide range of temporal windows (e.g., synaptic modification in neocortex)~\cite{Abbott-2000,Tri-2006}. This observation motivates the inclusion of an arbitrary Hebbian interaction range
in the TAH rule, which is the main focus of this work. We remark that a symmetric version of synaptic integration considering
the long range Hebbian interaction has been recently studied, which revealed a significantly different phase diagram of the network collective behavior from the learning rule considering only the nearest-neighbor pair~\cite{Huang-2021a,Huang-2021b}.
In contrast, the extended TAH rule brings the network out of equilibrium, yielding rich modes of network dynamics. In this work, we characterize thoroughly the spectral density of the non-normal synaptic coupling matrix
and the associated Jacobian matrix of linearized dynamics, which can reveal different collective modes supporting the network function.
Moreover, transition to chaos and sequence memory are also explored and discussed. 

The spectral density of this new type of non-normal matrix adds an 
important category of asymptotic spectrum into the non-Hermitian random matrix theory, displaying an anisotropic distribution of eigenvalues in the complex plane. In this sense, our study could open a new avenue to study 
the neural sequence generation and processing in more biological plausible setting with the random matrix theory. In addition, the synaptic integration can be decomposed into contributions of time-lagged correlations among
neural patterns, which shares the common features observed in complex systems like price fluctuations in financial markets and 
physiological time series in biology~\cite{EPL-2010,Nowak-2017}. Therefore, our analysis may also provide insights toward understanding a broad range of 
collective phenomena.

\section{Model}
We study a recurrent neural network composed of $N$ interacting neurons. The neural dynamics obeys the following rule,
\begin{equation}
	\tau\frac{\d r_i}{\d t} = -r_i +\sum_{j=1}^N J_{ij}\Phi(r_j)
	\label{eq:dynamic}
\end{equation}
where $r_i$ is the synaptic current reflecting the neural activity of the neuron $i$,
$\tau$ indicates the time constant of the dynamics ($\tau=0.01s$ here, the time is in seconds in this paper), and $\Phi$ is the non-linear activation function transferring the synaptic current into the firing activity ($\Phi=\tanh$ here).
For simulations, one can add an external input $I(t)$ (e.g., an instantaneous stimulus in our simulations)
to the right-hand-side of the dynamics equation.
The synaptic coupling between any two neurons (say $i$ and $j$) is given by 
\begin{equation}
	J_{ij} = \frac{1}{N}\sum_{\mu=1}^P\la c\xi^{\mu}_{i}\xi^{\mu}_{j}+\gamma\sum_{r=1}^d\xi^{\mu}_i\xi^{\mu+r}_j\ra,
	\label{eq:J}
\end{equation}
which contains the concurrent Hebbian term and pattern-separated (with a most distant separation of $d$ patterns) Hebbian term. The strength of the concurrent Hebbian term is specified by
$c$, and the strength of the pattern-separated term is specified by $\gamma$. As $d$ takes an arbitrary number and thus represents the depth of Hebbian interaction, we call this model deep-Hebbian network.
	
In this model, we have $P$ patterns, and each pattern follows the Rademacher distribution $P(\xi_i^\mu)=\frac{1}{2}\delta(\xi_i^\mu-1)+\frac{1}{2}\delta(\xi_i^\mu+1)$, 
in which the superscript $\mu$ denotes the index of pattern and the subscript $i$ denotes the index of neuron. 
We are interested in the situation of large $P$ and $N$, with fixed memory load $\alpha=\frac P N$.
When the pattern entries follow a standard Gaussian distribution, the statistics, e.g., the first and second moments, does not change (see the following rotated patterns as well), and thus does not affect the asymptotic spectrum.
Note that these patterns form a cyclic sequence with periodic boundary, corresponding to an ordered stimulus sequence in animal experiments~\cite{Miya-1988a}.

The synaptic coupling in our model is clearly asymmetric and there thus exists an out-of-equilibrium collective behavior.
The coupling asymmetry is ubiquitous in neocortex~\cite{Luo-2021}, e.g., the mechanism of spike-timing-dependent plasticity and other higher-order variants support 
this asymmetry~\cite{Abbott-2000,Tri-2006}. The presence of the asymmetry in the connectivity can generate a steady flow of neural activity, related to the 
sequence storage and retrieval~\cite{Haim-1986,Mezard-1988,Brunel-2020}. As the synaptic integration can occur in a broad range of temporal windows, we incorporate
an arbitrary window size (indicated by $d$), yet violating the coupling symmetry. Setting $(c,\gamma)=(1,0)$, we recovers the standard Hopfield model~\cite{Hopfield-1982,Amari-1977} where the coupling is symmetric and
an equilibrium distribution of neural activity is guaranteed (or a Lyapunov function exists). Setting $(c,\gamma,d)=(0,1,1)$, we recovers the standard TAH model~\cite{Haim-1986}.
Therefore, our model setting is more general than any previous models of associative memory, in either equilibrium or non-equilibrium context.

The coupling matrix in Eq.~\eqref{eq:J} can be recast in a compact form as
\begin{equation}
	\bJ = \frac{1}{N}\bs{\xi}^{\t} \X \bs{\xi}
	\label{eq:J_abstract}
\end{equation}
where $\bs{\xi}$ is the $P\times N$ pattern matrix, and $\X$ is a $P\times P$ circulant matrix with entries
\begin{equation}
	X_{\mu\nu} = c\delta_{\mu\nu}+\gamma\sum_{r=1}^{d} \delta_{\mu,[(\nu+r)\mod P]}.
\end{equation}
The circulant property is due to our cyclic-sequence setting, i.e.  $\bs{\xi}^{P+1}=\bs{\xi}^1$.
For example, if $P=5,d=2$, $\X$ reads
\begin{equation}
\left[
\begin{matrix}
	c&0&0&\gamma&\gamma\\
	\gamma&c&0&0&\gamma\\
	\gamma&\gamma&c&0&0\\
	0&\gamma&\gamma&c&0\\
	0&0&\gamma&\gamma&c
\end{matrix}
\right],
\end{equation}
where the number of $\gamma$ in each row of the matrix is conserved, equal to $d$.
We emphasize that our method in the following analysis does
not rely on the explicit form of $\X$. Therefore, despite our interest in deep-Hebbian network,
other models of collective phenomena expressed through correlation matrices can also be analogously treated by the same approach. 

The function of the network is to store and retrieve $P$ $N$-dimensional patterns $\{\bs{\xi}^1,\ldots,\bs{\xi}^P\}$. 
When the network
is stimulated by a state correlated with one of the stored patterns, the hint will trigger the retrieval,
which is marked by a significant overlap between neuron activations and stored patterns.  
 The overlap between 
the $\mu$-th pattern and the neuron activity at time $t$ is defined as 
\begin{equation}
	m_{\mu}(t) =\frac{1}{N}\sum_{j}\xi_{{j}}^{\mu}\Phi(r_j(t)).
\end{equation}

\begin{figure}
	\centering
	\includegraphics[bb=3 1 940 379,width=1.0\linewidth]{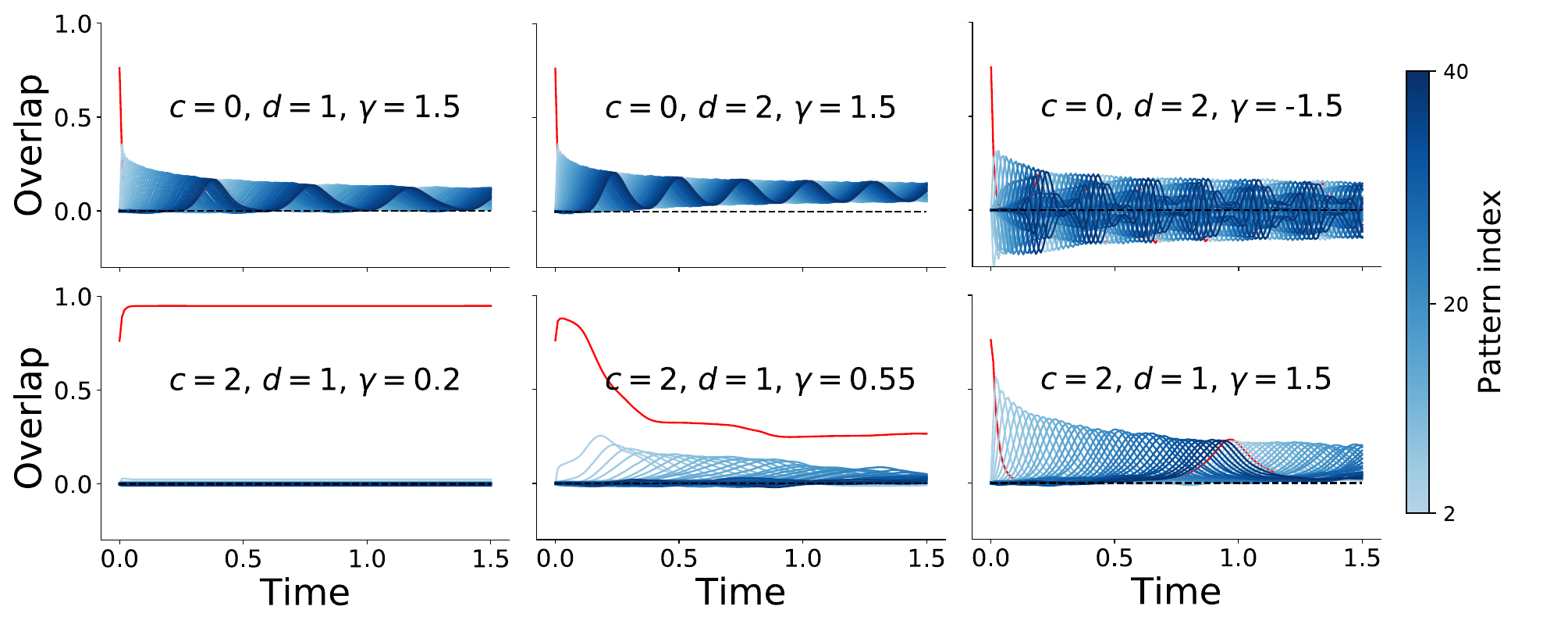}
	\caption{Different retrieval profiles. The overlaps are plotted against time $t$, and overlaps with different patterns are
	distinguished by their colors (see the color bar). In the simulation, we set $N=4000$ and $\alpha=0.01$, 
	and therefore there are in total forty patterns stored in the network. To trigger a retrieval (either one pattern or pattern sequence),
	we set the external input current
	as $\bs{I}(t)=\delta(t)\bs{\xi}^1$. The overlap with the first pattern $m_1(t)$ is highlighted in red. 
	The retrieval is static (bottom left panel) when $c$ dominates, and dynamical (top left panel and right panel) when $\gamma$ dominates. 
	A mixture of these two types can be found in the intermediate parameter regime (middle panel). 
	}
	\label{recall}
\end{figure}

\begin{figure}
\centering
\includegraphics[bb=5 11 803 515,width=0.9\linewidth]{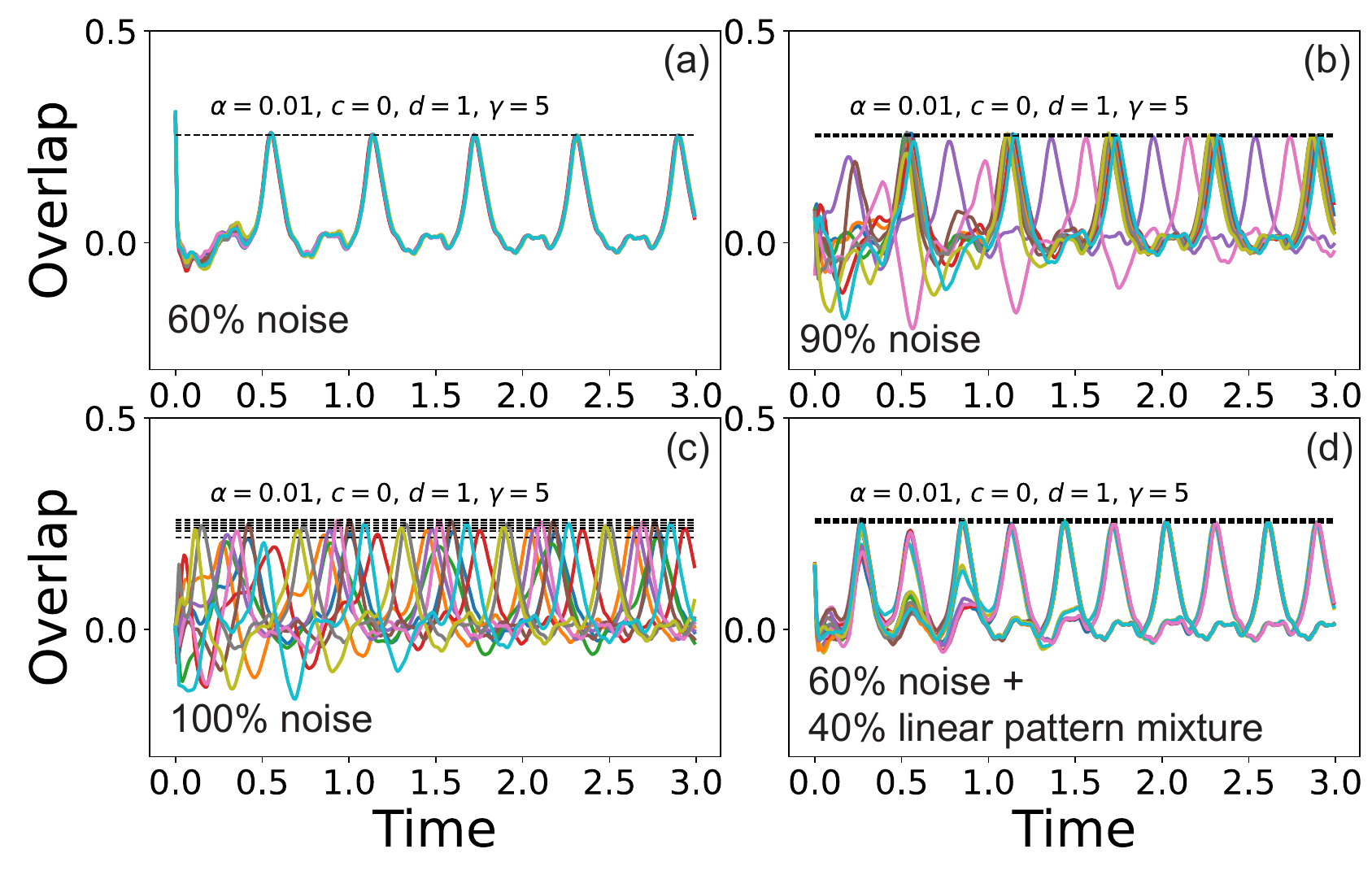}
\caption{The sequence replay behavior with noisy initialization. 
We show ten instances of the overlap profile initialized with the first pattern
in different noise realizations in each subfigure (solid color lines).
Each dashed black line shows the peak overlap amplitude. The overlap is calculated with the first pattern. The behavior of the overlap with other patterns is similar.
The connectivity is fixed ($N=6000$). All mirrored overlap profiles caused by the $\mathbb{Z}_2$ symmetry of the dynamics are flipped to their positive counterparts.
After a transient relaxation, a stable dynamics profile is observed.
(a) The initialization is $\mathbf{r}(0) = 0.4\boldsymbol{\xi}^1+0.6\mathbf{z}$, where $\mathbf{z}$ is i.i.d. standard Gaussian noise. 
The dynamics will fall into the same limit cycle. (b) The initialization 
is $\mathbf{r}(0) = 0.1\boldsymbol{\xi}^1+0.9\mathbf{z}$. Different noisy initializations lead to different limit cycles yet with the same peak overlap magnitude.
(c) The initialization is $\mathbf{r}(0) = \mathbf{z}$. This completely random 
initialization will also lead to the cyclic behavior, but displaying diverse peak overlap magnitudes with slightly small variance.
(d) The initialization is $\mathbf{r}(0) = 0.2\boldsymbol{\xi}^1+ 0.2\boldsymbol{\xi}^{30}+0.6\mathbf{z}$. In this case, 
the phase of the dynamics displays two types of cyclic behavior.} 
\label{basin}
\end{figure}

The retrieval behavior differs in different parameter regimes. In this paper, the memory overlap is averaged over thirty realizations of the pattern distribution (unless stated otherwise).
The behavior falls within two types---static and dynamical recall. In a static recall, the network activity gets
stuck in a stable fixed point correlated with some pattern (but not others), and thus $m_{\mu}(t)$ becomes stationary over time.
But in a dynamical recall, the activity leads to sequential memories of stored patterns (Fig.~\ref{recall}). We remark that due to the nature of our model
(see also previous works in the equilibrium counterpart~\cite{Huang-2021a} or the out-of-equilibrium case~\cite{Brunel-2020}),
the peak overlap magnitude can not achieve one (but a non-zero value), which we call sequence memory (see also the theoretical work~\cite{ACC-1998} which used 
the condition of vanishing overlap to determine the storage capacity in a simpler scenario). In Fig.~\ref{recall}, we also observe that 
the key model parameter $d$ tunes the sequence-replay period, which is very interesting for future neurophysiology experiment tests.
We also study how robust the recall behavior is when the network state initialization varies, which is shown in Fig.~\ref{basin}.
We conclude that as long as the memory load is small, the recall behavior is quite robust from different noisy initializations, and the large deep-Hebbian strength is able 
to maintain a relatively high magnitude of peak overlap (see also Fig.~\ref{fig:retrieval} for detailed analysis). We will analyze
the relationship between network structure and function in the remaining sections. 

Compared to other architectures (e.g., feedforward or symmetric networks), our rate model of recurrent networks shows richer properties of network dynamics only via simple deep-Hebbian construction (see next sections),
thereby being able to show sequence replay characteristics of cortical and hippocampus circuits, and moreover,
the introduced Hebbian depth $d$ can tune the replay period, which was not reported before, and in mathematics, higher values of $d$ (e.g., $d>1$) lead to an anisotropic spectrum of the connectivity matrix, which can be captured by our theory.
In addition, due to the pattern-separated Hebbian term, the network activity during replay can smoothly evolve in time and becomes transiently correlated in order with each of the patterns in the sequence.
This is a nice property in our recurrent rate network that helps the network activity to explore a broad dynamics regime.
The theoretical study in this paper would thus hopefully inspire future algorithmic designs in temporal sequence processing with our proposed recurrent rate networks, which we will demonstrate in forthcoming future works.
We next analyze the relationship between the recall dynamics and intrinsic property of the network structure.

\section{Linearized recurrent dynamics}
In this section, we first study the Jacobian spectrum of fixed points and transient states of low speed. The spectrum is then connected to the spectrum of the 
original non-normal coupling matrix ensemble.

To explore the stability of an arbitrary point in the phase space, including fixed points or transient states, we linearize the dynamics equation [Eq.~\eqref{eq:dynamic}]
as follows,
\begin{equation}
	\tau \frac{d\delta r_i}{dt} =\sum_{j=1}^{N} D_{ij}\delta r_j,
	\label{eq:linear}
\end{equation}
where $\delta r_i\equiv r_i - r^*_i$ denotes the displacement from the operating 
point $\mathbf{r}^*=\la r^*_1,\cdots,r^*_N\ra$, and $D_{ij} = -\delta_{ij}+J_{ij}\Phi^{\p}(r^*_j)$ specifies
the Jacobian matrix, in which $\Phi^{\p}(r^*_j)$
denotes the derivative $\left.\frac{\d \Phi(a)}{\d a}\right|_{a=r^*_j}$. Considering that the
operating point is a fixed point or a low-speed point, the linearized dynamics [Eq. \eqref{eq:linear}] can be solved as follows, 
\begin{equation}
	\delta \mathbf{r}(t) = \exp(\bD t/\tau)\delta \mathbf{r}(0) =\mathbf{R}\exp(\bs{\Lambda}t/\tau)\mathbf{L}^\da\delta \mathbf{r}(0) ,
\end{equation}
where we use the spectral decomposition $\bD = \mathbf{R} \bs{\Lambda} \mathbf{L}^\da$, where $\dagger$ denotes the conjugate
transpose operation. The instantaneous Jacobian matrix captures the behavior of the linearized dynamics around the operating point. The real-part eigenvalues of
the matrix determines the timescales of growth or decay of perturbation, and thus determines the local stability of the dynamics, while the imaginary parts control the oscillation frequency of the neural dynamics.
Through studying the spectrum of this
matrix, we can identify the key network parameters underlying the macroscopic behavior of the deep-Hebbian neural networks.

\subsection{Spectrum of Jacobian matrix}
We first consider the shifted Jacobian $\tilde{D}_{ij}\equiv D_{ij}+\delta_{ij}=J_{ij}\pp(r^*_j)$, for which the shifted Jacobian shares the same eigenvectors with the original Jacobian, but 
the spectrum has a translation along the real axis. 
By inserting Eq.~(\ref{eq:J_abstract}) into the shifted Jacobian and performing
the diagonalization $\X = \mathbf{U}^\da \bs{\Lambda} \mathbf{U}$, where $\dagger$ means the Hermitian conjugate operation
(the circulant matrix $\X$ can be diagonalized through a unitary transformation), we arrive at a simplified formula
\begin{equation}
	\tilde{\bD} = \frac{1}{N} \bs{\hx}^\da \bs{\Lambda} \bs{\hx} \bs{\pp},
	\label{eq:td}
\end{equation}
where $\bs{\pp}=\mbox{diag}\left[\pp(r^*_1),\cdots,\pp(r^*_N)\right]$, 
and $\bs{\hx} = \mathbf{U}\bs{\xi}$ denotes the rotated pattern. It is easy to check
the first two moments in statistics of the rotated patterns are respectively
$\langle \hx_i^{\mu}\rangle=0$, $\langle \overline{\hx_i^{\mu}} \hx_j^\nu\rangle=\delta_{ij}\delta_{\mu\nu}$,
which can be fully described by a standard complex Gaussian distribution in the large $N$ limit. 
Here $\bar{\cdot}$ denotes the complex conjugate operation, and $\langle \cdot\rangle$ denotes the disorder average over random instances of i.i.d. patterns, 
or random instances of the shifted matrix $\tilde{\bD}$ (given the operating point).

The spectral density is defined as the probability density $\tilde{\rho}(w)$ of finding an eigenvalue which coincides with a point $w$ 
on the complex plane. In the large $N$ limit, we assume that the self-averaging property holds for 
the density, i.e., the density of any single instance matches the asymptotic density after the matrix ensemble average is performed. 
The spectral density is intuitively defined as follows,
\begin{equation}
	\tilde{\rho}(w)=\frac{1}{N}\lb\sum_{i=1}^N \delta^{(2)}(w-\tilde{\lambda}_i)\rb,
	\label{eq:rho1}
\end{equation}
where $\tilde{\lambda}_{i}$ denotes the $i$-th eigenvalue of $\tilde{\bD}$, and $\delta^{(2)}(\cdot)$ defines the two 
dimensional Dirac delta function. In fact, the coupling matrix $\mathbf{J}$ or the shifted Jacobian is a non-Hermitian matrix, and thus the associated eigenvalues are distributed
on the complex plane (i.e., 2D space with two axes, one for real part and the other for imaginary part).
The spectral density of the Jacobian is recovered by the horizontal translation in the complex plane as
${\rho}(w)=\tilde{\rho}(w+1)$.
The standard procedure to calculate the density requires the computation of the eigenvalue potential $\phi$ as follows,
\begin{equation}
	\tilde{\rho}=-\frac{1}{4\pi}\nabla^2\phi,
\end{equation}
which is exactly the Poisson equation in a two-dimensional electrostatic problem~\cite{Stein-1988}. Therefore, the spectral density is equivalent to a two-dimensional charge distribution, and 
$\phi$ becomes an electrostatic potential, which is also called the eigenvalue potential in the random matrix theory.
A brief introduction of this mapping is given in Appendix~\ref{NHRMT}.

The real-valued potential can be written in a free-energy form as~\cite{Stein-1988}
\begin{equation}
	\phi(w) = -\frac{1}{N}\lim_{\epsilon\to 0^+}\lb\ln Z(w,\epsilon)\rb,
	\label{eq:phi_hermite}
\end{equation}
where the partition function $Z(w,\epsilon)=\operatorname{det}\left[\left| w\id-\tilde{\bD}\right|^2+|\epsilon|^2\id\right]$ in which $\id$ is an $N\times N$ identity matrix.
The additional infinitesimal variable $\epsilon$ ensures the matrix inside the determinant 
to be positive definite even when $w$ hits an eigenvalue of $\tilde{\bD}$. The regularization for avoiding the singularity can also be understood from
the mathematical definition of the Dirac delta function $\pi \delta^{(2)}\left(w-\tilde{\lambda}_i\right)=\lim _{|\epsilon|^2 \rightarrow 0} \frac{|\epsilon|^{2}}{\left(\left|w-\tilde{\lambda}_{i}\right|^{2}+|\epsilon|^{2}\right)^{2}}$.
In the electrostatic mapping, it is also convenient to introduce the electric field that is the very Green's function:
\begin{equation}
G(w) =-\frac{\partial \phi}{\partial w}=\frac{1}{N}\lb\mbox{Tr}\left[\la w\id-\tilde{\bD}\ra^{-1}\right]\rb,
\label{eq:green}
\end{equation} 
which is also called the resolvent in the random matrix theory. Therefore, the density can be estimated from $G$ as follows (see Appendix~\ref{NHRMT}),
\begin{equation}
	\tilde{\rho} = \frac{1}{\pi} \frac{\partial G}{\partial\bar w},
	\label{eq:rho2}
\end{equation} 
where the complex-valued $G$ is a function of $w$ and $\bar w$. 

A standard route to treat the disorder average in Eq.~\eqref{eq:phi_hermite} is the replica method, i.e.,
$\lb \ln Z\rb = \lim_{n\to 0} \frac{\lb Z^n\rb}{n}$ where $n$ copies of the original system are introduced,
and finally an analytical continuation $n\to 0$ is performed. However, the replica analysis for the current setting becomes complicated.
Instead, we adopt an annealed approximation, i.e., we assume that $\lb \ln Z^{-1}(w,\epsilon)\rb = \ln \lb Z^{-1}(w,\epsilon)\rb$, which nevertheless leads to accurate results
confirmed by an alternative Feynman diagram method (see Appendix~\ref{fey}) and numerical simulations.
A rigorous proof would be very interesting in future studies.

Next, we compute $\ln \lb Z^{-1}(w,\epsilon)\rb$. By using a multivariate complex Gaussian integral representation of
the determinant and the Hubbard-Stratonovich transformation, we can write 
\begin{equation}
	\begin{aligned}
		\phi&=-\frac{1}{N}\lb\ln\operatorname{det}\left[(\bz\id_N-\bJ^\da)(z\id_N-\bJ)+|\epsilon|^2\id\right]\rb\\
		& \propto  \frac{1}{N} \ln\left[\lim_{|\epsilon|\to 0} \int \d^{2N}\bs{z} \d^{2N} \bs{y} \exp\la -|\epsilon|^2 \bs{z}^\da \bs{z}-\bs{y}^\da \bs{y} -\i \bar w\bs{z}^\da  \bs{y}-\i w\bs{y}^\da  \bs{z}\ra \lb \exp \la \i \bs{z}^\da {\tilde{\bD}}^\da \bs{y} + \i \bs{y}^\da \tilde{\bD} \bs{z} \ra\rb\right],
		\label{eq:1HS}
	\end{aligned}
\end{equation}
where $\propto$ means up to an irrelevant factor that does not contribute to the leading order of $\phi$ in the large $N$ limit.
To carry out the disorder average in Eq.~\eqref{eq:1HS}, the local chaos hypothesis is used~\cite{LCH-1995}, which states that
large random neural networks is able to reach a steady-state where the network state is independent of the random coupling matrices.
In fact, the operating point is fixed in the current context, and thus 
the disorder average can be even done without the local chaos hypothesis, unless the steady-state distribution of network activity are taken into account.
To proceed, we have to introduce the following order parameters together with their hatted conjugate parameters, introduced through
applying the Fourier representation of the delta function,
\begin{equation}
	\begin{aligned}
		&u\equiv\frac{1}{N}\bs{z}^\da \bs{z}, &u^\prime\equiv\frac{1}{N}(\bs{\pp}\bs{z})^\da (\bs{\pp}\bs{z}), \\
		&t\equiv \frac{1}{N}\bs{y}^\da \bs{z}, &t^\prime\equiv \frac{1}{N}\bs{y}^\da (\bs{\pp}\bs{z}),\\
		&\bar t \equiv \frac{1}{N}\bs{z}^\da \bs{y}, &\bar{t^\prime} \equiv \frac{1}{N}(\bs{\pp}\bs{z})^\da \bs{y},\\
		&v\equiv \frac{1}{N}\bs{y}^\da \bs{y},\\
	\end{aligned}
\end{equation}
in which $u,v,u^\p$ are real positive parameters, and $t,t^\p$ are complex parameters. These order parameters have their own physics interpretations,
e.g., $\i t$ is exactly the Green's function, while $uv$ gives the eigenvector overlap function (see Appendix~\ref{app-d}).

After a lengthy calculation (see Appendix~\ref{app-c}), the potential can be recast into the following concise form,
\begin{equation}
	\phi \propto \frac{1}{N}\ln\left[\lim_{|\epsilon|^2\to 0} \int \mathcal{D}[\cdots] \exp\left[ N\la\phi_1+\phi_2+\phi_3\ra\right]\right],
	\label{eq:saddle_form}
\end{equation}
where $\int \mathcal{D}[\cdots]$ denotes the integration over all relevant order parameters and their hatted conjugated variables, and the sum of $\phi_1$, $\phi_2$ and $\phi_3$
is called the action in physics. More precisely,
\begin{equation}
	\begin{aligned}
		\phi_1 =& -|\epsilon|^2 u-v +\i\hat{u}u+\i\hat{v}v+\i\bar{\hat{t}} t+\i \bar{t}\hat{t}+ \i\hat{u}^\p u^\p+\i\otp t^\p+\i\bar{t^\p}\hat{t}^\p, \\
		\phi_2 =& -\frac{1}{N}\sum_{i=1}^N\ln\left[\bar{\hat{t}}\hat{t}-\hat{u}\hat{v}+(\pp_{ii})^2(\otp\hat{t}^\prime-\hat{u}^\prime\hat{v})+\pp_{ii}(\otp \hat{t}+\bar{\hat{t}}\hat{t}^\prime)\right],\\
		\phi_3 =&  -\alpha  \ln(u^\p v-\bar{t^\p} t^\p) -\i\bar w\bar t-\i wt \\
		&-\frac{\alpha}{P}\sum_{\mu=1}^P\ln\la k-ik\Lambda_{\mu}t^\p-\i k\overline{\Lambda_{\mu}}\bar{t^\p}+\overline{\Lambda_{\mu}}\Lambda_{\mu}\ra,
		\label{eq:phi123}
	\end{aligned}
\end{equation}
where $k = \left[u^\p v-\bar{t^\p} t^\p\right]^{-1}$. By applying the Laplace method in the large $N$ limit, $\phi$ is just the sum of the above three terms.
All relevant order parameters obey saddle-point equations whose solutions maximizing the action. Technical details for deriving the saddle point equations are given in Appendix~\ref{app-d}.

The non-normal matrix $\tilde{\bD}$ can be decomposed by using its right and left eigenvectors. The eigenvector overlap function captured by $\frac{uv}{\pi}$ determines the spectrum boundary (see Appendix~\ref{app-d}), i.e.,
$uv=0$ outside the spectrum, and $uv>0$ within the spectrum. Therefore, the value of $uv$ is related to whether the Green's function is analytic (the Cauchy-Riemann condition is satisfied) or not.

Taking the limit $|\epsilon|\to0$, the spectral density can be estimated by solving the following closed-form equations:
\begin{subequations}
	\label{density_equations}
	\begin{align}
		&C \frac{\alpha}{P}\sum_{\mu=1}^P \frac{|\Lambda_\mu|^2}{|\Lambda_{\mu}T-1|^2+|\Lambda_\mu|^2 u^\prime v} = 1,\\
		&C\frac{\alpha}{P}\sum_{\mu=1}^P \frac{\Lambda_\mu}{|\Lambda_{\mu}T-1|^2+|\Lambda_\mu|^2 u^\prime v}= Bw,
	\end{align} 
\end{subequations}
where $T\equiv \i t^\prime$ and $\Lambda_{\mu}$ is the eigenvalues of $\X$, which relies on the model parameters as follows,
\begin{equation}
\Lambda_{\mu} = c+\gamma \sum_{r=1}^d \exp(-2\pi \i r\mu/P).
\end{equation}
Details of the derivation are given in Appendix~\ref{app-d}.
The neural-state dependent auxiliary quantities are specified as follows,
\begin{subequations}
	\label{density_equation_rely}
	\begin{align}
		A(w,\hat{t}^\prime,\hat{u}^\prime \hat{v})& = \frac{1}{N}\sum_{i=1}^N\frac{1}{|\Phi_{ii}^\p\hat{t}^\p+\bar w|^2-(\Phi_{ii}^\p)^2 \hat{u}^\p \hat{v}},\\
		B(w,\hat{t}^\prime,\hat{u}^\prime \hat{v})& = \frac{1}{N}\sum_{i=1}^N\frac{\pp_{ii}}{|\Phi_{ii}^\p\hat{t}^\p+\bar w|^2-(\Phi_{ii}^\p)^2 \hat{u}^\p \hat{v}},\\
		C(w,\hat{t}^\prime,\hat{u}^\prime \hat{v})& = \frac{1}{N}\sum_{i=1}^N\frac{(\pp_{ii})^2}{|(\Phi_{ii}^\p\hat{t}^\p+\bar w)|^2-(\Phi_{ii}^\p)^2 \hat{u}^\p \hat{v}},
	\end{align}
\end{subequations}
where $\pp_{ii}$ is determined by the choice of activation function and the operating point.
Although $A$ does not appear in Eq.~\eqref{density_equations}, it is useful for transforming one physics variable to another one for the sake of the following analysis.
Note that the order parameters $T$ and $u^\prime v$ can be expressed by
$w,\hat{t}^\prime,\hat{u}^\prime \hat{v}$ via the saddle-point condition,
\begin{subequations}
	\label{density_equation_rely2}
	\begin{align}
		T = B\bar w+C\hat{t}^\prime,\\
		u^\prime v = -C^2 \hat{u}^\prime\hat{v}.
	\end{align}
\end{subequations}

Considering Eq.~\eqref{density_equation_rely2} and Eq.~\eqref{density_equation_rely}, one can
immediately find the Eq. \eqref{density_equations} is actually a closed-form equation
of $w,\hat{t}^\prime$, and $\hat{u}^\prime \hat{v}$. We then determine the boundary of the spectrum. 
By setting $uv\to 0$, 
we have the following boundary equations (see details in Appendix~\ref{app-d})
\begin{subequations}
	\label{boundary_equations}
	\begin{align}
		&C \frac{\alpha}{P}\sum_{\mu=1}^P \frac{|\Lambda_\mu|^2}{|\Lambda_{\mu}\la B\bar w+C\hat{t}^\prime\ra-1|^2} = 1,\\
		&C\frac{\alpha}{P}\sum_{\mu=1}^P \frac{\Lambda_\mu}{|\Lambda_{\mu}\la B\bar w+C\hat{t}^\prime\ra-1|^2}= Bw.
	\end{align} 
\end{subequations}
After solving Eq.~\eqref{density_equations} in terms of $\hat{t}^\p$, $\hat{u}^\prime$ and $\hat{v}$, one can recover the Green's
function $G$ and $uv$ through the saddle-point condition, i.e.,
\begin{subequations}
	\label{density_equation_rely3}
	\begin{align}
		G = A\bar w+B\hat{t}^\prime,\\
		u v = -AC \hat{u}^\prime\hat{v}.
	\end{align}
\end{subequations}
We conclude that $G$ leads to the density by using Eq.~\eqref{eq:rho2}, 
and $uv$ reports the eigenvector overlap function by $O(w,\bar w) = uv/\pi$ (see the derivation in Appendix~\ref{app-d}).


\subsection{Spectrum of synaptic coupling matrix}
The Jacobian spectrum can be greatly simplified by assuming $\bs{\pp}$ is an identity matrix. In this case, 
the spectrum of the Jacobian reduces to the spectrum of the synaptic coupling matrix.
In this simplification, we have $A=B=C$. Hence, Eq.~\eqref{density_equations} reduces to
\begin{subequations}
	\label{density_equations_connectivity}
	\begin{align}
		\frac{\alpha}{P}\sum_{\mu=1}^P \frac{|\Lambda_{\mu}G|^2+|\Lambda_{\mu}|^2uv}{|\Lambda_{\mu}G-1|+|\Lambda_\mu|^2 uv}=1,\\
		\frac{\alpha}{P}\sum_{\mu=1}^P \frac{\Lambda_{\mu}}{|\Lambda_{\mu}G-1|+|\Lambda_\mu|^2 uv}=w.
	\end{align}
\end{subequations}
This result coincides with that derived by a diagrammatic method (see Appendix~\ref{fey}), which confirms our annealed computation in turn.
The spectrum boundary obeys the following equations,
\begin{subequations}
\label{boundary_equations_connectivity}
	\begin{align}
		\frac{\alpha}{P}\sum_{\mu=1}^P \frac{|\Lambda_{\mu}G|^2}{|\Lambda_{\mu}G-1|}=1,\\
		\frac{\alpha}{P}\sum_{\mu=1}^P \frac{\Lambda_{\mu}}{|\Lambda_{\mu}G-1|}=w.
	\end{align}
\end{subequations}

We remark that only the case of $c=0,d=1$ can have an analytic closed-form expression of the spectrum.
The spectra of both the connectivity and Jacobian matrices bear disk-like or annulus-like shapes~\cite{Ring-1997,Zee-2001}. Technical details to obtain the 
closed form results for the connectivity are given in Appendix~\ref{SRa}.  
Our work 
extends this single ring law to the Jacobian matrix (see details in Appendix~\ref{SRa}), as the outer boundary of the
spectrum is given by
\begin{equation}
 R_{\rm{out}} = \gamma\sqrt{\alpha\lb (\pp)^2\rb_r+\lb \pp\rb_r^2},
 \label{eq:outring}
\end{equation}
where $\lb \pp\rb_r\equiv \frac{1}{N}\sum_{i=1}^N \pp_{ii}$ and $\lb (\pp)^2\rb_r\equiv \frac{1}{N}\sum_{i=1}^N (\pp_{ii})^2$
denote the empirical moments of the diagonal elements of $\bs{\pp}$. The subscript $r$ means that the moments can be computed by using the distribution of synaptic currents. 

\begin{figure}[htbp]
	\centering
	\includegraphics[bb=4 5 775 553,width=0.9\linewidth]{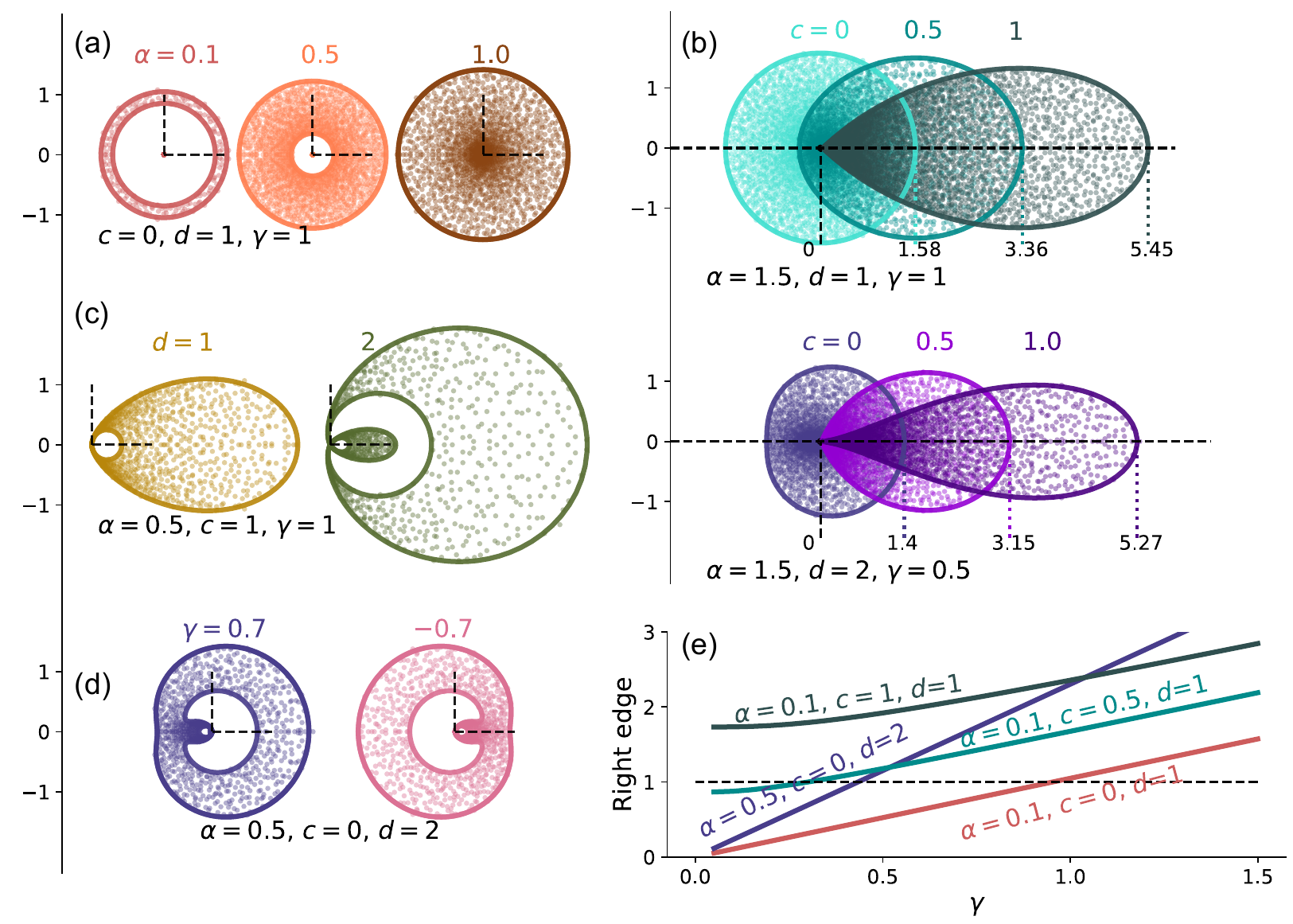}
	\caption{The spectrum of the connectivity matrix.
	(a)-(d): Spectrum of the connectivity with different model parameters.
	Colored dots denote eigenvalues of the connectivity generated by the numerical
	diagonalization ($N=4000$), and colored lines are the spectrum boundary predicted by our theory [see
	Eq.~\eqref{boundary_equations_connectivity}]. The perpendicular dashed lines in (a), (c), and (d)
	mark the origin and the unit length in real and imaginary axises of the spectra. (e)
	The right edge of the spectrum, i.e., the maximum real part of the eigenvalues, plotted against $\gamma$.   }
	\label{fig:parameter_change}
\end{figure}
\section{Spectrum, phase diagram, chaos and sequence retrieval}
In this section, we investigate the effects of model parameters (i.e., $\alpha$, $c$, $d$ and $\gamma$) on the spectrum of the connectivity matrix,
and moreover the Jacobian matrix, which determines the stability and time scales of the linearized dynamics.

The loading rate $\alpha$ constrains the rank
 of the connectivity matrix, i.e., $\mbox{rank}(\mathbf{J})=\min(N,\alpha N)$, whose derivation uses elementary (e.g., Sylvester) rank inequality. Therefore, 
 if $\alpha<1$, there appear exactly  $(1-\alpha)N$ eigenvalues localized at the origin
 [Fig.~\ref{fig:parameter_change} (a)]. In addition, the spectrum is affected in the other two manners. 
 (i) When $\alpha<1$, there emerge voids within the spectrum. The number of the voids depends 
 on the model parameters, especially the Hebbian length $d$. As $\alpha$ decreases, the voids are enlarged,
 and all non-zero eigenvalues finally collapse onto the outer spectrum-boundary 
 as $\alpha$ gets close (but not equal) to zero. (ii) The spectrum becomes significantly extended in both
 horizontal and longitudinal directions when $\alpha$ gets larger, roughly grows as $\sqrt{\alpha}$ when $\alpha$ is large.

 \begin{figure}[htbp]
	\centering
	\includegraphics[bb=2 2 810 666,width=0.9\linewidth]{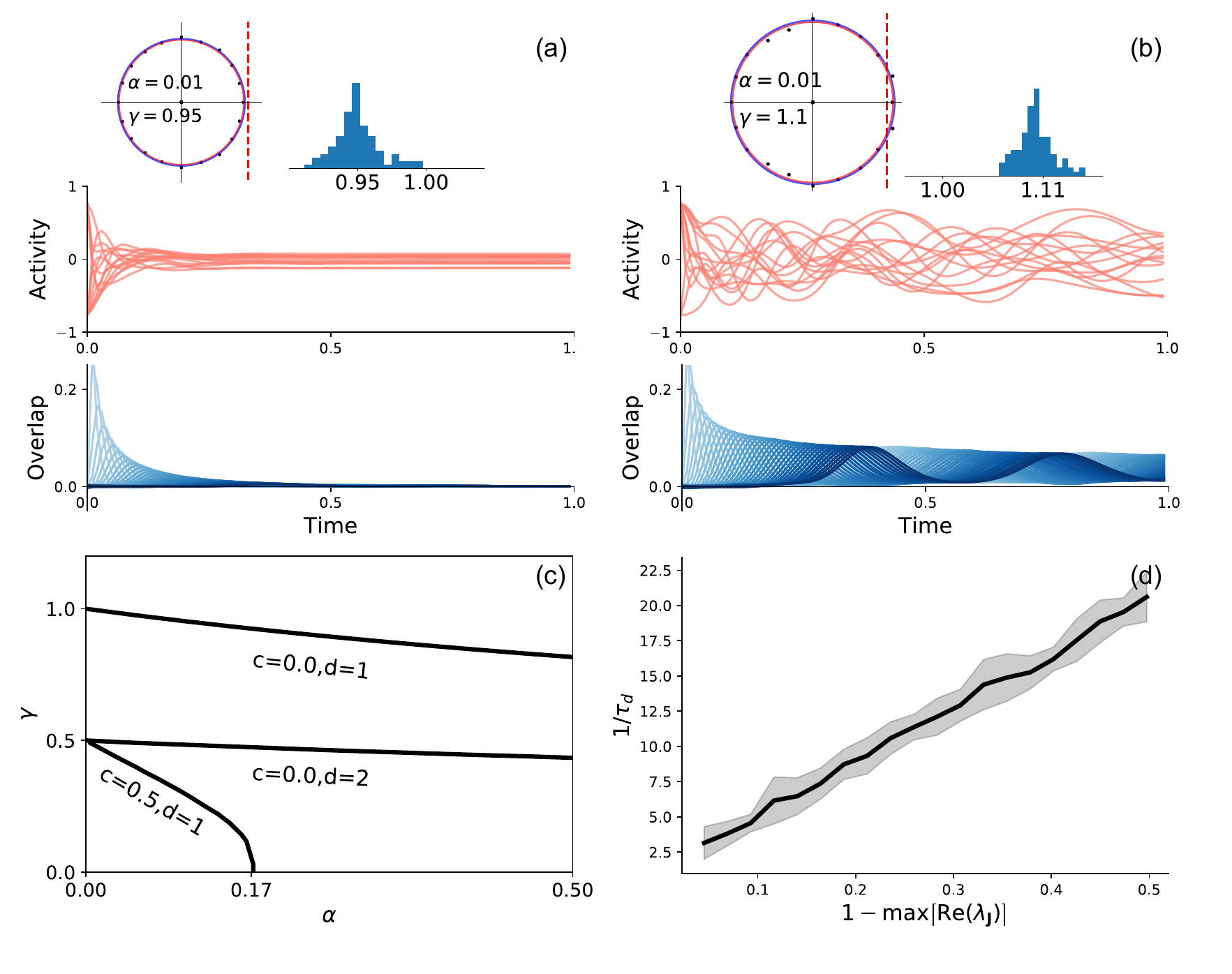}
	\caption{Spectrum of the connectivity matrix determines the linear stability and time scales of network dynamics around the null fixed point. 
	Simulations are carried out on networks of size $N=4000$.
	(a-b) Right edge of the spectrum determining the linear stability. Solid circles in
	each plot denote the spectrum boundary predicted by Eq.~\eqref{boundary_equations_connectivity}, while the black dots 
	are numerical eigenvalues, and the red dashed lines indicates the critical line $\mbox{Re}(\lambda)=1$. The histograms
	show the distribution of the right edge across 100 random realizations of the connectivity. The bottom sub-figure
	shows the dynamics of network initialized with the first stored pattern. Ten representative activity trajectories are
	shown, and the overlap $m_{\mu}(t)$ with stored patterns in sequence are distinguished by
	colors from light blue to dark blue. (c) The phase boundary
	shown by the lines delimits the linear stable phase (under each line) and unstable phase (above each line)
	under different model parameters. The stability here only describes the null (zero-activity) fixed point. (d) When the system is stable at the null fixed point, the inverse overlap-decay-time-scale
	is proportional to the distance between the right edge and the critical line, i.e.,
	$1-\max[\operatorname{Re}(\lambda_{\mathbf{J}})]$. The time scale of the overlap decay is numerically
	calculated as $\mbox{argmin}_t (\sum_{\mu} m_{\mu}(t)<0.1\sum_{\mu} m_{\mu}(0))$. Other model parameters are $c=0$ and $d=1$.}
	\label{fig:connectivity_dynamic}
\end{figure}

 We then study the effects of the Hebbian length $d$. The number of the voids coincides with the value of $d$. 
For $d=2$ [Fig.~\ref{fig:parameter_change} (c)], a small-sized void is embedded
in a larger void, and the small one is much smaller in area compared to the large 
one. As $d>2$, extra voids emerge in the spectrum, but their areas would become much smaller. 
 
 In contrast to Fig.~\ref{fig:parameter_change} (a), the isotropic property is broken in Fig.~\ref{fig:parameter_change} (b), due to the non-zero values of $c$.
 The role played by $c$ is to stretch the spectrum to the positive horizontal direction, 
shaping an anisotropic spectrum. When $c\gg\gamma$, the overall spectrum will collapse to the horizontal axis.
If we multiply the same factor to $c$ and $\gamma$ at the same time, the spectrum is just 
scaled with that factor. Therefore it is the ratio of $c$ and $\gamma$ that 
determines the shape of the spectrum, instead of their respective specific values. The role played by 
$\gamma$ competes with the role of $c$, suppressing the horizontal stretch.  As a result, the 
spectrum shows a distinct shape in Fig.~\ref{fig:parameter_change} (d). Figure \ref{fig:parameter_change}(e) 
shows how the right edge of the spectrum, the maximum real part of the eigenvalues, is modulated
by the model parameters. An approximately linear increase of the right edge with $\gamma$ when $\gamma$ is large, and the slope is
related to the value of $d$. 

In essence, there exist two kinds of symmetry/symmetry-breaking in the spectrum. 
When $d=1$ and $c=0$, the whole spectrum is rotationally symmetric [Fig.~\ref{fig:parameter_change}(a)],
and both inner and outer boundaries form an annulus. For this special case, the radius of the inner annulus is given by
$\gamma(1-\alpha)^{3/2}$, while the radius of the outer annulus is given by $\gamma(1+\alpha)^{1/2}$ (see Appendix~\ref{SRa}).
However, 
changing $c$ or $d$ will break this rotational symmetry. On the other hand,
the gauge symmetry is kept when $d=1$, i.e., the spectrum is invariant under $\gamma\to -\gamma$. 
Because $\Lambda_{\mu} = c+\gamma \exp(-2\pi\mathrm{i}\mu/P)$, it is easy to see any function 
involving the summation of all $\Lambda_{\mu}$, especially Eq.~\eqref{boundary_equations_connectivity}, 
is invariant under the transformation. Nevertheless, once $d>1$, this symmetry 
would be broken [Fig.~\ref{fig:parameter_change} (d)], which reshapes the memory retrieval profile as well (Fig.~\ref{recall}).

\begin{figure}[htbp]
	\centering
	\includegraphics[bb=2 2 900 666,width=0.9\linewidth]{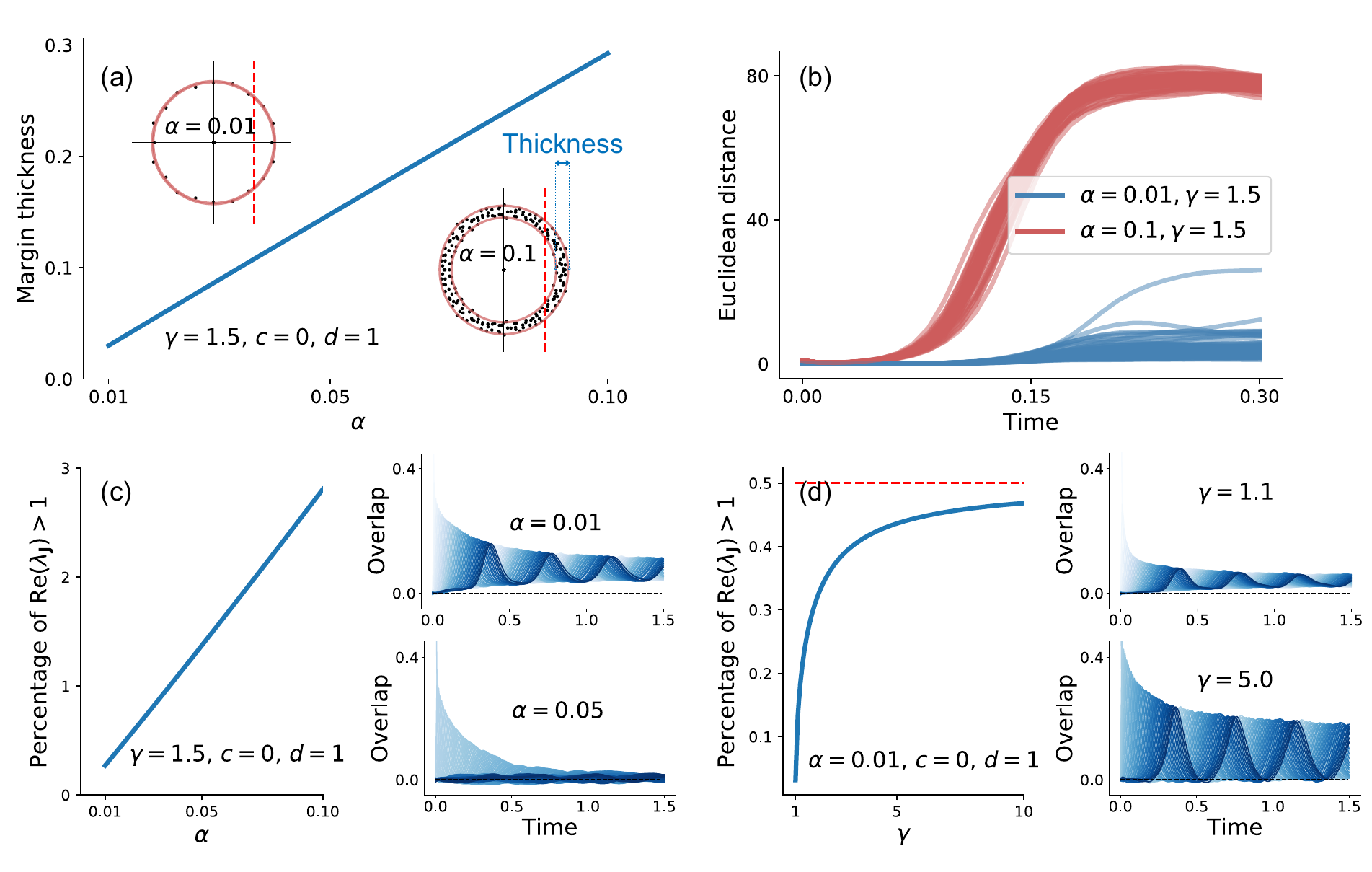}
	\caption{Effects of the connectivity spectrum on the sequence retrieval behavior. Simulations are carried out on networks of size $N=4000$.
	(a) The margin thickness of the spectrum (indicated by the blue double arrow)
	plotted against $\alpha$.
	(b) Evolution of the Euclidean distance between two initially-nearby trajectories . 
	All trajectories are initialized near the null fixed point. Different lines with 
	the same color indicate different initializations. (c,d) The number of eigenvalues (real part) exceeding one versus $\alpha$ and $\gamma$, respectively.
	The right sub-figures show the sequence retrieval behavior. The overlaps with patterns in sequence are
distinguished by colors from light blue to dark blue. The red dashed line in (d) shows the
	upper bound $0.5\alpha\times100\%$ (see the main text). }
	\label{fig:retrieval}
\end{figure}
 
Taking a step further, we ask what structural properties of the spectrum determine the dynamical behavior. 
The right edge of the boundary determines the linear stability and
time scales of the dynamics around the null fixed point (all neurons are silent), i.e., if $\max{\mbox{Re}(\lambda_{\mathbf{J}}})<1$, the null
point
is a stable fixed point of the dynamics, and all trajectories in its neighborhood (covering the entire state space) will converge to the fixed point, where
the time scale of the relaxation is determined by the distance from the right edge to the critical line ($\mbox{Re}(\lambda_{\bJ})=1$, see also Fig.~\ref{fig:connectivity_dynamic} (d)].
Notice that the activities for some neurons may exhibit a transient amplification before the following decay [Fig.~\ref{fig:connectivity_dynamic} (a)], which is an intrinsic 
collective property of the non-normal random connectivity matrix we consider here. The transient amplification bears important computational benefits, which attracts recent interests of studying recurrent neural networks~\cite{Tim-2014,Coding-2020,Tim-2021}.
Because of this transient amplification, memory retrieval in sequence is possible, as shown by the transient overlap profile in Fig.~\ref{fig:connectivity_dynamic} (a).
Therefore, we conjecture that the null phase also has the computational function, as the length of retrieval sequences can 
be freely adjusted by the distance from the right edge of the connectivity-spectrum to the critical line. Note that the time scale of the overlap decay is inversely proportional to the distance from the critical line [Fig.~\ref{fig:connectivity_dynamic} (d)].
When the right edge of the spectrum slightly cross the critical line, the sequential retrieval of the entire ordered set of patterns becomes possible [Fig.\ref{fig:connectivity_dynamic} (b)],
provided that $\alpha$ is small enough.
By considering the linear stability, we draw the phase diagram of the recurrent dynamics in Fig.~\ref{fig:connectivity_dynamic} (c).
The non-zero values of $c$ would suppress the null phase, playing a similar role to the increasing Hebbian length.
\begin{figure}[htbp]
	\centering
	\includegraphics[bb=5 8 852 554,width=0.9\linewidth]{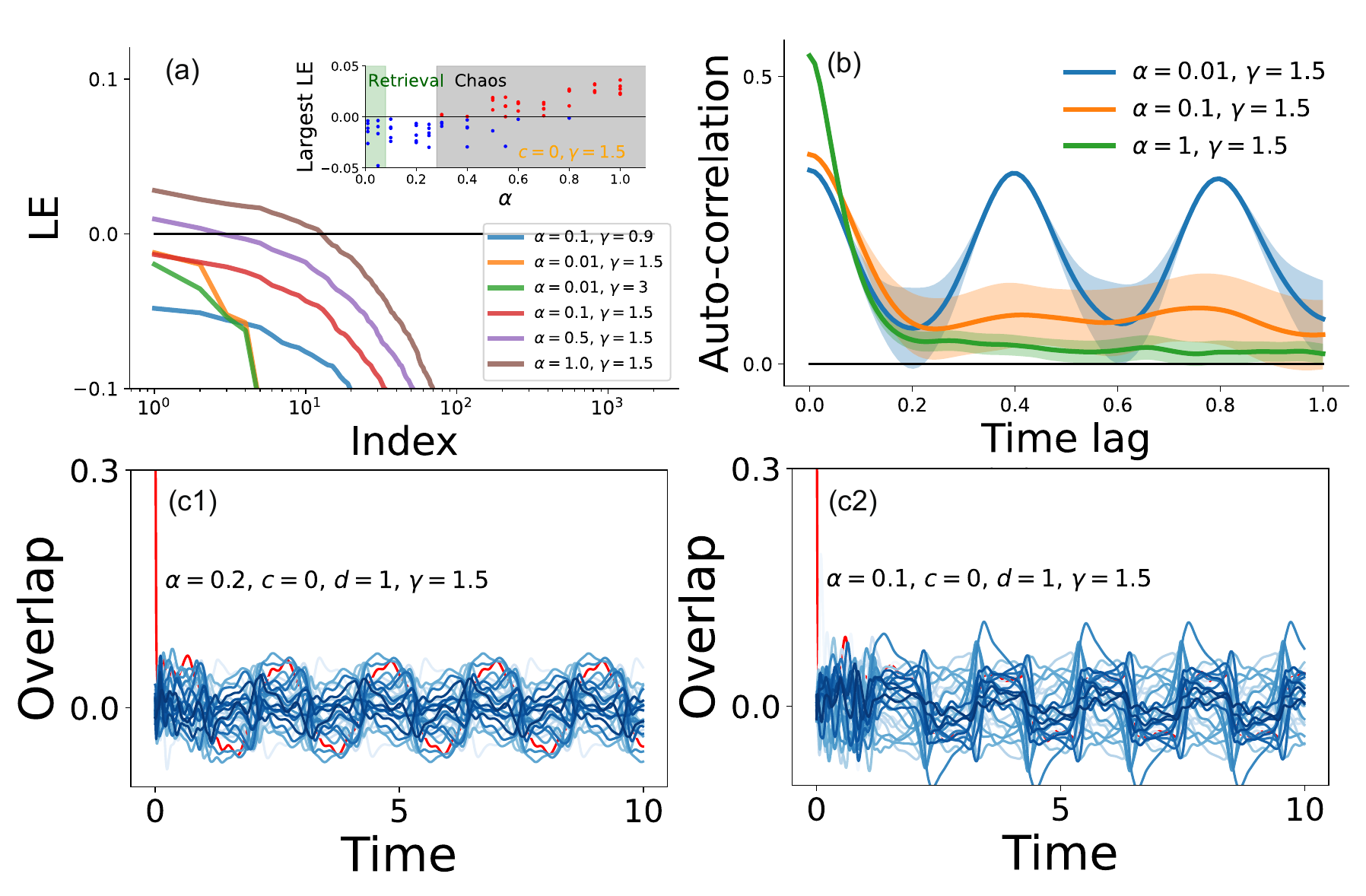}
	\caption{Lyapunov spectra of network dynamics and associated auto-correlations. 
	Other parameters are $N=3000,c=0$, and $d=1$. (a) Numerical Lyapunov spectra of the dynamics. 
	Chaotic behavior emerges once the largest Lyapunov exponent exceeds zero. 
         As $\alpha$ increases, more
	exponents exceed zero. The inset shows individual estimates (five instances) of the largest LE, 
	where the retrieval behavior disappears before the chaos sets in. In between, there emerge non-trivial fixed points or limit cycles.
	(b) Auto-correlation of the firing rate, which is defined as
	$\frac{1}{N}\sum_i\lb\phi(r_i(t))\phi(r_i(t+t_{\ell}))\rb_{\bJ}$, where $t_\ell$ is the time lag. 
	In a steady state, the auto-correlation is independent of $t$. (c) The overlap dynamics before the chaos sets in. The displayed overlap is associated to patterns taken from the sequence with a fixed interval, due to the fact that
	the number of stored patterns is quite large when $\alpha$ gets large. The overlaps with patterns in sequence are
distinguished by colors from light blue to dark blue. }
	\label{fig:longtimedynamic}
\end{figure}

We next explore the linear unstable region [the part above the lines in Fig.~\ref{fig:connectivity_dynamic} (c)]. First,
we define the margin thickness as the interval between the right edge
of the outer boundary and the right edge of the inner boundary. Withe increasing value of $\alpha$, the margin thickness 
also increases [Fig.~\ref{fig:retrieval} (a)]. A larger $\alpha$ implies more 
unstable directions in the network dynamics around the null fixed point, which is supported by the numerical simulations of trajectory-distance behavior [Fig.~\ref{fig:retrieval} (b)].
For a small $\alpha$, the whole 
spectrum nearly condenses onto its boundary. By utilizing the rotational symmetry of the spectrum, we get the ratio of the number of eigenvalues [$\operatorname{Re}(\lambda)$] exceeding one
to the total number of all eigenvalues as follows,
\begin{equation}
\Re\approx\alpha\frac{\operatorname{arccos}\la\frac{1}{\gamma\sqrt{1+\alpha}}\ra}{\pi}<\frac{\alpha}{2},
\end{equation}
where $\alpha/2$ serves as an upper bound shown in Fig.~\ref{fig:retrieval} (d). Note that the memory patterns can be retrieved in sequence even in the long-time limit, which is intimately related to
the thin spectrum-margin. As the margin becomes dense, 
the persistence of the retrieval behavior
fades away, instead replaced by a short-lived retrieval [Fig.~\ref{fig:retrieval}(c), where the network structure is posited above the linear stability line, and the network dynamics after 
this short-lived retrieval would enter a chaotic state.

To further study the network dynamics in the long-time limit, we calculate the full spectrum of the Lyapunov exponents (LEs)~\cite{Lya-1990,Engel-2020}. 
LEs are a set of exponents organized in descending order, describing the growth rates of the perturbations 
along different directions. They are defined as the logarithms of the eigenvalues of the Oseledets
matrix, which is defined by $\lim_{t\to\infty}\left[\la\exp\la\frac{1}{\tau}\int_0^t \mathbf{D}(t^\prime)\d t^\prime\ra\ra^\mathrm{T}\exp\la\frac{1}{\tau}\int_0^t \mathbf{D}(t^\prime)\d t^\prime\ra\right]^{\frac{1}{2t}}$.
More details are given in the Appendix~\ref{LEsec}.
First of all, the positive largest Lyapunov exponent implies chaos [Fig.~\ref{fig:longtimedynamic}(a)].
Through a rough numerical estimate, this transition-to-chaos seems to occur after the proliferation of critical points.
However, we remark that this may be due to the finite size system effects of the simulations. As $N$ increases, the transition to chaos may
coincide with the instability of the null fixed point, like that in the previous work~\cite{chaos-1988}.
We shall come to this issue in the conclusion part.
Consistent with this picture, our simulations of auto-correlation show that
the auto-correlation may display oscillatory behavior (related to the sequence replay) or long-term residual correlation, which suggests that the network dynamics becomes much slower than that in the 
deep chaotic regime (large $\alpha$) where the auto-correlation rapidly decays to zero. 
The rough numerical estimate in the inset of Fig.~\ref{fig:longtimedynamic}(a) also shows that the retrieval phase is limited to the small-$\alpha$ regime (see thin spectrum margin in Fig.~\ref{fig:retrieval}).
We thus conclude that approaching the edge of chaos from below has rich types of dynamics, showing computational benefits of our model.
\begin{figure}[htbp]
	\centering
	\includegraphics[bb=9 8 845 687,width=0.9\linewidth]{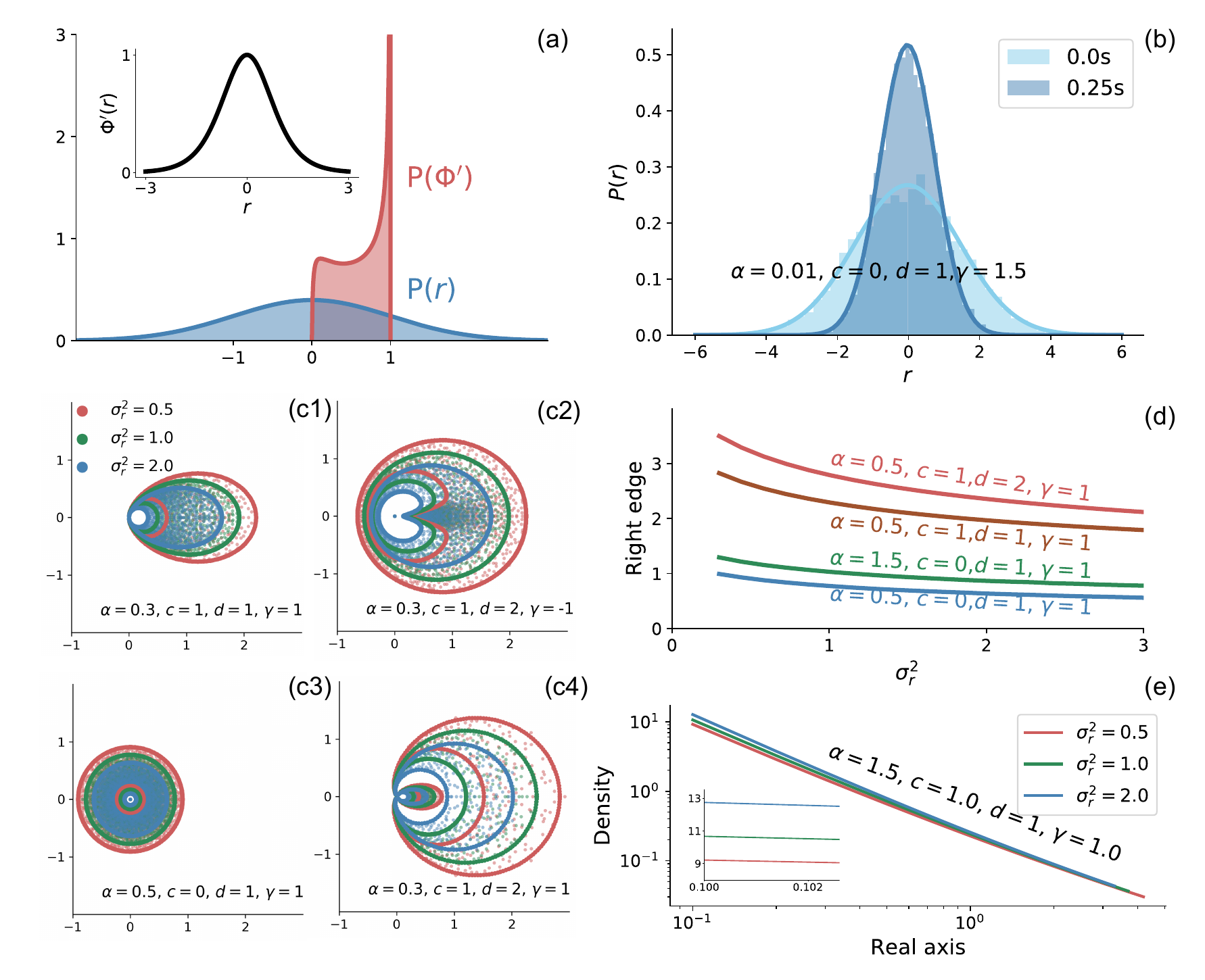}
	\caption{Spectrum of the shifted Jacobian matrix with the Gaussian synaptic-current assumption. Simulations are carried out on networks of size $N=4000$. 
	(a) The transformation from the distribution of the synaptic currents $\mathbf{r}$. We assume that the
	synaptic current follows a zero-mean Gaussian, shown in the blue line. The histogram is the numerical 
	samples. The distribution of $\pp(r)$ is shown in red. The black curve in the inset shows the derivative
	of the activation function, given by $1-\tanh^2(r)$. (b)  The synaptic current is initialized as a Gaussian
	distributed random variable at the time step equal to 0.0 second, and the current after 0.25 seconds is still
	Gaussian distributed. The histograms are obtained from real simulations, 
	and the solid lines are Gaussian distributions parameterized by the variance of the synaptic currents. 
	(c1-c4)  Spectrum of shifted Jacobian with the Gaussian distributed synaptic currents of different variances. 
	Solid lines are predictions of Eq.~\eqref{boundary_equations} , and the dots are numerical eigenvalues. (d) 
	The right edge of the spectrum plotted against the variance of synaptic currents. (e) The density of the spectrum
	with different variances along the real axis. The density difference for three values of variance close to the origin point is shown in the inset.
	}
	\label{fig:jacobi_spectrum1}
\end{figure}

We next turn to the spectrum of the Jacobian matrix, which depends on the specific neural state. We already show that the Jacobian matrix at $\mathbf{r}=0$ (the null fixed point) 
reduces to the connectivity matrix.
We focus on the spectrum of the shifted-Jacobian matrix and its function implication on the network dynamics.
Note that, when $N\to \infty$, the synaptic current can be assumed to follow
 some probability distribution, and then the distribution of the structured matrix $\bs{\pp}$ is determined by the probability distribution
 transformation illustrated in Fig.\ref{fig:jacobi_spectrum1} (a). In other words, the corresponding spectrum 
  depends uniquely on this distribution of $\bs{\pp}$ rather than single realizations of the neural activity.

\begin{figure}[htbp]
	\centering
	\includegraphics[bb=3 6 935 558,width=0.9\linewidth]{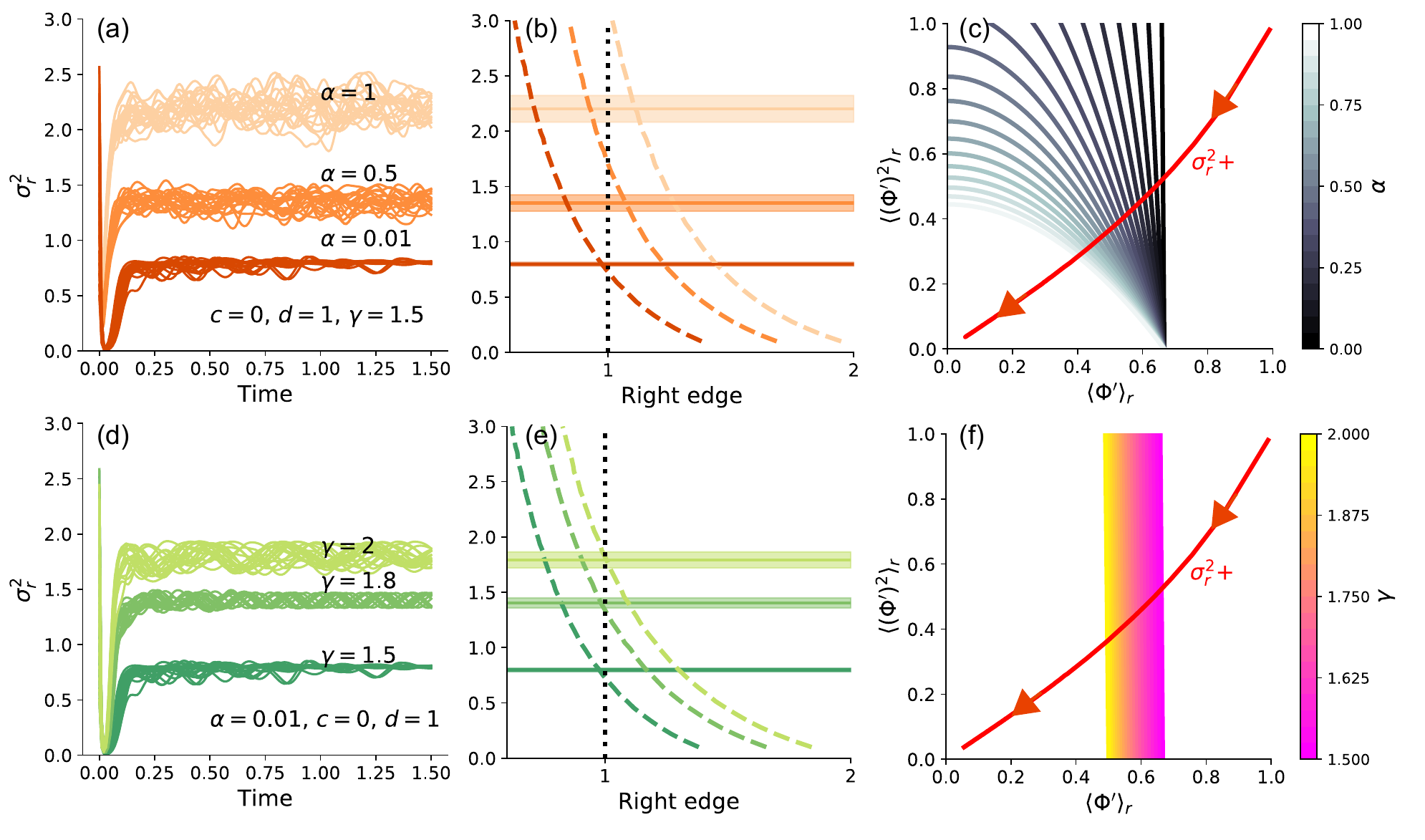}
	\caption{Right edge of the shifted-Jacobian spectrum in the steady state. Simulations are carried out on networks of size $N=4000$. (a) Evolution of the synaptic-current variance over time. 
	Colors indicate different model parameters, and different lines with the same color show the trajectories with different initializations.
	After a short relaxation, the variance becomes steady. (b) The current variance tunes the right edge of the spectrum [the dashed lines 
	obtained via Eq.~ \eqref{eq:outring} ]. The color scheme is the same with (a). The solid lines indicate the trajectory-averaged variance 
	of synaptic currents in the steady state, and the dashed line shows the linear stability. The dotted line shows the boundary of linear stability.
	(c) The phase diagram of the network dynamics 
	in the space spanned by the first two moments of $\pp(r)$. On the left hand side of each lines (except for the red line), the right edge 
	of the spectrum is below one. The red line shows the phase region where the synaptic current follows the zero-mean Gaussian statistics,
	i.e., $\lb(\pp)^2\rb_{r} = \int \d r (\pp(r))^2 \frac{1}{\sqrt{2\pi\sigma_r^2}}\exp\la-\frac{r^2}{2\sigma_r^2}\ra$, $\lb\pp\rb_{r} = \int \d r \pp(r) \frac{1}{\sqrt{2\pi\sigma_r^2}}\exp\la-\frac{r^2}{2\sigma_r^2}\ra$, 
	where $\sigma_r^2$ is the current variance. The arrow indicates the direction of increasing variances. (d-f) is similar to (a-c), focusing on the
	change of $\gamma$ instead of $\alpha$.} 
	\label{fig:jacobi_spectrum2}
\end{figure}

According to the local chaos hypothesis~\cite{LCH-1972,LCH-1995}, when the network is sufficiently large, the network activity is decoupled from the specific 
connectivity matrix, and thus the synaptic current of each neuron can be independently modeled by a Gaussian distribution with zero mean and $\sigma_r^2$ variance,
i.e., $r_i\sim\mathcal{N}(0,\sigma_r^2)$. A numerical verification is shown in Fig.~\ref{fig:jacobi_spectrum1} (b).
However, this hypothesis seems to break in the case of $c\neq0$ and $\alpha>>1$.
Note that the right edge of the shifted-Jacobian is always
less than that of the connectivity matrix [Fig.~\ref{fig:jacobi_spectrum1} (d)], because the derivative of the activation function $\Phi'(r)=1-\tanh^2(r)$ is bounded in $[0,1]$. In addition,
as the variance increases, the right peak of $P(\Phi')$ becomes lower [Fig.~\ref{fig:jacobi_spectrum1} (a)].

We draw some representative examples of the shifted-Jacobian spectrum with different variances in Fig.~\ref{fig:jacobi_spectrum1} (c). The shape preserves the same
features in the connectivity spectrum, e.g., the gauge invariance is broken for $d>1$. We also study how the variance influences the right edge of the spectrum in Fig.~\ref{fig:jacobi_spectrum1} (d), and find that
the spectrum becomes more contract as the variance grows. Setting the variance zero recovers the connectivity spectrum. As a result, for the model whose 
right edge of the connectivity spectrum is much larger than one, nontrivial stable fixed points can only be found in the neural state space with extremely high synaptic-current variances, which 
can shift the right edge of the Jacobian spectrum to the left hand side of the critical line (along the real axis). However, finding a nontrivial fixed point requires one to solve
$\mathbf{r}^*=\mathbf{J}\Phi(\mathbf{r}^*)$, which can be numerically accessible, but counting the number of (stable in all directions, 
or partially stable in some directions) fixed points is a challenging problem.

Interestingly, the spectral density of the shifted-Jacobian remains unchanged with increasing variances [Fig.~\ref{fig:jacobi_spectrum1} (e)], especially at the region far from the origin.
There appears a slight difference close to the origin for the spectral density. To study how the Jacobian spectrum determines the dynamics behavior, we simulate networks of $4\ 000$ neurons [Fig.~\ref{fig:jacobi_spectrum2} (a, d)].
The dynamics starts from different random initializations with different variances of synaptic current, and soon reaches a steady state, where the variance becomes nearly unchanged over time, suggesting
a relatively stable shape of the Jacobian spectrum.

We show the right edge of the shifted-Jacobian spectrum versus the current variance in Fig.~\ref{fig:jacobi_spectrum2} (b, e), from which we can see how increasing $\alpha$ impairs the network function of sequence retrieval.
As $\alpha$ increases, the right edge moves from the left hand side to the right hand side of the critical boundary, and meanwhile the spectrum margin thickness also grows, thereby destabilizing the sustained overlap profile like that in 
Fig.~\ref{fig:retrieval} (c). However, as $\gamma$ increases, the right edge of the Jacobian spectrum would also cross the critical boundary, but the thin margin of the spectrum ($\alpha=0.01$) greatly limits the emergence of unstable directions
that would destroy the memory traces, thereby protecting the memory to some extent.

We finally study the phase diagram of network dynamics in the synaptic current space when $c=0$ and $d=1$, characterized
by the first two moments of $\pp(r)$ [Fig.~\ref{fig:jacobi_spectrum2} (c, f)]. Each line in Fig.~\ref{fig:jacobi_spectrum2} (c, f)
is determined by setting the left hand side of Eq.~\eqref{eq:outring} equal to one as follows,
\begin{equation}
1=\gamma \sqrt{\alpha\langle (\pp)^2\rangle_{r}+\langle \pp\rangle_{r}^2},
\label{eq:jacobi_stable}
\end{equation}
which demonstrates that the region under each line is linearly stable. Technical details to derive Eq.~\eqref{eq:jacobi_stable} are given in the Appendix~\ref{GSR}.

\section{Conclusion}
In this work, we develop a random matrix theory of asymmetric associative memory networks of arbitrary Hebbian length. The network function such as sequence retrieval is related to
the spectral properties of the connectivity matrix and the Jacobian matrix. First, we calculate the spectrum of the Jacobian $\mathbf{\tilde{D}} = \bs{\xi}^T \mathbf{X}\bs{\xi}\bs{\pp}$, which
consists of the structured matrix $\mathbf{X}$ and an arbitrary diagonal (activity-dependent) matrix $\bs{\pp}$. By this analysis, we extend the well-known single ring theorem~\cite{Ring-1997,Belin-2017}.
We recover the spectrum of the connectivity matrix by setting $\bs{\pp}=\id$, which reveals several novel types of spectrum in non-Hermitian random matrix ensembles. In particular,
some types of spectra break the rotational symmetry about the origin, being of nested $d$-hole shapes as well ($\alpha<1$), or break the gauge invariance ($d>1$).
We then study how the model parameters affect the geometric shape of the connectivity spectrum and analyze 
the linear stability of the null fixed point, which is determined by the right edge of the spectrum. Approaching the critical boundary from the left-hand side, the network operates with a transient memory, and the time scales are related to
the distance towards the right edge of the spectrum, while the chaos state emerges after the critical line is crossed.
Even if the null fixed point becomes unstable, a small-$\alpha$ regime limits the number of unstable directions, allowing for sequence replay as well.
We thus conclude that in a finite-size system approaching the edge of chaos from below has also the computational benefits for sequence memory.

Our theoretical analysis not only applies to a broad class of associative memory models of sequence retrieval, but also applies to mathematical modeling of time-lagged correlations (of arbitrary time delays) in financial or biological time series data~\cite{Nowak-2017,EPL-2010}.
Our study also inspires several promising directions. First, a complete understanding of the edge of chaos in our model requires developing dynamical mean-field theory based on the functional path integral, such as recent theoretical works of random recurrent neural networks~\cite{Ostojic-2018,Helias-2018}.
This theory may fully characterize the time scales of the network dynamics beyond the linear stability line, and furthermore clarify whether chaotic attractors emerge continuously. In particular,
the topological complexity, counting the number of fixed points, is thus an important physics quantity to be explored in future works, e.g., via the Kac-Rice formula~\cite{Tou-2013}.
Taking a step further, considering both the sequence length and the number of sequences is also very interesting, and
the theoretical capacity of the sequence storage may be also calculated, such as in the previous works~\cite{ACC-1998,Brunel-2020} in simpler settings.
Second, the neural circuits in a biological brain are always not fully connected. It is thus interesting to combine the network sparseness in the connectivity with the sequence retrieval function of the network.
Lastly, the spectrum of our non-Hermitian matrix ensemble belongs to an intriguing category with isotropic or anisotropic properties depending on the specific model parameters. Moreover, the density within the spectrum is highly non-uniform with the nested
$d$-hole structure.  The non-normal property also causes the transient amplification, especially when the network setting falls within the critical regime. Thus, it is also very interesting in future works to show
how these characteristics are modified when more biological constrains are considered, such as cell types, sparseness, and excitation-inhibition balance.

\begin{acknowledgments}
This research was supported
by the National Natural Science Foundation of China for Grant number 12122515 (H.H.). 
\end{acknowledgments}
\appendix

\section{Introduction of spectral density for non-Hermitian random matrices}
\label{NHRMT}
\subsection{Spectral density and the Green's function}
The spectral density of an $N\times N$ non-Hermitian matrix $\bJ$ at $z$ on the complex plane is defined as
\begin{equation}
\rho(z) \equiv \frac{1}{N}\sum_{i=1}^N \delta^{(2)}(z-\lambda_i),
\label{spe1}
\end{equation}
where $\lambda_{i}$ are the eigenvalue of $\bJ$. To proceed, we introduce the Green's function $G$ or the so-called resolvent in random matrix theory as follows,
\begin{equation}
G(z) = \frac{1}{N}\la\operatorname{Tr}\frac{1}{\id_N z-\bJ}\ra,
\label{spe2}
\end{equation} 
where $\id_N$ is an $N\times N$ identity matrix. Transforming the summation over eigenvalues to the integral with the spectral density, we have 
\begin{equation}
G(z) = \frac{1}{N}\la\sum_i \frac{1}{z-\lambda_i}\ra = \frac{1}{N}\int \d^2\lambda\frac{\rho(\lambda)}{z-\lambda},
\label{spe3}
\end{equation}
where the integral is done over the complex plane. Eigenvalues are not real-valued any more for a non-Hermitian matrix.
 Then, we carry out a contour integral of $G(z)$ along an arbitrary closed path $\partial \mathcal{C}$, assuming that no eigenvalues of $\bJ$ 
 lie on the path, and we get
\begin{equation}
\begin{aligned}
\int_{\partial \mathcal{C}} \d^2 {z} G(z) &=  \frac{1}{N}\int_{\partial \mathcal{C}} \d^2 {z} \la\sum_i \frac{1}{z-\lambda_i}\ra\\ 
& = 2\pi \i \frac{1}{N}\sum_{\lambda\in \mathcal{C}} 1
=2\pi \i \int_\mathcal{C} \d^2 z \rho(z),
\label{spe4}
\end{aligned}
\end{equation}
where we have used the residue theorem, and $\mathcal{C}$ denotes the region bounded by $\partial\mathcal{C}$. 
We then apply the complex version of Gauss's law as follows,
\begin{equation}
\int_{\partial \mathcal{C}} \mathrm{d}^2 z\ G(z) = \i \int_{\mathcal{C}} \mathrm{d}^{2} z\left(\frac{\partial G}{\partial x}+\i \frac{\partial G}{\partial y}\right)=2\i \int_{\mathcal{C}} \mathrm{d}^{2} z\ \frac{\partial G}{\partial \bar z},
\label{spe5}
\end{equation}
where $x$ and $y$ is respectively the real part and imaginary part of $w$, and the Wirtinger derivatives are defined as follows, 
\begin{subequations}
	\begin{align}
		\frac{\partial }{\partial z} &= \frac{1}{2}\la\frac{\partial }{\partial x}-\i \frac{\partial }{\partial y}\ra,\\
		\frac{\partial }{\partial \bz} &= \frac{1}{2}\la\frac{\partial }{\partial x}+\i \frac{\partial }{\partial y}\ra.
	\end{align}
\end{subequations}
Hereafter, $\bz$ denotes the complex conjugate of the complex number $z$.
Because $\mathcal{C}$ is arbitrarily chosen, by comparing Eq.~\eqref{spe4} and Eq.~\eqref{spe5}, we have
\begin{equation}
\rho(z) = \frac{1}{\pi}\frac{\partial G}{\partial \bz} = \frac{1}{2\pi} \la\frac{\partial G}{\partial x}+\i\frac{\partial G}{\partial y}\ra,
\label{spe6}
\end{equation}
which indicates that if $G(w)$ is an analytic function of $w$, the density vanishes and vice versa. Equation \eqref{spe6} thus establishes the 
relationship between the spectral density $\rho$ and the resolvent $G$.

Next, we give a concrete example to show how to use Eq.~\eqref{spe6} to calculate the spectral density from the Green's function. 
In the classic circular law for the fully asymmetric random matrix~\cite{Stein-1988}, where $J_{ij}$ and $J_{ji}$ follows independently a Gaussian distribution,
the spectral density is bounded by $x^2+y^2=1$ on the complex plane, 
and the Green's function is given by
\begin{equation}
G(z) = 
\begin{cases}
\frac{1}{z}, &\mbox{$z$ outside the circle}\\
\bz, &\mbox{$z$ inside the circle}
\end{cases}
\end{equation}
where the Green's function can be derived by using the following replica method (see Appendix~\ref{app-c}) or Feynman diagrammatic techniques (see Appendix~\ref{fey}).
We immediately have
\begin{equation}
\rho(z) = \frac{1}{\pi}\frac{\partial G(z,\bz)}{\partial \bz}=
\begin{cases}
0, &\mbox{$z$ outside the circle}\\
1/\pi. &\mbox{$z$ inside the circle}
\end{cases}
\end{equation}

\subsection{Poisson equation and eigenvalue potential}
\label{EP}
From the Gauss's law, one can define an electrostatic potential to turn the eigenvalue spectrum problem into a two-dimension classical electrostatic problem.
The eigenvalue potential is constructed via the following way
\begin{equation}
\frac{\partial{\phi}}{\partial z} = -G.
\label{pos1}
\end{equation}
Using the Wirtinger derivatives, we arrive at
\begin{equation}
4\frac{\partial^2 \phi}{\partial z\partial \bz}\equiv \nabla^2 \phi = -4\pi \rho(z),
\label{pos2}
\end{equation}
which suggests that $\rho$ is the two-dimensional charge distribution, and the spectral density is now related to the computation of the potential.
Moreover, the Green's function $G$ plays a role like that of an electric field in two dimensional space. The explicit expression of the potential consistent with
the Green's function reads as follows~\cite{Stein-1988,HH-2022},
\begin{equation}
\phi = -\frac{1}{N}\operatorname{Tr}\ln(z\id_N-\bJ)-\frac{1}{N}\operatorname{Tr}\ln(\bz\id_N-\bJ^\da)
\label{pos3}
\end{equation}
After using the mathematical identities $\operatorname{Tr}\ln\mathbf{A} = \ln\operatorname{det}\mathbf{A}$, $\operatorname{det}\mathbf{A}^\t=\operatorname{det}\mathbf{A}$, and $\operatorname{det}(\mathbf{A}\mathbf{B})=\operatorname{det}\mathbf{A}\operatorname{det}\mathbf{B}$, 
one can rewrite Eq.~(\ref{pos3}) into a compact form as follows
\begin{equation}
\phi(z) = -\frac{1}{N}\la\ln\operatorname{det}\left[(\bz\id_N-\bJ^\da)(z\id_N-\bJ)\right]\ra.
\label{pos4}
\end{equation}
Finally, the procedure to compute the spectral density can be summarized below,
\begin{enumerate}
	\item Calculate the potential $\phi(z,\bz)$,
	\item Get the Green's function through $G(z,\bz)=-\frac{\partial \phi}{\partial z}$,
	\item Calculate the spectral density via $\rho(z)=\frac{1}{\pi}\frac{\partial G}{\partial \bz}$.
\end{enumerate}

\section{Useful formulas of complex Gaussian integral}
In this section, we introduce three types of complex Gaussian integral useful for our calculation of the spectral density of non-Hermitian random matrices.
\subsection{Standard form}
The first type is given below,
\begin{equation}
\int \mathrm{d}^{2N } \bs{z} \exp \left(-\bs{z}^{\dagger} \mathbf{B} \bs{z}\right),
\label{eq:basic}
\end{equation}
where $\boldsymbol{z}$ is a complex vector of length $N$, and $\int \mathrm{d}^{2N }\bs{z}$ denotes the integral over the whole complex plane.
$\boldsymbol{B}$ is a positive definite Hermitian matrix. By a unitary diagonalization of $\mathbf{B}$, we get 
\begin{equation}
\mathbf{U}^{\dagger} \mathbf{B} \mathbf{U}=\operatorname{diag}\left(b_{1}, \cdots, b_{N}\right).
\end{equation}
Hence, we can introduce the new integral variable $\boldsymbol{z}'\equiv\mathbf{U}^\dagger \boldsymbol{z}$. 
We then rewrite Eq.~\eqref{eq:basic} as
\begin{equation}
\begin{aligned}
\int \mathrm{d}^{2N } \bs{z} \exp \left(-\bs{z}^{\dagger} \mathbf{B} \bs{z}\right)&=\int \mathrm{d}^{2 N} \boldsymbol{z}^{\prime} \exp \left(-\sum_{i} b_{i}\left|z^{\prime}_{i}\right|^{2}\right)=\prod_{i} \frac{\pi}{b_{i}} \\
&=\frac{\pi^{N}}{\operatorname{det} \mathbf{B}}.
\label{eq:basic_result}
\end{aligned}
\end{equation}
Equation~\eqref{eq:basic_result} suggests that one can turn a determinant to a complex Gaussian integral, i.e.,
\begin{equation}
\begin{aligned}
\operatorname{det} \mathbf{B}^{-1}= \frac{1}{\pi^N}\int \mathrm{d}^{2N } \bs{z} \exp \left(-\bs{z}^{\dagger} \mathbf{B} \bs{z}\right).
\label{eq:basic_result2}
\end{aligned}
\end{equation}

\subsection{With Hermitian linear term}
Now we consider a bit more complex case, where an Hermitian linear term is added to the exponential term in  Eq.~\eqref{eq:basic}:
\begin{equation}
\int \mathrm{d}^{2N } \bs{z} \exp \left(-\bs{z}^{\dagger} \mathbf{B} \bs{z}+\boldsymbol{j}^\dagger \bs{z} +\bs{z}^\dagger \boldsymbol{j}\right).
\label{eq:basic2}
\end{equation}
The linear term is Hermitian because $(\boldsymbol{j}^\dagger \bs{z} +\bs{z}^\dagger \boldsymbol{j})^\da = \boldsymbol{j}^\dagger \bs{z} +\bs{z}^\dagger \boldsymbol{j}$.
We first rewrite the exponential term in Eq.~\eqref{eq:basic2} as 
\begin{equation}
-\bs{z}^{\dagger} \mathbf{B} \bs{z}+\boldsymbol{j}^\dagger \bs{z} +\bs{z}^\dagger \boldsymbol{j} = -(\bs{z}-\mathbf{B}^{-1}\boldsymbol{j})^\dagger \mathbf{B}(\bs{z}-\mathbf{B}^{-1}\boldsymbol{j})+\boldsymbol{j}^\dagger \mathbf{B}^{-1}\boldsymbol{j}.
\label{eq:quadratic}
\end{equation}
Then, with the change of the integral variable $\bs{z}\gets\bs{z}-\mathbf{B}^{-1}\boldsymbol{j}$, 
Eq.~\eqref{eq:basic2} can be transformed into the form of Eq.~\eqref{eq:basic}, leading to the integral result given below,
\begin{equation}
\frac{\pi^N}{\operatorname{det}\mathbf{B}}\exp(\boldsymbol{j}^\dagger \mathbf{B}^{-1}\boldsymbol{j}),
\end{equation}
which implies the complex Hubbard-Stratonovich transformation as follows,
\begin{equation}
\exp(\boldsymbol{j}^\dagger \mathbf{B}^{-1}\boldsymbol{j})=\frac{\operatorname{det}\mathbf{B}}{\pi^N}\int \mathrm{d}^{2N } \bs{z} \exp \left(-\bs{z}^{\dagger} \mathbf{B} \bs{z}+\boldsymbol{j}^\dagger \bs{z} +\bs{z}^\dagger \boldsymbol{j}\right).
\label{eq:HS}
\end{equation}

\subsection{With non-Hermitian linear term}
We next consider the added term is a non-Hermitian term as follows,
\begin{equation}
\int \mathrm{d}^{2N } \bs{z} \exp \left(-\bs{z}^{\dagger} \mathbf{B} \bs{z}+\boldsymbol{j}^\dagger \bs{z} +\bs{z}^\dagger \boldsymbol{j}^{\prime}\right).
\label{eq:basic3}
\end{equation}
In this case, we can treat $\boldsymbol{z}$ and $\boldsymbol{z}^\dagger$ as independent integral variables and shift them separately as
\begin{equation}
\begin{aligned}
&\boldsymbol{z}^\dagger \to \boldsymbol{z}^\dagger+\boldsymbol{j}^\dagger,\\
&\bs{z}\to\bs{z}+\boldsymbol{j}^{\prime}.
\end{aligned}
\end{equation}
Then the linear term becomes Hermitian, and we can repeat the manipulation in Eq.~\eqref{eq:quadratic}. Finally, the integral is worked out as follows,
\begin{equation}
\int \mathrm{d}^{2N } \bs{z} \exp \left(-\bs{z}^{\dagger} \mathbf{B} \bs{z}+\boldsymbol{j}^\dagger \bs{z} +\bs{z}^\dagger \boldsymbol{j}^{\prime}\right) = \frac{\pi^N}{\operatorname{det}\mathbf{B}}\exp(\boldsymbol{j}^\dagger \mathbf{B}^{-1}\boldsymbol{j}^\prime).
\label{eq:basic3_result}
\end{equation}

\section{Details of the annealed calculation}
\label{app-c}
In the annealed approximation, we have
\begin{equation}
	N\phi(w)\equiv-\lim_{|\epsilon|^2\to0}\lb\ln Z(w,\epsilon)\rb=\lim_{|\epsilon|^2\to0}\lb\ln Z^{-1}(w,\epsilon)\rb \approx \lim_{|\epsilon|^2\to0}\ln \lb Z^{-1}(w,\epsilon)\rb,
\end{equation}
where $Z(w,\epsilon)$ is defined in the main text.
Alternatively, we have
\begin{equation}
	\exp \la N\phi\ra = \lim_{|\epsilon|^2\to0} \lb Z^{-1}(w,\epsilon) \rb.
	\label{eq:expnp}
\end{equation}
With the help of Eq.~\eqref{eq:basic_result2}, we can transform $Z^{-1}$ into the complex Gaussian integral by introducing $N$ new complex
variables $\{z_i\}_{i=1}^N$. More precisely,
\begin{equation}
	Z^{-1}(w,\epsilon) = \frac{1}{\pi^{2N}} \int \d^{2N} \bs{z} \exp\la-\bs{z}^\da \la w\id-\tilde{\bD}\ra^\da \la w\id-\tilde{\bD}\ra\bs{z}-|\epsilon|^2 \bs{z}^\da\bs{z}\ra.
\end{equation}
Using Eq.~\eqref{eq:HS}, we introduce a new complex vector $\bs{y}$ to linearize the quadratic term as follows,
\begin{equation}\label{HSZ}
	Z^{-1}(w,\epsilon) = \frac{1}{\pi^{2N}} \int \d^{2N} \bs{z} \d^{2N} \bs{y} \exp\la-|\epsilon|^2 \bs{z}^\da\bs{z}-\bs{y}^\da\bs{y}-\i\bs{z}^\da \la w\id-\tilde{\bD}\ra^\da \bs{y}-\i\bs{y}^\da\la w\id-\tilde{\bD}\ra\bs{z}\ra.
\end{equation}
 Inserting the explicit form of $\tilde{\bD}$ into Eq.~\eqref{HSZ}, we get
 \begin{equation}
 \begin{aligned}
 \exp\la N\phi\ra &\propto \lim_{|\epsilon|^2\to0} 
 \int \d^{2N}\bs{z} \d^{2N}\bs{y} \exp\la -|\epsilon|^2 \bs{z}^\da \bs{z}-\bs{y}^\da \bs{y} -\i\bs{z}^\da \bar w \bs{y}-\i\bs{y}^\da w \bs{z}\ra \\
 &\times \lb \exp \la \i \frac{1}{N} \bs{z}^\da (\bs{\hx}  \bs{\pp})^\da \bs{\Lambda}^\da \bs{\hx} \bs{y} + \i \frac{1}{N}\bs{y}^\da \bs{\hx}^\da \bs{\Lambda} (\bs{\hx}  \bs{\pp}) \bs{z} \ra\rb,
\label{eq:2}
 \end{aligned}
 \end{equation}
 where $\propto$ means an irrelevant prefactor is omitted, and this prefactor does not affect the final result of saddle point equations and the spectrum properties.
\subsection{Calculating the disorder average}
We first introduce the following $P$-dimensional order parameters
\begin{equation}
\begin{aligned}
\bs{Z}= \frac{1}{N}(\bs{\hx}  \bs{\pp}) \bs{z},\\
\bs{Y} = \frac{1}{N}\bs{\hx} \bs{y}.
\end{aligned}
\end{equation}
By inserting the following Fourier expression of $\delta$ function $\delta^{(2)}(w)=\int\frac{\d^2\hat{w}}{\pi^2}e^{\i(w\bar{\hat{w}}+\bar w\hat{w})}$,
\begin{equation}
\begin{aligned}
\d^{2P} \bs{Z}\d^{2P} \bs{\hz} \exp\la \i\bs{\hz}^\da(N\bs{Z}-(\bs{\hx} \bs{\pp}) \bs{z})+\i(N\bs{Z}^\da-\bs{z}^\da(\bs{\hx}  \bs{\pp})^\da)\bs{\hz}\ra\\
\d^{2P} \bs{Y}\d^{2P} \bs{\hy} \exp\la \i\bs{\hy}^\da(N\bs{Y}-\bs{\hx}  \bs{y})+\i(N\bs{Y}^\da-\bs{y}^\da(\bs{\hx})^\da)\bs{\hy}\ra\\
\end{aligned}
\end{equation}
into Eq.~\eqref{eq:2}, we immediately arrive at
\begin{equation}
\begin{aligned}
 \exp\la N\phi\ra \propto \lim_{|\epsilon|^2\to0} 
\int \D&\exp\la -|\epsilon|^2 \bs{z}^\da \bs{z}-\bs{y}^\da \bs{y} -\i\bs{z}^\da \bar w \bs{y}-\i\bs{y}^\da w \bs{z}+\i N\bs{Y}^\da\bs{\Lambda} \bs{Z}+\i N\bs{Z}^\da \bs{\Lambda}^\da \bs{Y}\ra\\
&\times \exp\la \i N\bs{\hz}^\da \bs{Z}+\i N\bs{\hy}^\da \bs{Y}+\i N\bs{Z}^\da \bs{\hz}+\i N\bs{Y}^\da \bs{\hy}\ra\\
&\times \lb\exp\la-\i\bs{\hz}^\da(\bs{\hx}  \bs{\pp}) \bs{z}-\i\bs{\hy}^\da\bs{\hx} \bs{y}-\i\bs{z}^\da(\bs{\hx} \bs{\pp})^\da \bs{\hz}-\i\bs{y}^\da\bs{\hx}^\da \bs{\hy}\ra\rb,
\label{eq:3}
\end{aligned}
\end{equation}
 where $\D$ indicates all relevant integral measures (over $\bs{z},\bs{y},\bs{Z},\bs{Y},\bs{\hat{Z}},\bs{\hat{Y}}$), and $\lb\ldots\rb$ means the disorder average over the rotated patterns.
 The average term can be calculated as an integral with the standard complex Gaussian measure as follows,
 \begin{equation}
 \begin{aligned}
 &\inta \\
 & = \prod_{\mu i}\int \dd \hx_{\mu i} \exp\la-\bar{\hx}_{\mu i} \hx_{\mu_i}-\i\bar{\hz}_{\mu}\hx_{\mu i}\pp_{ii} z_i-\i\bar{\hy}_{\mu}\hx_{\mu i} y_i-\i\pp_{ii} \bar{z}_{i}\bar{\hx}_{\mu i} \hz_{\mu}-\i \bar{y}_{i}\bar{\hx}_{\mu i} \hy_{\mu}\ra\\
 & = \prod_{\mu i}\exp\la-\la\bar{\hz}_{\mu} \pp_{ii} z_i+\bar{\hy}_{\mu} y_i\ra\la \pp_{ii} \bar{z}_i \hz_{\mu}+\bar{y}_i \hy_{\mu}\ra\ra\\
 & = \exp\la -\bs{\hz}^\da \bs{\hz} (\bs{\pp} \bs{z})^\da (\bs{\pp} \bs{z}) - \bs{\hy}^\da \bs{\hy} \bs{y}^\da \bs{y}-\bs{\hz}^\da \bs{\hy} \bs{y}^\da (\bs{\pp}\bs{z}) -\bs{\hy}^\da \bs{\hz} (\bs{\pp}\bs{z})^\da \bs{y}\ra,
 \end{aligned}
 \end{equation}
where Eq.~\eqref{eq:basic2} is used to get the final equality.
Plugging the average term into Eq.~\eqref{eq:3}, we have
\begin{equation}
\begin{aligned}
\exp\la N\phi\ra \propto \lim_{|\epsilon|^2\to0} 
\int \D &\exp\la -|\epsilon|^2 \bs{z}^\da \bs{z}-\bs{y}^\da \bs{y} -\i\bs{z}^\da \bar w\bs{y}-\i\bs{y}^\da w \bs{z}+\i N\bs{Y}^\da\bs{\Lambda} \bs{Z}+\i N\bs{Z}^\da \bs{\Lambda}^\da \bs{Y}\ra\\
&\times \exp\la \i N\bs{\hz}^\da \bs{Z}+\i N\bs{\hy}^\da \bs{Y}+\i N\bs{Z}^\da \bs{\hz}+\i N\bs{Y}^\da \bs{\hy}\ra\\
&\times\exp\la -\bs{\hz}^\da \bs{\hz} (\bs{\pp} \bs{z})^\da (\bs{\pp} \bs{z}) - \bs{\hy}^\da \bs{\hy} \bs{y}^\da \bs{y}-\bs{\hz}^\da \bs{\hy} \bs{y}^\da (\bs{\pp}\bs{z}) -\bs{\hy}^\da \bs{\hz} (\bs{\pp}\bs{z})^\da \bs{y}\ra.
\label{eq:4}
\end{aligned}
\end{equation}

 \subsection{Introducing order parameters}
 To proceed, we introduce the following order parameters:
 \begin{subequations}
 \begin{align}
 &u\equiv\frac{1}{N}\bs{z}^\da \bs{z},\\
 &v\equiv \frac{1}{N}\bs{y}^\da \bs{y},\\
 &t\equiv \frac{1}{N}\bs{y}^\da \bs{z},\\
 &\bar{t} \equiv \frac{1}{N}\bs{z}^\da \bs{y},\\
 &u^\prime\equiv\frac{1}{N}(\bs{\pp}\bs{z})^\da (\bs{\pp}\bs{z}),\\
 &t^\prime\equiv \frac{1}{N}\bs{y}^\da (\bs{\pp}\bs{z}),\\
 &\bar{t^\p} \equiv \frac{1}{N}(\bs{\pp}\bs{z})^\da \bs{y}.
 \end{align}
 \end{subequations}
By inserting the Fourier representations of Dirac delta functions defining these order parameters into Eq.~\eqref{eq:4},
\begin{subequations}
\begin{align}
&\d u \d \hat{u} \exp\la \i\hat{u}(Nu-\bs{z}^\da \bs{z})\ra,\\
 &\d v \d \hat{v} \exp\la \i\hat{v}(Nv-\bs{y}^\da \bs{y})\ra,\\
&\dd t\dd \hat{t} \exp\la \i\bar{\hat{t}}(Nt-\bs{y}^\da \bs{z})+\i(N\bar t-\bs{z}^\da \bs{y})\hat{t}\ra,\\
&\d u^\prime \d \hat{u}^\prime \exp\la \i\hat{u}^\prime(Nu^\prime -(\bs{\pp}\bs{z})^\da (\bs{\pp}\bs{z}))\ra,\\
 &\dd t^\prime\dd \hat{t}^\prime \exp\la \i\overline{\hat{t}^\p}(Nt^\prime-\bs{y}^\da (\bs{\pp}\bs{z}))+\i(N\bar{t^\p}-(\bs{\pp}\bs{z})^\da \bs{y})\hat{t}^\prime\ra,
 \end{align}
 \end{subequations}
 we then have
\begin{equation}
\begin{aligned}
\int &\D \lim_{|\epsilon|^2\to0} \\
&\exp\la -|\epsilon|^2 Nu-Nv -\i N\bar w \bar t-\i Nw t+\i N\hat{u}u+\i N\hat{v}v+\i N\bar{\hat{t}} t+\i N\bar t\hat{t}+ \i N\hat{u}^\p u^\p+\i N\overline{\hat{t}^\p} t^\p+\i N\bar{t^\p}\hat{t}^\p \ra\\
&\times\exp\la \i N\bs{Y}^\da\bs{\Lambda} \bs{Z}+\i N\bs{Z}^\da \bs{\Lambda}^\da \bs{Y}\ra\\
&\times \exp\la \i N\bs{\hz}^\da \bs{Z}+\i N\bs{\hy}^\da \bs{Y}+\i N\bs{Z}^\da \bs{\hz}+\i N\bs{Y}^\da \bs{\hy}\ra\\
&\times\exp\la -Nu^\p\bs{\hz}^\da \bs{\hz}  - Nv\bs{\hy}^\da \bs{\hy} -Nt^\p\bs{\hz}^\da \bs{\hy}  -N\bar{t^\p}\bs{\hy}^\da \bs{\hz} \ra\\
&\times\int \d^{2N} \bs{z}\d^{2N} \bs{y}\exp\la-\i\hat{u}\bs{z}^\da \bs{z}-\i\hat{v}\bs{y}^\da \bs{y}-\i\bar{\hat{t}} \bs{y}^\da \bs{z}-\i\hat{t}\bs{z}^\da \bs{y}\ra\\
&\times\exp\la -\i\hat{u}^\p (\bs{\pp}\bs{z})^\da(\bs{\pp}\bs{z})- \i\overline{\hat{t}^\p} \bs{y}^\da (\bs{\pp}\bs{z})-\i\hat{t}^\prime (\bs{\pp}\bs{z})^\da \bs{y}\ra.
\label{eq:5}
\end{aligned}
 \end{equation}
\subsection{Simplification by the integral over auxiliary complex domains}
 First, we notice that
the integral in the last line of Eq.~\eqref{eq:5} can be rewritten as
\begin{equation}	
		\int \d^{2N} \bs{z}\d^{2N} \bs{y} 
		\exp\left(-
		\left[
		\begin{matrix}
			\bs{z}^\da&\bs{y}^\da
		\end{matrix}
		\right]
		\left[
		\begin{matrix}
			\i\hat{u}\id+ \i\hat{u}^\prime \bs{\pp}&\i\bar{\hat{t}}\id+\i\overline{\hat{t}^\p}\bs{\pp}\\
			\i\hat{t}\id+\i\hat{t}^\prime\bs{\pp}&\i\hat{v}\id
		\end{matrix}
		\right]
		\left[
		\begin{matrix}
			\bs{z}\\\bs{y}
		\end{matrix}
		\right]
		\right).
		\label{eq:6}
\end{equation}
Using Eq.~\eqref{eq:basic3}, one work out the above integral [Eq.~\eqref{eq:6}] as follows,
\begin{equation}
	\begin{aligned}
		&= \frac{\pi^{2N}}{\det \left[
			\begin{matrix}
				\i\hat{u}\id+ \i\hat{u}^\prime \bs{\pp}&\i\bar{\hat{t}}\id+\i\overline{\hat{t}'}\bs{\pp}\\
				\i\hat{t}\id+\i\hat{t}^\prime\bs{\pp}&\i\hat{v}\id
			\end{matrix}
			\right]}\\ 
		&\propto \exp\left(-\sum_i\ln(\bar{\hat{t}}\hat{t}-\hat{u}\hat{v}+(\pp_{ii})^2(\overline{\hat{t}^\p}\hat{t}^\p-\hat{u}^\prime\hat{v})+\pp_{ii}(\overline{\hat{t}^\p} \hat{t}+\bar{\hat{t}}\hat{t}^\prime))\right).
	\end{aligned}
\end{equation}
Equation~\eqref{eq:5} becomes then
\begin{equation}
\begin{aligned}
 \int &\D \lim_{|\epsilon|^2\to0} \\
&\exp\la -|\epsilon|^2 Nu-Nv -\i N\bar w \bar t-\i Nw t+\i N\hat{u}u+\i N\hat{v}v+\i N\bar{\hat{t}} t+\i N\bar t\hat{t}+ \i N\hat{u}^\p u^\p+\i N\overline{\hat{t}^\p} t^\p+\i N\bar{t^\p}\hat{t}^\p \ra\\
&\times\exp\la \i N\bs{Y}^\da\bs{\Lambda} \bs{Z}+\i N\bs{Z}^\da \bs{\Lambda}^\da \bs{Y}\ra\\
&\times \exp\la \i N\bs{\hz}^\da \bs{Z}+\i N\bs{\hy}^\da \bs{Y}+\i N\bs{Z}^\da \bs{\hz}+\i N\bs{Y}^\da \bs{\hy}\ra\\
&\times\exp\la -Nu^\p\bs{\hz}^\da \bs{\hz}  - Nv\bs{\hy}^\da \bs{\hy} -Nt^\p\bs{\hz}^\da \bs{\hy}  -N(t^\p)^*\bs{\hy}^\da \bs{\hz} \ra\\
 &\times\exp\left(-\sum_i\ln(\bar{\hat{t}}\hat{t}-\hat{u}\hat{v}+(\pp_{ii})^2(\overline{\hat{t}^\p}\hat{t}^\p-\hat{u}^\prime\hat{v})+\pp_{ii}(\overline{\hat{t}^\p} \hat{t}+\bar{\hat{t}}\hat{t}^\prime))\right).\\
 \label{eq:7}
 \end{aligned}
 \end{equation}

We find that the last two lines of Eq.~\eqref{eq:7} is a Gaussian integral over $\bs{\hz}$ and $\bs{\hy}$, which can be rewritten as 
\begin{equation}
 \begin{aligned}
  \int \d^{2P} \bs{\hz}\d^{2P} \bs{\hy}& \exp\left(-
 \left[
 \begin{matrix}
 \bs{\hz}^\da&\bs{\hy}^\da
 \end{matrix}
 \right]
 \left[
 \begin{matrix}
 Nu^\p\id&N\bar{t^\p}\id\\
 Nt^\p\id&Nv\id
 \end{matrix}
 \right]
 \left[
 \begin{matrix}
 \bs{\hz}\\\bs{\hy}
 \end{matrix}
 \right]
 \right)\\
 \times&\exp
 \la
 \left[
 \begin{matrix}
 \bs{\hz}^\da&\bs{\hy}^\da
\end{matrix}
 \right]
 \left[
 \begin{matrix}
 \i N\bs{Z}\\ \i N\bs{Y}
 \end{matrix}
 \right]
 +
 \left[
 \begin{matrix}
 \i N\bs{Z}^\da&\i N\bs{Y}^\da
 \end{matrix}
 \right]
 \left[
 \begin{matrix}
 \bs{\hz}\\\bs{\hy}
 \end{matrix}
 \right]
 \ra\\
 =&\frac{\pi^{2P}\exp\left( \left[
 \begin{matrix}
 \i N\bs{Z}^\da&\i N\bs{Y}^\da
 \end{matrix}
 \right]\left[
 \begin{matrix}
 Nu^\p\id&N\bar{t^\p}  \bs{I}\\
 Nt^\p\id&Nv\bs{I}
 \end{matrix}
 \right]^{-1}\left[
 \begin{matrix}
 \i N\bs{Z}\\\i N\bs{Y}
 \end{matrix}
 \right] \right)}{\det\left[
 \begin{matrix}
 Nu^\p\id&N\bar{t^\p}  \id\\
 Nt^\p\id&Nv\id
 \end{matrix}
 \right]}\\
 & = \frac{\pi^{2P}\exp\left( -\left[
 \begin{matrix}
 \bs{Z}^\da&\bs{Y}^\da
 \end{matrix}
 \right]\left[
 \begin{matrix}
 Nkv\id&-Nk\bar{t^\p}  \id\\
 -Nkt^\p\id&Nku^\p \id
 \end{matrix}
 \right]\left[
 \begin{matrix}
 \bs{Z}\\\bs{Y}
 \end{matrix}
 \right] \right)}{N^{2P}(u^\p v-\bar{t^\p} t^\p)^{P}},\\
 & \propto {\exp\left( -\alpha N \ln(u^\p v-\bar{t^\p} t^\p)-\left[
 \begin{matrix}
 \bs{Z}^\da&\bs{Y}^\da
 \end{matrix}
 \right]\left[
 \begin{matrix}
 Nkv\id&-Nk\bar{t^\p}  \id\\
 -Nkt^\p\id&Nku^\p\id
 \end{matrix}
 \right]\left[
 \begin{matrix}
 \bs{Z}\\\bs{Y}
 \end{matrix}
 \right] \right)},
 \label{eq:9}
 \end{aligned}
 \end{equation}
 where $k\equiv 1/\det\left[\begin{matrix}u^\p&\bar{t^\p}\\t^\p&v\end{matrix}\right]=\frac{1}{u^\p v-\bar{t^\p} t^\p}$.
 Substituting Eq.~\eqref{eq:9} into Eq.~\eqref{eq:7}, and rewriting the quadratic terms into the form of matrix product, we have
\begin{equation}
\begin{aligned}
\int &\D \exp\la -|\epsilon|^2 Nu-Nv -\i N\bar w \bar t-\i Nw t+\i N\hat{u}u+\i N\hat{v}v+\i N\bar{\hat{t}} t+\i N\bar t\hat{t}+ \i N\hat{u}^\p u^\p+\i N\overline{\hat{t}^\p} t^\p+\i N\bar{t^\p}\hat{t}^\p \ra\\
 &\times\exp\left(-\sum_i\ln(\bar{\hat{t}}\hat{t}-\hat{u}\hat{v}+(\pp_{ii})^2(\overline{\hat{t}^\p}\hat{t}^\p-\hat{u}^\prime\hat{v})+\pp_{ii}(\overline{\hat{t}^\p} \hat{t}+\bar{\hat{t}}\hat{t}^\prime))\right)\\
&\times\exp\left( -\alpha N \ln(u^\p v-\bar{t^\p} t^\p)\right) \\
&\times\int \d^{2P} \bs{Z} \d^{2P} \bs{Y} \exp
 \la 
 -\left[
\begin{matrix}
 \bs{Z}^\da&\bs{Y}^\da
 \end{matrix}
 \right]
 \left[
 \begin{matrix}
 Nkv\id&-Nk\bar{t^\p} \id-\i\bL\\
 -Nkt^\p\id-\i\bL^\da&Nku^\p\id
 \end{matrix}
 \right]
 \left[
 \begin{matrix}
 \bs{Z}\\\bs{Y}
 \end{matrix}
 \right] 
 \ra.
 \label{eq:10}
 \end{aligned}
 \end{equation}
 The last integral in Eq.~\eqref{eq:10} is worked out by
 \begin{equation}
 \begin{aligned}
 =&{\pi^{2P}}\left\{\det \left[
 \begin{matrix}
 Nkv\id&-Nk\bar{t^\p}  \id-\i\boldsymbol{\Lambda}\\
 -Nkt^\p\id-\i\bL^\da&Nku^\p\id
 \end{matrix}
 \right]\right\}^{-1}\\
 =&\frac{\pi^{2P}}{N^{2P}}\left\{\prod_{\mu}\left[k^2 u^\p v-(k\bar{t^\p}+\i\Lambda_{\mu})(kt^\p+\i\overline{\Lambda_{\mu}})\right]\right\}^{-1}\\
 \propto& \exp\la-\sum_{\mu}\ln\la k-\i k\Lambda_{\mu}t^\p-\i k\overline{\Lambda_{\mu}}\bar{t^\p}  +\overline{\Lambda_{\mu}}\Lambda_{\mu}\ra\ra.
 \end{aligned}
 \end{equation}
Finally we have 
\begin{equation}
	\begin{aligned}
		\exp\la N\phi\ra&\propto\lim_{|\epsilon|^2\to0} \\
		\int &\D \exp\la -|\epsilon|^2 Nu-Nv +\i N\hat{u}u+\i N\hat{v}v+\i N\bar{\hat{t}} t+\i N\bar t\hat{t}+ \i N\hat{u}^\p u^\p+\i N\overline{\hat{t}^\p} t^\p+\i N\bar{t^\p}\hat{t}^\p \ra\\
		&\times\exp\left(-\sum_i\ln(\bar{\hat{t}}\hat{t}-\hat{u}\hat{v}+(\pp_{ii})^2(\overline{\hat{t}^\p}\hat{t}^\p-\hat{u}^\prime\hat{v})+\pp_{ii}(\overline{\hat{t}^\p} \hat{t}+\bar{\hat{t}}\hat{t}^\prime))\right)\\
		&\times\exp\left( -\alpha N \ln(u^\p v-\bar{t^\p} t^\p)\right)\times\exp\la -\i N\bar w\bar t-\i Nw t\ra \\
		&\times\exp\la-\sum_{\mu}\ln\la k-\i k\Lambda_{\mu}t^\p-\i k\overline{\Lambda_{\mu}}\bar{t^\p}  +\overline{\Lambda_{\mu}}\Lambda_{\mu}\ra\ra\\
		&\equiv\int \D \exp(N\psi), \label{eq:11}
	\end{aligned}
\end{equation}
where we denote an action $\psi$ as follows,
\begin{equation}
	\begin{aligned}
		\psi&\equiv -|\epsilon|^2 u-v +\i \hat{u}u+\i \hat{v}v+\i \bar{\hat{t}} t+\i \bar t\hat{t}+ \i \hat{u}^\p u^\p+\i \overline{\hat{t}^\p} t^\p+\i \bar{t^\p}\hat{t}^\p \\
		&-\frac{1}{N}\sum_i\ln(\bar{\hat{t}}\hat{t}-\hat{u}\hat{v}+(\pp_{ii})^2(\overline{\hat{t}^\p}\hat{t}^\p-\hat{u}^\prime\hat{v})+\pp_{ii}(\overline{\hat{t}^\p} \hat{t}+\bar{\hat{t}}\hat{t}^\prime))\\
		&-\alpha  \ln(u^\p v-\bar{t^\p} t^\p) -\i \bar w\bar t-\i w t-\frac{\alpha}{P}\sum_{\mu}\ln\la k-\i k\Lambda_{\mu}t^\p-\i k\overline{\Lambda_{\mu}}\bar{t^\p}  +\overline{\Lambda_{\mu}}\Lambda_{\mu}\ra.
	\end{aligned}
\end{equation}
 
 \section{Saddle point equations}
 \label{app-d}
 The Laplace method to estimate the integral $\int\D$ requires that
\begin{equation}
	\frac{\partial \psi}{\partial \left[\cdots\right]} =0,
\end{equation}
i.e., the partial derivatives with respect to all order parameters vanish, yielding the so-called saddle point equations.
$\lim_{|\epsilon|^2\to0} $ is carried out after the saddle point equations are derived.

\subsection{Derivatives with respect to hatted variables}
 The derivatives with respect to hatted variables lead to the following results:
 \begin{subequations}
 \begin{align}
 &\frac{\partial \psi}{\partial \hat{u}}= \i u+A\hat{v} = 0,\label{sd1}\\
 &\frac{\partial \psi}{\partial \hat{v}}= \i v+A\hat{u}+C\hat{u}^\p=0,\label{sd2}\\
 &\frac{\partial \psi}{\partial \hat{u}^\p} = \i u^\p+C\hat{v}=0,\label{sd3}\\
 &\frac{\partial \psi}{\partial \hat{t}} = \i \bar t-A\bar{\hat{t}}-B\overline{\hat{t}^\p}=0,\label{sd4}\\
 &\frac{\partial \psi}{\partial \bar{\hat{t}}} = \i t-A\hat{t}-B\hat{t}^\p=0,\label{sd5}\\
 &\frac{\partial \psi}{\partial \hat{t}^\p} = \i \bar{t^\p}-B\bar{\hat{t}}-C\overline{\hat{t}^\p}=0,\label{sd6}\\
 &\frac{\partial \psi}{\partial \overline{\hat{t}^\p}}= \i t^\p-B\hat{t}-C\hat{t}^\p=0.\label{sd7}
 \end{align}
 \end{subequations}
 where the neural-state dependent coefficients are given by
 \begin{subequations}
 \label{abc}
 \begin{align}
 A&\equiv\frac{1}{N}\sum_i \frac{1}{q_i},\\
 B&\equiv\frac{1}{N}\sum_i \frac{\pp_{ii}}{q_i},\\
 C&\equiv\frac{1}{N}\sum_i \frac{(\pp_{ii})^2}{q_i},
 \end{align}
 \end{subequations}
 where
 \begin{equation}
 q_i \equiv \bar{\hat{t}}\hat{t}-\hat{u}\hat{v}+(\pp_{ii})^2(\overline{\hat{t}^\p}\hat{t}^\prime-\hat{u}^\prime\hat{v})+\pp_{ii}(\otp \hat{t}+\bar{\hat{t}}\hat{t}^\prime).
 \end{equation}
 \subsection{Derivatives with respect to non-hatted variables}
 The derivatives with respect to non-hatted variables proceed as follows,
 \begin{subequations}
	 \begin{align}
		 \frac{\partial \psi}{\partial u} &= -|\epsilon|^2 +\i\hat{u} = 0, \label{sd8}\\
		 \frac{\partial \psi}{\partial u^\prime} &=\i\hat{u}^\p-\alpha  kv\mathcal{I} = 0,\label{sd9}\\
		\frac{\partial \psi}{\partial v} &=-1+\i\hat{v}-\alpha  ku^\prime\mathcal{I}=0,\label{sd10}\\
		 \frac{\partial \psi}{\partial t} &= -\i w+\i\bar{\hat{t}} = 0,\label{sd11}\\
		 \frac{\partial \psi}{\partial \bar t} &= -\i \bar w+\i\hat{t} = 0,\label{sd12}\\
		 \frac{\partial \psi}{\partial t^\p} &= \i\otp+\alpha  k \mathcal{I}\bar{t^\p}+\i\alpha k\mathcal{I}^\p = 0,\label{sd13}\\
		 \frac{\partial \psi}{\partial \bar{t^\p}} &= \i\hat{t}^\p+\alpha  k \mathcal{I}t^\p+\i\alpha k\overline{\mathcal{I}^\p} =0, \label{sd14}
		 \end{align}
	 \end{subequations}
 where 
  \begin{subequations}
  	\begin{align}
 	 &\mathcal{I}\equiv \frac{1}{P}\sum_{\mu=1}^P\frac{\overline{\Lambda_{\mu}}\Lambda_{\mu}}{\mathcal{Q}_{\mu}},\\
 	 &\mathcal{I}^\p \equiv \frac{1}{P}\sum_{\mu=1}^P\frac{\Lambda_{\mu}}{\mathcal{Q}_{\mu}},\\
 	  &\mathcal{Q}_{\mu}\equiv k-\i k\Lambda_{\mu}t^\p-\i k\overline{\Lambda_{\mu}}\bar{t^\p}  +\overline{\Lambda_{\mu}}\Lambda_{\mu}.
 	 	\end{align}
 	 \end{subequations}
 
 \subsection{Interpretation of order parameters}
According to the electrostatic mapping in Sec.~\ref{EP}, we have
\begin{equation}
	\begin{aligned}
	G &= -\frac{\partial \phi}{\partial w} =  -\frac{\partial \psi_0}{\partial w}=\i t,\\
	\bar G &= -\frac{\partial \phi}{\partial \bar w} =  -\frac{\partial \psi_0}{\partial \bar w}=\i \bar t.
	\label{meaning1}
	\end{aligned}
\end{equation}
where $\psi_0$ is the value of $\psi$ at the maximum. Hence, the physical meaning of $\i t$
is exactly the Green's function. Equations~\eqref{sd4} and \eqref{sd5} suggests that $\i\bar t=\overline{\i t}$. Note that $t$ and $\bar t$ are no longer complex conjugates with each other when evaluated at the saddle point,
because the integral is deformed to be over contours in the complex plane~\cite{Galla-2020}.
Equations~\eqref{sd6} and \eqref{sd7} imply that $\bar{t^\p}$ is not a complex conjugate of $t^\prime$, and $\i\bar {t^\p} = \overline{\i t^\p}$.

By considering Eqs.~\eqref{sd1}, \eqref{sd3} and \eqref{sd10}, we have
\begin{equation}
	A = u(1-C\alpha k\mathcal{I}).
	\label{sde16}
\end{equation}
Similarly, considering Eqs.~\eqref{sd2}, \eqref{sd8} and \eqref{sd9}, we have
\begin{equation}
	A|\epsilon|^2 = v(1-C\alpha k\mathcal{I}).
	\label{sde17}
\end{equation}
It is immediately found that
\begin{equation}
	\frac{v}{u}=|\epsilon|^2.
	\label{sde15}
\end{equation}

If we assume that $\mathbf{L}$ and $\mathbf{R}$ are the left and right eigenvectors of the non-Hermitian matrix $\bJ$, respectively,
	we then have the spectral decomposition $\bJ=\mathbf{R}\boldsymbol{\Lambda}\mathbf{L}^\da$, which has the biorthogonal property $\mathbf{L}^\da\mathbf{R}=\id$.
	Note that $\mathbf{L}^\da\mathbf{L}\neq\id$ and $\mathbf{R}^\da\mathbf{R}\neq\id$. 
	We then define the eigenvector overlap matrix $O_{\alpha\beta}=(\mathbf{L}^\da\mathbf{L})_{\alpha\beta}(\mathbf{R}^\da\mathbf{R})_{\beta\alpha}$, which describes the correlation between left and right
	eigenvectors. Then,
	the one-point correlation function of eigenvectors is defined as~\cite{Chalker-1998}
	\begin{equation}
	 O\left( {w,\bar w} \right)=\lb\frac{1}{N}\sum_{\alpha=1}^N O_{\alpha\alpha}\delta^{(2)}(w-\lambda_{\alpha})\rb.
	\end{equation}
 The correlator is related to the eigenvalue potential through the relationship
$O\left( {w,\bar w} \right)=\frac{1}{\pi}\lim_{|\epsilon|\to 0}\frac{\partial\phi}{\partial \epsilon}\frac{\partial\phi}{\partial\bar{\epsilon}}$~\cite{Belin-2017}.
Then we can calculate the eigenvector overlap function as follows,
\begin{equation}
	\begin{aligned}
		O(w,\bar w) &= \frac{1}{\pi} \frac{\partial \phi}{\partial \epsilon} \frac{\partial \phi}{\partial \bar \epsilon}\\
		&=\frac{1}{\pi} \frac{\partial \psi}{\partial \epsilon} \frac{\partial \psi}{\partial \bar\epsilon}\\
		&=\frac{1}{\pi} |\epsilon|^2 u^2\\
		&=\frac{1}{\pi} uv,
	\end{aligned}
\label{meaning2}
\end{equation}
where $\epsilon\to0^+$ is implied. Hence, the product $uv$ is exactly the eigenvector overlap function.

Considering Eq.~\eqref{sd11} and Eq.~\eqref{sd12}, we also have
\begin{equation}
	\begin{aligned}
	\bar{\hat{t}} = w,\\
	\hat{t} = \bar w. \\
	\end{aligned}
\label{meaning3}
\end{equation}

 \subsection{Equations determining the spectral boundary and density within boundary}
The condition for vanishing eigenvector overlap function determines the boundary of the spectrum~\cite{Burda-2011}.
One can also verify that once $uv=0$ holds, 
$\frac{\partial G}{\partial \bar w}=0$. Therefore,
\begin{equation}
	uv  
	\begin{cases}
		=0 & \mbox{outside the boundary,}\\
		>0 &\mbox{inside the boudary.}
	\end{cases}
\end{equation}
Equation~\eqref{sde17} suggests that when $|\epsilon|^2\to 0$,
 \begin{itemize}
	 \item $v=0$,
	 \item $C\alpha k \mathcal{I}=1$.
 \end{itemize} 
Therefore, the boundary curve is determined by the above two constraints. 
 \subsubsection{Spectral density inside the boundary}
Now we consider the simplified saddle equations inside the boundary. Following the above analysis, we have
\begin{equation}
	\alpha Ck\mathcal{I}=1.
	\label{bd1}
\end{equation}
Based on Eq.~\eqref{sd6} and Eq.~\eqref{sd13}, we remove $\otp$ and get
\begin{equation}
 \la C\alpha k\mathcal{I}-1\ra\bar{t^\p}+\i C\alpha k \mathcal{I}^\p-\i B\bar{\hat{t}} =0,
\end{equation}
and taking into account $C\alpha k \mathcal{I}=1$, we have 
\begin{equation}
	\alpha C k \mathcal{I}^\p=B\bar{\hat{t}}.
	\label{bd2}
\end{equation}
Equation~\eqref{sd8} suggests that $\hat{u}=0$. Then we find from Eq.~\eqref{sd7},
\begin{equation}
	\i t^\p=B\hat{t}+C\hat{t}^\p.
	\label{bd3}
\end{equation}
By multiplying Eq.~\eqref{sd2} with Eq.~\eqref{sd3}, we get
\begin{equation}
	\begin{aligned}
	u^\p v &= -AC\hat{u}\hat{v}-C^2\hat{u}^\p\hat{v}\\
	& = -C^2 \hat{u}^\p\hat{v}.
	\label{bd4}
	\end{aligned}
\end{equation}
Finally, we summarize the closed-form equations for determining the order parameters inside the boundary as follows,
\begin{subequations}
	 \label{bd5}
	 \begin{align}
		 	C\alpha k\mathcal{I} = 1,\\
		 	C\alpha k\mathcal{I}^\p = Bw,\\
		 	G^\p = B\bar w+C\hat{t}^\prime,\\
			u^\prime v = -C^2 \hat{u}^\prime\hat{v},
		 \end{align} 
 \end{subequations}
where we define $G^\p = \i t^\p$ in parallel with $G=\i t$, and have used Eq.~\eqref{meaning3}. The other notations in Eq.~\eqref{bd5}
are specified as follows,
\begin{subequations}
	\begin{align}
		k\mathcal{I} &= \frac{1}{P}\sum_{\mu} \frac{|\Lambda_{\mu}|^2}{|\Lambda_{\mu}G^\prime-1|^2+u^\prime v |\Lambda_\mu|^2},\\
		k\mathcal{I}^\prime &= \frac{1}{P}\sum_{\mu} \frac{\Lambda_\mu}{|\Lambda_{\mu}G^\prime-1|^2+u^\prime v |\Lambda_\mu|^2},\\
		 B& =\frac{1}{N} \sum_i\frac{\pp_{ii}}{|(\Phi_{ii}^\p\hat{t}^\p+\hat{t})|^2-(\Phi_{ii}^\p)^2 \hat{u}^\p \hat{v}},\\
		C& = \frac{1}{N}\sum_i\frac{(\pp_{ii})^2}{|(\Phi_{ii}^\p\hat{t}^\p+\hat{t})|^2-(\Phi_{ii}^\p)^2 \hat{u}^\p \hat{v}}.
		\end{align}
	\end{subequations}
	
  Equation~\eqref{bd5} can be separated into six real-valued equations which involve in six real-valued variables 
$\{G^\p,\hat{t}^\p,u^\p v,\hat{u}^\p\hat{v}\}$ (a complex variable is equivalent to two real variables).
Since $w$ is a location on the complex plane, it can be arbitrary chosen as long as it falls within the boundary. 
Once we find a solution $\{G^\p,\hat{t}^\p,u^\p v,\hat{u}^\p\hat{v}\}$ from Eq.~\eqref{bd5}, we can get 
the Green's function [via Eq.~\eqref{sd5} and $G=\i t$] and also gets $uv = -AC \hat{u}^\p\hat{v}$ by using Eq.~\eqref{sd1} and Eq.~\eqref{sd2}.
Numerical methods for solving Eq.~\eqref{bd5} can be found in the Appendix~\ref{app-F}.

\subsubsection{Boundary curve}
Adding the constraint $v = 0$ to Eq.~\eqref{bd5} leading to the boundary condition for the spectrum in the complex plane:
\begin{subequations}
	\label{bd6}
	\begin{align}
		C\alpha k\mathcal{I} = 1,\\
		C\alpha k\mathcal{I}^\p = Bw,\\
		G^\p = B\bar w+C\hat{t}^\prime,
	\end{align} 
\end{subequations}
where
\begin{subequations}
	\begin{align}
		k\mathcal{I} &= \frac{1}{P}\sum_{\mu} \frac{|\Lambda_{\mu}|^2}{\left|\Lambda_{\mu}G^\prime-1\right|^2},\\
		k\mathcal{I}^\prime &= \frac{1}{P}\sum_{\mu} \frac{\Lambda_\mu}{\left|\Lambda_{\mu}G^\prime-1\right|^2},\\
		B& = \frac{1}{N}\sum_i\frac{\pp_{ii}}{|(\Phi_{ii}^\p\hat{t}^\p+\hat{t})|^2},\\
		C& = \frac{1}{N}\sum_i\frac{(\pp_{ii})^2}{|(\Phi_{ii}^\p\hat{t}^\p+\hat{t})|^2}.
	\end{align}
\end{subequations}
We conclude that the set of $w$ satisfying Eq.~\eqref{bd6} forms the boundary of the spectrum, i.e., the boundary curve.
 
 \subsubsection{Special example: spectrum of the coupling matrix}
 Setting $\pp_{ii}=1$, the Jacobian matrix reduces to the coupling matrix.
 From Eq.~\eqref{abc}, we find
\begin{equation}
	A=B=C=\frac{1}{q},
\label{con1}
\end{equation}
where 
\begin{equation}
	 q = |\hat{t}^\p+\hat{t}|^2-\hat{u}^\p\hat{v}.
\end{equation}
By using Eq.~\eqref{sd3}, Eq.~\eqref{sd2}, Eq.~\eqref{sd6} and Eq.~\eqref{sd7}, one obtain the following compact equation,
\begin{equation}
	u^\p v -\bar{t^\p} t^\p = C+(B^2-AC)\bar{\hat{t}} \hat{t}.
\end{equation}
Using Eq.~\eqref{con1} and recalling the definition $k \equiv (u^\p v -\bar{t^\p} t^\p)^{-1}$, we have
\begin{equation}
	q=k.
\label{con2}
\end{equation}
Using Eq.~\eqref{con2}, we simplify Eq.~\eqref{bd5} to 
\begin{subequations}
	\label{con3}
	\begin{align}
		&\frac{\alpha}{P}\sum_{\mu=1}^P \frac{|\Lambda_\mu G|^2 +|\Lambda_\mu|^2uv}{|\Lambda_{\mu}G-1|^2+|\Lambda_\mu|^2uv} = 1,\\
			&	\frac{\alpha}{P}\sum_{\mu=1}^P \frac{\Lambda_{\mu}}{|\Lambda_{\mu}G-1|^2+|\Lambda_\mu|^2uv} = w.
	\end{align}
\end{subequations}
Moreover, the boundary equations must include the condition $uv=0$, yielding
\begin{subequations}
	\label{con4}
	\begin{align}
		&\frac{\alpha}{P}\sum_{\mu=1}^P \frac{|\Lambda_\mu G|^2 }{|\Lambda_{\mu}G-1|^2} = 1,\\
		&	\frac{\alpha}{P}\sum_{\mu=1}^P \frac{\Lambda_{\mu}}{|\Lambda_{\mu}G-1|^2} = w,
	\end{align}
\end{subequations}
which coincides with the results independently obtained
by using the Feynman diagrammatic method, which we will introduce in Sec.~\ref{fey}.
\section{Closed-form solutions of spectrum}
\label{SRa}
In general, Eq.~\eqref{bd6} can only be solved numerically, 
but closed-form solutions can be found when the spectrum is rotationally symmetric. 
One simplest case is $c=0,d=1$, where the eigenvalues of $\X$ form a circle
on the complex plane as
\begin{equation}
	\label{simlbd}
	\begin{aligned}
		\Lambda_{\mu} &= \gamma\exp(-2\pi \i \mu/P)\\
		& = \gamma \cos(2\pi\mu/P)-\i\gamma\sin(2\pi\mu/P),
	\end{aligned}	
\end{equation}
which results in the rotational symmetry of the spectrum, as we shall show later.
\subsection{Single ring law}
We first consider the spectrum of the connectivity. 
Equation~\eqref{con4} can be further calculated as follows,
\begin{equation}
	\begin{aligned}
		1&=\frac{\alpha}{P}\sum_{\mu=1}^{P}\frac{\overline{\Lambda_{\mu}}\Lambda_{\mu}\overline{G} G}{\overline{\Lambda_{\mu}}\Lambda_{\mu}\overline{G} G-\Lambda_{\mu}G-\overline{\Lambda_{\mu}} \overline{G}+1},\\
		& = \frac{\alpha}{P}\sum_{\mu=1}^{P} \frac{\gamma^2 |G|^2}{\gamma^2 |G|^2-2a\gamma\cos(2\pi\mu/P)-2b\gamma\sin(2\pi\mu/P)+1},\\
		& = \frac{\alpha}{P}\sum_{\mu=1}^{P} \frac{\gamma^2 |G|^2}{\gamma^2 |G|^2-2\gamma |G|\sin(2\pi\mu/P+\arctan(b/a))+1},\\
		& = \frac{\alpha}{2\pi}\int_{0}^{2\pi} \d x\frac{\gamma^2 |G|^2}{\gamma^2 |G|^2-2\gamma |G|\sin(x+\arctan(b/a))+1},\\
		& = \frac{\alpha}{2\pi}\int_{0}^{2\pi} \d x\frac{\gamma^2 |G|^2}{\gamma^2 |G|^2-2\gamma |G|\sin x+1},\\
		& = \alpha \frac{\gamma^2 |G|^2}{|\gamma^2 |G|^2-1|},
	\end{aligned}
\end{equation}
where we have decomposed $G = a+b\i$ into its real and imaginary parts. 
The reasonable values of $|G|^2$ are given by
\begin{equation}
	|G|^2=
	\begin{cases}
		\frac{1}{\gamma^2(1\pm\alpha)}&\alpha<1,\\
		\frac{1}{\gamma^2(1+\alpha)}&\alpha>1.\\
	\end{cases}
\end{equation}
Then by substituting $z(\theta)=e^{\i\theta}$, the eigenvalues of $\X$ can
be re-parameterized as $\Lambda(\theta) = \gamma z(\theta)$, and for any function
$\mathcal{F}$, we have
$\lim_{P\to\infty}\frac{1}{P}\sum_{\mu=1}^{P}\mathcal{F}(\Lambda_\mu)=\frac{1}{2\pi}\int_0^{2\pi}\d\theta\mathcal{F}(\Lambda(\theta))$.
Thus the second equation of Eq.~\eqref{con4} can be further simplified by
\begin{equation}\label{sreq1}
	\begin{aligned}
		w &= \frac{\alpha}{2\pi} \int_{0}^{2\pi} \d \theta \frac{\gamma z}{\gamma^2 |G|^2-\gamma G z-\gamma \overline{G} \overline{z}+1 }\\
		& = \frac{\alpha}{2\pi \i} \int_{|z|=1}\d z \frac{1}{z}\frac{\gamma z}{\gamma^2 |G|^2-\gamma G z-\gamma \overline{G}/z+1 }\\
		& = \frac{\alpha}{2\pi \i} \int_{|z|=1}\d z \frac{\gamma z}{(\gamma^2 |G|^2+1)z-\gamma G z^2 -\gamma \overline{G}  }\\
		& = \frac{\alpha}{2\pi \i} \int_{|z|=1}\d z \frac{\frac{1}{G} z }{-(z-1/(\gamma G))(z-\gamma \overline{G})}\\
		&=
		\begin{cases}
			\frac{\alpha}{\gamma^2|G|^2-1}\frac{1}{G} &\gamma |G|>1,\\
			 \frac{\alpha\gamma^2|G|^2}{1-\gamma^2|G|^2}\frac{1}{G}&\gamma|G|<1.
		\end{cases}
	\end{aligned}
\end{equation}
In the above Eq.~(\ref{sreq1}), we have used $\overline{z}=\frac{1}{z}$ and carried out the contour integral to obtain the last equality.
Therefore the inner and outer boundaries of the spectrum are given below.
\begin{itemize}
	\item When $\alpha<1$, the radii of the outer and inner boundaries are respectively given below
	\begin{equation}
		\begin{aligned}
			R_{\mathrm{out}} &= \frac{\gamma \alpha }{\sqrt{1+\alpha}-\frac{1}{\sqrt{1+\alpha}}}= \gamma\sqrt{1+\alpha},\\
			R_{\mathrm{in}} &= \frac{\gamma \alpha}{\frac{1}{(\sqrt{1-\alpha})^3}-\frac{1}{\sqrt{1-\alpha}}} = \gamma\la\sqrt{1-\alpha}\ra^3.
		\end{aligned}
	\end{equation}
	\item When $\alpha>1$, the inner boundary vanishes and the radius of the outer boundary is given by 
	\begin{equation}
		\begin{aligned}
			R_{\mathrm{out}} &= \frac{\gamma \alpha }{\sqrt{1+\alpha}-\frac{1}{\sqrt{1+\alpha}}} = \gamma\sqrt{1+\alpha}.
		\end{aligned}
	\end{equation}
\end{itemize}

We next derive the analytic spectral density. The first equation of Eq.~\eqref{con3} is simplified as
\begin{equation}
	\label{con3sim}
	\begin{aligned}
		1&=\frac{\alpha}{P}\sum_{\mu=1}^{P}\frac{\overline{\Lambda_{\mu}}\Lambda_{\mu}\overline{G} G+\overline{\Lambda_{\mu}}\Lambda_{\mu} uv}{\overline{\Lambda_{\mu}}\Lambda_{\mu}\overline{G} G-\Lambda_{\mu}G-\overline{\Lambda_{\mu}} \overline{G}+1+\overline{\Lambda_{\mu}}\Lambda_{\mu} uv}\\
		& = \frac{\alpha}{P}\sum_{\mu=1}^{P} \frac{\gamma^2 |G|^2+\gamma^2 uv}{\gamma^2 |G|^2-2a\gamma\cos(2\pi\mu/P)-2b\gamma\sin(2\pi\mu/P)+1+\gamma^2 uv}\\
		& = \frac{\alpha}{P}\sum_{\mu=1}^{P} \frac{\gamma^2 |G|^2+\gamma^2 uv}{\gamma^2 |G|^2-2\gamma |G|\sin(2\pi\mu/P+\arctan(b/a))+1+\gamma^2 uv}\\
		& = \frac{\alpha}{2\pi}\int_{0}^{2\pi} \d x\frac{\gamma^2 |G|^2+\gamma^2 uv}{\gamma^2 |G|^2-2\gamma |G|\sin(x+\arctan(b/a))+1+\gamma^2 uv}\\
		& = \frac{\alpha}{2\pi}\int_{0}^{2\pi} \d x\frac{\gamma^2 |G|^2+\gamma^2 uv}{\gamma^2 |G|^2-2\gamma |G|\sin x+1+\gamma^2 uv}\\
		& = \alpha \frac{\gamma^2 |G|^2+\gamma^2 uv}{\sqrt{(\gamma^2 |G|^2+1+\gamma^2 uv)^2-4\gamma^2|G|^2}}.
	\end{aligned}
\end{equation}
The second equation of Eq.~\eqref{con3} is then simplified as 
\begin{equation}\label{sreq2}
	\begin{aligned}
		w &= \frac{\alpha}{2\pi} \int_{0}^{2\pi} \d \theta \frac{\gamma z}{\gamma^2 |G|^2-\gamma G z-\gamma \overline{G} \overline{z}+1+uv }\\
		& = \frac{\alpha}{2\pi \i} \int_{|z|=1}\d z \frac{1}{z}\frac{\gamma z}{\gamma^2 |G|^2-\gamma G z-\gamma \overline{G} 1/z+1+uv }\\
		& = \frac{\alpha}{2\pi \i} \int_{|z|=1}\d z \frac{\gamma z}{(\gamma^2 |G|^2+1+uv)z-\gamma G z^2 -\gamma \overline{G}  }.
	\end{aligned}
\end{equation}
We then denote $z_{\pm}$ as the roots of $(\gamma^2 |G|^2+1+uv)z-\gamma G z^2 -\gamma \overline{G} $. We can obtain
\begin{equation}
	\begin{aligned}
		z_{\pm} &= \frac{\gamma^2 |G|^2+1+uv\pm \sqrt{\la\gamma^2 |G|^2+1+uv\ra^2-4\gamma^2 |G|^2}}{2\gamma G}\\
		& = \frac{\gamma^2 |G|^2+1+uv\pm \alpha\gamma^2(|G|^2+uv)}{2\gamma G}\\
		& = \frac{\gamma^2(1\pm\alpha)(|G|^2+uv)+1}{2\gamma G},
	\end{aligned}
\end{equation}
which leads immediately to  
$w = \frac{\alpha}{2\pi \i}\int_{|z|=1}\d z \frac{\gamma z}{-\gamma G(z-z_-)(z-z_+)}$.

In order to complete the contour integral over $z$ in Eq.~(\ref{sreq2}), three cases 
should be considered (one of $z_{\pm}$ is in the contour or both
of $z_{\pm}$ are in the contour), and thus the corresponding integral results are listed as follows, 
\begin{equation}
	\label{wpm}
	\begin{aligned}
		w_{\pm}& = \frac{\alpha}{G} \frac{z_{\pm}}{\mp(z_+-z_-)}
		=\mp\frac{\alpha}{G}\la\frac{1\pm\alpha}{2\alpha}+\frac{1}{2\gamma^2\alpha(|G|^2+uv)}\ra,\\
		w_0 &= w_++w_- = -\frac{\alpha}{G},
	\end{aligned}
\end{equation}
where the third result is trivial and can be discarded.
Then by introducing $f_{\pm}\equiv w_{\pm}G$, we rewrite Eq.~\eqref{wpm} as
\begin{equation}
	\gamma^2(|G|^2+uv) =\mp\frac{1}{\la 2f_{\pm}+\alpha\pm 1\ra}.
	\label{G2uv}
\end{equation}

A reorganization of Eq.~\eqref{con3sim} gives rise to
\begin{equation}
	\begin{aligned}
		(\gamma^2 |G|^2+1+\gamma^2 uv)^2-4\gamma^2|G|^2&= \alpha^2( \gamma^2 |G|^2+\gamma^2 uv)^2.
	\end{aligned}
\end{equation}
By using Eq.~\eqref{G2uv}, we have 
\begin{equation}
\left(\mp\frac{1}{ 2f_{\pm}+\alpha\pm 1}+1\right)^2-4\gamma^2\frac{f_\pm^2}{|G|^2}= \alpha^2\left( \mp\frac{1}{ 2f_{\pm}+\alpha\pm 1}\right)^2,
\end{equation}
which is a cubic equation of $f_{\pm}$, i.e.,
\begin{equation}
	\label{cubic}
	4\gamma^2 f_\pm^3+4\gamma^2(\alpha\pm 1)f_{\pm}^2+(\gamma^2(\alpha\pm1)^2-|w|^2)f_\pm-|w|^2\alpha=0.
\end{equation}
The only physical solution is given by $f_-$. 
On one hand, $|w_+|>1$ is excluded from the contour integral. 
On the other hand, $f$ should always be positive, because $f$ is
actually the radial cumulative density, which can be seen from
\begin{equation}
	\label{radial}
	\begin{aligned}
		\rho &= \frac{1}{\pi}\frac{\partial G}{\partial \overline{w}}\\
		& = \frac{1}{w\pi}\frac{\partial f}{\partial \overline{w}}\\
		&  = \frac{1}{w\pi}\frac{\partial |w|}{\partial \overline{w}}\frac{\partial f(|w|)}{\partial |w|}\\
		&  = \frac{1}{w\pi}\frac{1}{2}\la\frac{\partial }{\partial x}+\i\frac{\partial }{\partial y}\ra\sqrt{x^2+y^2}\frac{\partial f(|w|)}{\partial |w|}\\
		& = \frac{1}{2\pi |w|}\frac{\partial f(|w|)}{\partial |w|}.
	\end{aligned}
\end{equation}
Therefore, to calculated the density,
we need to solve the cubic equation first, 
and then calculate the density by using Eq.~\eqref{radial}. 
\subsection{Generalized single ring law}
\label{GSR}
In this section, we analyze the Jacobian spectrum when $c=0$ and $d=1$. 
For convenience, we recast Eq.~\eqref{bd6} as follows, 
\begin{equation}
	\label{simbd6}
	\begin{aligned}
		&C \frac{\alpha}{P} \sum_{\mu=1}^{P} \frac{\left|\Lambda_{\mu}\right|^{2}}{\left|\Lambda_{\mu}\left(B \overline{w}+C \hat{t}^{\prime}\right)-1\right|^{2}}=1,\\
		&C \frac{\alpha}{P} \sum_{\mu=1}^{P} \frac{\Lambda_{\mu}}{\left|\Lambda_{\mu}\left(B \overline{w}+C \hat{t}^{\prime}\right)-1\right|^{2}}=B w.
	\end{aligned}
\end{equation}
By inserting Eq.~\eqref{simlbd} into the first equation of Eq.~\eqref{simbd6}, we have 
\begin{equation}
	\begin{aligned}
		1&=C \frac{\alpha}{P} \sum_{\mu=1}^{P} \frac{ \gamma^2}{\left|\Lambda_{\mu}T-1\right|^{2}} \\
		&=  C\frac{\alpha}{P}\sum_{\mu}^{P} \frac{\gamma^2 }{\gamma^2 |T|^2-2\mbox{Re}T\gamma\cos(2\pi\mu/P)-2\mbox{Im}T\gamma\sin(2\pi\mu/P)+1}\\
		&=  C\frac{\alpha}{2\pi }\int_{0}^{2\pi} \d \theta \frac{\gamma^2 }{\gamma^2 |T|^2-2\gamma|T|\sin \theta+1}\\
		& = \frac{\alpha \gamma^2 C}{|\gamma^2 |T|^2-1|},
	\end{aligned}
\end{equation}
where $T = B\overline{w}+C\hat{t}^\p$, and the coefficient $C$ is defined in Eq.~\eqref{abc}. Therefore 
\begin{equation}
	|T|^2 =
	\begin{cases}
		\frac{1-\alpha\gamma^2 C}{\gamma^2}&\gamma|T|<1,\\
		\frac{1+\alpha\gamma^2 C}{\gamma^2}&\gamma|T|>1,	
	\end{cases}
\end{equation}
and
\begin{equation}
	\label{simC}
	C =
	\begin{cases}
		-\frac{\gamma^2|T|^2-1}{\alpha \gamma^2}&\gamma|T|<1,\\
		+\frac{\gamma^2|T|^2-1}{\alpha \gamma^2}&\gamma|T|>1.	
	\end{cases}
\end{equation}
Then, the second equation of Eq.~\eqref{simbd6} is calculated below,
\begin{equation}
	\label{simw}
	\begin{aligned}
		w&=\frac{C}{B} \frac{\alpha}{P} \sum_{\mu=1}^{P} \frac{\Lambda_{\mu}}{\left|\Lambda_{\mu}T-1\right|^{2}}\\
		&= \frac{C}{B}\frac{\alpha}{2\pi} \int_{0}^{2\pi} \d \theta \frac{\gamma z}{\gamma^2 |T|^2-\gamma T z-\gamma \overline{T} \overline{z}+1 }\\
		& = \frac{C}{B}\frac{\alpha}{2\pi \i} \int_{|z|=1}\d z \frac{1}{z}\frac{\gamma z}{\gamma^2 |T|^2-\gamma T z-\gamma \overline{T}/z+1 }\\
		& = \frac{C}{B}\frac{\alpha}{2\pi \i} \int_{|z|=1}\d z \frac{\gamma z}{(\gamma^2 |T|^2+1)z-\gamma T z^2 -\gamma \overline{T}  }\\
		& = \frac{C}{B}\frac{\alpha}{2\pi \i} \int_{|z|=1}\d z \frac{\frac{1}{T} z }{-(z-1/(\gamma T))(z-\gamma \overline{T})}\\
		& = 
		\begin{cases}
			\frac{\alpha C}{B}\frac{\gamma^2}{1-\gamma^2|T|^2}\overline{T}&\gamma|T|<1,\\
			\frac{\alpha C}{B}\frac{1}{\gamma^2|T|^2-1}\frac{1}{T}&\gamma|T|>1,
		\end{cases}\\
		& = 
		\begin{cases}
			\frac{1}{B}\overline{T}&\gamma|T|<1,\\
			\frac{1}{B\gamma^2}\frac{1}{T}&\gamma|T|>1,
		\end{cases}
	\end{aligned}
\end{equation}
where we have used Eq.~\eqref{simC} to derive the last line. 
It is easy to see that $T$ and $\overline{w}$ are collinear, 
i.e. they have the same angle on the complex plane. Note that
$\hat{t}^\p$ is also collinear with $\overline{w}$,
because $\hat{t}^\p=\frac{T}{C}-\frac{B}{C}\overline{w}$. 
By taking the modulus on both sides of Eq.~\eqref{simw}, we arrive at
\begin{equation}
	B = 
	\begin{cases}
		\frac{|T|}{|w|} &\gamma|T|<1,\\
		\frac{1}{\gamma^2 |T||w|} &\gamma|T|>1.
	\end{cases}
\end{equation}
In addition,
\begin{equation}
	\begin{aligned}
		\hat{t}^\p
		&=\begin{cases}
			0&\gamma|T|<1,\\
			\frac{\alpha}{|T|}&\gamma|T|>1.
		\end{cases}		
	\end{aligned}
\end{equation}
Note that $\gamma|T|<1$ corresponds to the outer boundary. Then, we rewrite Eq.~\eqref{abc}
by assuming the synaptic current $r$ is drawn from some distribution, e.g.,
Gaussian distribution in the main text, as follows
\begin{equation}
	\label{simBC}
	\begin{aligned}
		B = \lb\frac{\pp(r)}{\la\pp(r)|\hat{t}^\p|+|w|\ra^2}\rb_{r},\\
		C = \lb\frac{(\pp(r))^2}{\la\pp(r)|\hat{t}^\p|+|w|\ra^2}\rb_{r},
	\end{aligned}
\end{equation}
where $\lb\cdot\rb_{r}$ denotes the average over the distribution of $r$.
Then we get a surprisingly simple result for the outer boundary as
\begin{equation}
	|w|^2 = \alpha\gamma^2\lb (\pp)^2\rb_{r}+\gamma^2\lb \pp\rb_{r}^2,
\end{equation}
which results in the radius of the outer boundary:
\begin{equation}
	R_{{\rm out}} = \gamma\sqrt{\alpha\lb (\pp)^2\rb_{r}+\lb \pp\rb_{r}^2}.
\end{equation}
Unfortunately, the radius of the inner boundary does not have a closed-form result, 
but can be solved numerically by an iteration of Eq.~\eqref{simBC}.

\section{Numerical methods for solving saddle point equations}
\label{app-F}
\subsection{Calculating the boundary curve}
We first give a detailed introduction of how to calculate the boundary curve from Eq.~\eqref{bd6}. 
One can simplify Eq.~\eqref{bd6} by inserting the third equation of Eq.~\eqref{bd6} into the first two equations,
and then write the remaining two equations as follows
 \begin{equation}
 \begin{aligned}
 F_1(\boldsymbol{x})=0,\\
 F_2(\boldsymbol{x})=0,\\
 F_3(\boldsymbol{x})=0,\\
 \end{aligned}
 \end{equation}
 where $F_1$ is the first equation of Eq.~\eqref{bd6} and $F_2$ (resp. $F_3$) is the real (resp. imaginary)
 part of the second equation in Eq.~\eqref{bd6}. The notation $\boldsymbol{x}\in\mathbb{R}^4$ corresponds
 to $\{ {\rm Re}(w), {\rm Im}(w), {\rm Re}(\hat{t}^\p),{\rm Im}(\hat{t}^\p)\}$. 

 Now we have three real equations and four real variables to be determined, 
 and thus finding the solution amounts to finding the intersection curve of three surfaces in four-dimensional space.
 We use the marching method~\cite{Marching-1987,Marching-1988,Marching-1990} to carry out this numerical analysis. 
 This method makes use of local gradient information, consisting of two steps---Newton and tangent steps. 
 \subsubsection{Newton step}
Given an initial point in the vicinity of the intersection curve, the newton step bring the starting point closer
to the intersection curve. The update rule is given by
 \begin{equation}
 \boldsymbol{x}_{t+1} = \boldsymbol{x}_{t}+\Delta_t \boldsymbol{x},
\label{nt1}
 \end{equation}
 where the Newton increment is set to be the linear combination of all gradients
 \begin{equation}
 \Delta_t \boldsymbol{x} = \sum_{a=1}^3 \mathcal{D}_t^a \nabla F_a(\boldsymbol{x}_t).
\label{nt2}
 \end{equation}
The linear coefficient can be found from the following constraint
 \begin{equation}
 \Delta_t \boldsymbol{x} \cdot \nabla F_a(\boldsymbol{x}_t) = -F_a(\boldsymbol{x}_t),
\label{nt3}
 \end{equation}
 which finds the direction which approaches three surfaces at the same time.
 By writing Eq. \eqref{nt3} into the component form:
 \begin{equation}\label{eqf5}
 \sum_{b,i} \mathcal{D}_t^b \la\nabla F_b(\boldsymbol{x})\ra_{i} \la\nabla F_a(\boldsymbol{x}_t)\ra_{i} = -F_a(\boldsymbol{x}_t),
 \end{equation}
we find it is convenient to define the following gradient matrix 
 \begin{equation}
 \mathbf{g}_t\equiv 
 \la
 \begin{matrix}
 \frac{\partial F_1}{\partial x_1}&\frac{\partial F_1}{\partial x_2}&\frac{\partial F_1}{\partial x_3}&\frac{\partial F_1}{\partial x_4}\\
 \frac{\partial F_2}{\partial x_1}&\frac{\partial F_2}{\partial x_2}&\frac{\partial F_2}{\partial x_3}&\frac{\partial F_2}{\partial x_4}\\
 \frac{\partial F_3}{\partial x_1}&\frac{\partial F_3}{\partial x_2}&\frac{\partial F_3}{\partial x_3}&\frac{\partial F_3}{\partial x_4}\\
 \end{matrix}
 \ra.
 \end{equation}
Defining $\bs{F}_t=\{F_a(\bs{x}_t)\}$, we can then rewrite Eq.~\eqref{eqf5} into a compact form as
 \begin{equation}
 \mathbf{g}_t \mathbf{g}_t^\mathrm{T} \boldsymbol{{\cal D}}_t = -\boldsymbol{F}_t.
 \label{nt4}
 \end{equation}
 Hence the linear coefficient is solved as
 \begin{equation}
  \boldsymbol{{\cal D}}_t = -\frac{\boldsymbol{F}_t}{\mathbf{g}_t \mathbf{g}_t^\mathrm{T}}.
 \end{equation}
 
 \subsubsection{Tangent step}
 Given an initial point on the intersection curve, the tangent step will make the point move along
 the direction of the local tangent of the intersection curve.
 The tangent direction should be orthogonal to all three gradients, specified by
 \begin{equation}
 \boldsymbol{T}=
 \left|
 \begin{matrix}
 \boldsymbol{e}_1&\boldsymbol{e}_2&\boldsymbol{e}_3&\boldsymbol{e}_4\\
 g_{11}&g_{12}&g_{13}&g_{14}\\
 g_{21}&g_{22}&g_{23}&g_{24}\\
 g_{31}&g_{32}&g_{33}&g_{34}\\
 \end{matrix}
 \right|
 \end{equation}
where $\bs{e}_1,\bs{e}_2,\bs{e}_3$, and $\bs{e}_4 $ are standard bases in $\mathbb{R}^4$. It is easy to 
check $\bs{T}$ is indeed orthogonal to the three gradients.
Therefore the tangent step is given by
 \begin{equation}
 \boldsymbol{x}_{t+1} = \boldsymbol{x}_{t}+ \frac{l}{||\boldsymbol{T}||}\boldsymbol{T},
 \end{equation}
where $l$ is the step length. The pseudocode is given in the algorithm~\ref{alg1}.

\begin{algorithm}[H]
	\caption{Calculation of the boundary curve from Eq.~\eqref{bd6}}
	\label{alg1}
	\begin{algorithmic}[1]
		\Require Give an initial state $\bs{x}$
		\State Initialize $t=0$
		\Repeat		
	     \Repeat 
	     \State update $\bs{x}$ by the Newton step
	    \Until converge
	    \State $\bs{x}_t = \bs{x}$
	    \State $t\gets t+1$
	    \State update $\bs{x}$ by the tangent step $\bs{x}\gets \bs{x}+ \frac{l}{||\boldsymbol{T}||}\boldsymbol{T}$
		\Until the solutions $\bs{x}_0, \bs{x}_1, \bs{x}_2, \cdots $ form a closed curve
		\Ensure the solutions $\bs{x}_0, \bs{x}_1, \bs{x}_2, \cdots $.
	\end{algorithmic}  
\end{algorithm}
\subsection{Calculation of the spectral density and overlap function inside the boundary}
Here, we introduce how to solve Eq.~\eqref{bd5}.
We first reduce Eq.~\eqref{bd5} by inserting the last two equations into the first two equations,
and then write the remaining equations as follows, 
\begin{equation}
	\begin{aligned}
		F_1(\boldsymbol{x})=0,\\
		F_2(\boldsymbol{x})=0,\\
		F_3(\boldsymbol{x})=0,\\
	\end{aligned}
\end{equation}
where $F_1$ is the first equation of Eq.~\eqref{bd5} and $F_2$ (resp. $F_3$) is the real (resp. imaginary)
part of the second equation of Eq.~\eqref{bd5}. In addition, 
$\boldsymbol{x}\in\mathbb{R}^3$ corresponds to $\{\mathrm{Re}(\hat{t}^\p),\mathrm{Im}(\hat{t}^\p),\hat{u}^\p\hat{v}\}$. 
We use the marching method to solve these equations, with the pseudocode given in the algorithm~\ref{alg2}.
\begin{algorithm}[H]
	\caption{Calculation of the density and overlap function from Eq.~\eqref{bd5}}
	\label{alg2}
	\begin{algorithmic}[1]	
		\Require Given a location $w$ inside the boundary and an initial point $\bs{x}$
		\State Initialize $\bs{w}= [w,w+\delta,w-\delta,w+\i\delta,w-\i\delta]$, $\delta$ is a small number for numerical differentiation
		\State Initialize $i=0$
		\State Initialize point $\bs{x}_0 = \bs{x}$
		\For{$i<5$}
		\State $w\gets \bs{w}[i]$
		\State $\bs{x}\gets \bs{x}_0$
		\Repeat
		\State update $\bs{x}$ by the Newton step
		\Until converge
		\If{$i==0$}
		\State $\bs{x}_0\gets \bs{x}$
		\EndIf
		\State compute $G_i = A\bar w+B\hat{t}^\prime$ 
		\State compute $O_i = -\frac{1}{\pi} AC \hat{u}^\p \hat{v}$ 
		\EndFor
		\State Calculate $\frac{\partial G}{\partial \operatorname{Re}{w}} = \frac{G_1-G_2}{2\delta}$ ,$\frac{\partial G}{\partial \operatorname{Im}{w}} = \frac{G_3-G_4}{2\delta}$
		\State $\rho = \frac{1}{2\pi}\operatorname{Re}\la \frac{\partial G}{\partial \operatorname{Re}{w}}+\i\frac{\partial G}{\partial \operatorname{Im}{w}}\ra$
		\Ensure return the density $\rho$ and the overlap function $O_0$
	\end{algorithmic}  
\end{algorithm}

\section{Numerical method for calculating Lyapunov exponents}
\label{LEsec}
The evolution of the perturbation $\delta\mathbf{r}$ is given by
\begin{equation}
\tau\frac{\d \delta\mathbf{r}}{\d t} = \mathbf{D}(t)\delta\mathbf{r},
\end{equation}
where $\mathbf{D}(t)$ is the Jacobian of the system at time $t$.
Hence the long time perturbation can be expressed into the exponential form as
\begin{equation}\label{G2}
\delta \mathbf{r}(T) = \delta \mathbf{r}(0)\exp\la\int_0^T \mathbf{D}(t)/\tau\d t\ra.
\end{equation}
The perturbation will grow exponentially along the directions with positive exponents,
and shrink along the directions with negative exponents. 
Lyapunov exponents (LEs) are a set of exponents $\ell_1,\cdots,\ell_N$ in the descending order. 
The collection of LEs is called the Lyapunov spectrum. They
describe the growth rate of the volume of the perturbation spanning the tangent space. 
 
The maximum Lyapunov exponent (MLE) $\ell_1$ is commonly
used as the criterion for determining whether a system is chaotic or not.
The dynamics is chaotic if $\ell_1>0$, indicating that nearby trajectories diverge exponentially fast.
The MLE can be calculated as follows,
\begin{equation}\label{G3}
\ell_1 = \lim_{T\to\infty} \frac{1}{T}\lim_{||\delta\mathbf{r}(0)||\to0}\ln\frac{||\delta\mathbf{r}(T)||}{||\delta\mathbf{r}(0)||},
\end{equation}
where the two limits can not be exchanged.
The full Lyapunov spectrum provides additional insights into the long time behavior of the dynamics.
The LEs are the logarithms of the eigenvalues of the following Oseledets matrix~\cite{Lya-1990,Engel-2020,PRX-2022}
$\lim_{t\to\infty}\left[\la\exp\la\frac{1}{\tau}\int_0^t \mathbf{D}(t^\prime)\d t^\prime\ra\ra^\mathrm{T}\exp\la\frac{1}{\tau}\int_0^t \mathbf{D}(t^\prime)\d t^\prime\ra\right]^{\frac{1}{2t}}$.
By this definition, inserting Eq.~\eqref{G2} into Eq.~\eqref{G3}, one can find the maximal value of LEs is the very MLE.
However, the long-term Jacobian would become ill-conditioned, i.e., the ratio between the largest and smallest singular values diverges with time.

In practice, we directly estimate the shrinkage or expansion of the volume of the tangent space. More precisely, we first initialize a matrix $\mathbf{Q}$ by an identity, and then run the discretized dynamics
of the original continuous one $\tau\frac{\d\mathbf{r}}{\d t}=-\mathbf{r}+\mathbf{J}\phi(\mathbf{r})$ as
\begin{equation}
\mathbf{r}_{a+1} = F(\mathbf{r}_a),
\label{eq:discrete_dynamic}
\end{equation}
where $a=1,2,\ldots,T$ indicates the discrete time step with a fixed small step size $\epsilon$, $F(\cdot)$ defines the discrete mapping, 
and the associated Jacobian $\mathbf{D}_a=\id+\frac{\epsilon}{\tau}\mathbf{D}(a\epsilon)$. The Jacobian at the state $\mathbf{r}_a$ can then be used to update the $\mathbf{Q}$ matrix.
After performing the QR decomposition of the matrix $\mathbf{Q}$, we can relate the $i$-th diagonal element of $\mathbf{R}$ in the QR decomposition to the $i$-th Lyapunov exponent by a time average as follows~\cite{Lya-1990,Engel-2020},
\begin{equation}
 \ell_i=\frac{1}{T}\sum_{a=1}^T\ln\mathbf{R}_{ii}(a),
\end{equation}
where $\mathbf{R}_{ii}$ describes the expansion or contraction of $\mathbf{Q}$. An initial relaxation of the original dynamics is required before the calculation of the Lyapunov spectrum,
as the dynamics has to converge to an attractor state. The pseudocode for the estimation of LEs is given in the algorithm~\ref{alg3}.

\begin{algorithm}
	\caption{Estimation of the Lyapunov spectrum $\bs{\ell}$}
	\label{alg3}
	\begin{algorithmic}[1]
		\renewcommand{\algorithmicrequire}{\textbf{Input:}}
	\renewcommand{\algorithmicensure}{\textbf{Output:}}	
		\Require Model parameters
		\State Run the dynamics Eq.~\eqref{eq:discrete_dynamic} for a sufficiently long time, and record the final state as $\mathbf{r}$
		\State Initialize $a=1$
		\State Initialize $\mathbf{Q}=\id$
		\State Initialize $\bs{\ell}=\bs{0}$
		\For{$a<=T$}
		\State $\mathbf{r}\gets F(\mathbf{r})$
		\State $\mathbf{Q}\gets\mathbf{D}_a(\mathbf{r})\mathbf{Q}$
		\State $\mathbf{Q},\mathbf{R}\gets\operatorname{QR\ decomposition}$ of $\mathbf{Q}$
		\State $\ell_i\gets\ell_i+\ln\mathbf{R}_{ii}$ 
		\State $a\gets a+1$
		\EndFor
		\State $\bs{\ell}\gets\bs{\ell}/T$
		\State Sort $\bs{\ell}$ in descending order
		\Ensure return $\bs{\ell}$
	\end{algorithmic}  
\end{algorithm}

\section{Feynman diagrammatic techniques for calculating the eigen-spectrum of non-Hermitian synaptic coupling matrices}
\label{fey}

The eigenvalue spectrum of non-Hermitian coupling matrix can also be derived by adopting Feynman diagrammatic techniques. We show the technical details to reproduce
the results obtained by applying the annealed approximation, as a cross-checking of the spectrum formula. 

The synaptic coupling between any two neurons $i$ and $j$ in our model is given by
	\begin{equation}
	{J_{ij}} = \frac{1}{N}\sum\limits_{\mu  = 1}^P {\left[ {c\xi _i^\mu \xi _j^\mu  + \gamma \sum\limits_{r = 1}^d {\xi _i^\mu \xi _j^{\mu  + r}} } \right]} ,\label{def_J}
	\end{equation}
which contains the concurrent Hebbian term and pattern-separated Hebbian term. The strength of the concurrent Hebbian term is specified by
$c$, and the strength of the pattern-separated term is specified by $\gamma$. 
	
In this model, we have $P$ patterns, and each pattern follows an independent Rademacher distribution,
i.e., $p\left( {\xi _i^\mu } \right) = \frac{1}{2}\delta \left( {\xi _i^\mu  - 1} \right) + \frac{1}{2}\delta \left( {\xi _i^\mu  + 1} \right)$, 
in which the superscript $\mu$ denotes the index of pattern and the subscript $i$ denotes the index of neuron. The pattern entries can be Gaussian i.i.d. random variables, which does not 
change the distribution of the following rotated patterns.
We are interested in the situation of large $P$ and $N$, with the fixed memory load
	\begin{equation}
	\alpha = \frac P N.\label{def_alpha}
	\end{equation}
Note that these patterns form a cyclic sequence with periodic boundary, corresponding to an ordered stimulus sequence in animal experiments~\cite{Miya-1988a}.
	
The asymmetric couplings in Eq.~(\ref{def_J}) can be recast into the matrix form:
	\begin{equation}
	\bJ = \frac{1}{N}{{\bm{\xi}} ^T}{\X}\bm{\xi} ,\label{def_J_mat}
	\end{equation}
in which the matrix $\X$ is a $P\times P$ circulant matrix defined by
	\begin{equation}
	{X_{\mu \nu }} = c{\delta _{\mu \nu }} + \gamma \sum\limits_{r = 1}^d {{\delta _{\mu ,[( {\nu  +r} )\bmod P]}}} .\label{def_X}
	\end{equation}
For a circulant matrix, the $m$-th eigenvalue of $\X$ is given by~\cite{Gray-2005} 
	\begin{equation}
		{\Lambda _m} = c + \gamma \sum_{r = 1}^d \cos\Bigl(2\pi\frac{mr}{P}\Bigr) .\label{eigenvalue}
	\end{equation}
Note that the non-Hermitian matrix $\X$ can be diagonalized using a unitary matrix $\mathbf{U}$ as $\X=\mathbf{U}^\dagger\bm{\Lambda}\mathbf{U}$. Hence, we
introduce a rotated memory pattern $\bhx$ ($\forall \mu$) defined as
	\begin{equation}
	 \bhx = \mathbf{U}\bx.
	\end{equation}
Considering the case of $P\to\infty$, and using the central limit theorem, we have
	\begin{equation}
	\left\langle \hat{\xi}_i^\mu  \right\rangle  = \left\langle \sum_{\nu  = 1}^P {U_{\mu\nu} \xi _i^\nu }  \right\rangle = 0,
	\end{equation}
	and
	\begin{equation}
	\left\langle \bar{{{\hat \xi }}}_i^{\mu}\hat \xi _i^\mu  \right\rangle  = \left\langle {\sum_{\nu  = 1}^P {\bar{U}_{\mu\nu} \xi _i^\nu } \sum_{\rho  = 1}^P {U_{\mu\rho} \xi _i^\rho } } \right\rangle
	= \sum_{\nu  = 1}^P {{{\left( {U_{\mu\nu} \bar{U}_{\mu\nu} } \right)}^2}{{\left( {\xi _i^\nu } \right)}^2}}  = 1,
	\end{equation}
where the average is done over the distribution of the original pattern. We conclude that the 
distribution of the rotated memory patterns is an i.i.d. standard complex Gaussian distribution, which greatly helps our analytical computation.
The coupling matrix can thus be rewritten as
	\begin{equation}
	\bJ = \frac{1}{N}\bhx^\dag \bm{\Lambda}\bhx.
	\end{equation}
	
In the following derivation, we denote an $N\times P$ matrix $\cX=\bhx^\dagger$. Then the coupling matrix is recast as 
	\begin{equation}
	\bJ = \frac{1}{N}\cX\bm{\Lambda} \cX^\dagger ,
	\end{equation}
and the only non-vanishing second-order moments are given as follows,
	\begin{equation}
	\left\langle {{\Xi_{i\mu}}\bar{\Xi}_{i\mu}} \right\rangle  = 1.
	\end{equation} 
	
The density of eigenvalues for the random matrix $\bJ$ is defined as follows,
	\begin{equation}
		\rho(z)=\left\langle \frac{1}{N}\sum_{k = 1}^N \delta^{(2)} \left( z - \lambda_k \right) \right\rangle_{\bJ}.
	\end{equation}
The density is related to the Green's function~\cite{HH-2022}:
	\begin{equation}
		G\left( {z,\bar z} \right) = \left\langle \frac{1}{N}\sum\limits_{k = 1}^N \frac{1}{z - \lambda _k}  \right\rangle_{\bJ} 
		= \left\langle \frac{1}{N}\Tr\frac{1}{z\id_N - \bJ} \right\rangle _{\bJ},
	\end{equation}
	and we can get the density from the Green's function via
	\begin{equation}
		\rho \left( {x,y} \right) = \frac{1}{\pi }\frac{\partial }{{\partial \bar z}}\left\langle \frac{1}{N}\sum_{k = 1}^N \frac{1}{z - \lambda _k}  \right\rangle _{\bJ} 
		= \frac{1}{\pi }\frac{\partial }{{\partial \bar z}}G\left( {z,\bar z} \right),
	\end{equation}
which actually expresses the Gauss law in two dimension. The spectral density is clearly related to the nonholomorphic behavior of the Green's function~\cite{HH-2022}.

In order to calculate the density of eigenvalues of this non-Hermitian coupling matrix,
we first define an auxiliary matrix ${\cal J}$ as
	\begin{equation}
	{{\cal J}} = \left( {\begin{array}{*{20}{c}}
		\bJ&0\\
		0&{{\bJ^\dag }}
		\end{array}} \right),\label{extend_J}
	\end{equation}
and the quaternionic Green's function (resolvent) can be defined as
	\begin{equation}
	{{\cal G}}\left( Q \right) = \left\langle {\frac{1}{{Q \otimes\id_N - {{\cal J}}}}} \right\rangle , 
	\end{equation}
where $\otimes$ denotes the Kronecker product, $\id_N$ denotes an $N\times N$ identity matrix, and 
	\begin{equation}
	Q = \left( {\begin{array}{*{20}{c}}
		z&{\i\bar w}\\
		{\i w}&{\bar z}
		\end{array}} \right),
	\end{equation}
	is a quaternion in the $2\times2$ matrix form. For an $i\times j$ matrix $A$ and an $m\times n$ matrix $B$, their
	Kronecker product $\otimes$ is 
	defined as 
	\begin{equation}
		A \otimes B = \left(
		\begin{array}{cccc}
			A_{11}B & A_{12}B & \dots & A_{1j}B\\
			A_{21}B & A_{22}B & \dots & A_{2j}B\\
			\dots & \dots & \dots &\dots\\
			A_{i1}B & A_{i2}B & \dots & A_{ij}B\\
		\end{array}
		\right),
	\end{equation}
	which forms a new matrix with dimension $(im) \times (jn)$.
	
Using this quaternionic Green's function, we can define a Green's function of $2\times 2$ matrix form:
	\begin{equation}
	G\left( Q \right) = \frac{1}{N}{\rm{bTr_2}}{{\cal G}}\left( Q \right),\label{matrix_G}
	\end{equation} 
	where $\rm{bTr}_2$ denotes a block trace operation taking the
	trace separately for four $N\times N$ block matrices in a $2N\times 2N$ matrix, yielding
	a $2\times 2$ matrix. 
Using the block inverse formula:
	\begin{equation}
		{\left( {\begin{array}{*{20}{c}}
					A&B\\
					C&D
			\end{array}} \right)^{ - 1}} = \left( {\begin{array}{*{20}{c}}
				{{A^{ - 1}} + {A^{ - 1}}B{{\left( {D - CA^{-1}B} \right)}^{ - 1}}C{A^{ - 1}}}&{ - {A^{ - 1}}B{{\left( {D - {C^{ - 1}}AB} \right)}^{ - 1}}}\\
				{ - {{\left( {D - CA^{-1}B} \right)}^{ - 1}}C{A^{ - 1}}}&{{{\left( {D - CA^{-1}B} \right)}^{ - 1}}}
		\end{array}} \right),\label{inverse-matrix}
	\end{equation}
we obtain the explicit form of $G(Q)$ as follows,
	\begin{equation}
		G\left( Q \right) = \left( {\begin{array}{*{20}{c}}
				{\frac{1}{N}\left\langle {{\rm{Tr}}\frac{{\bar z{\id_N} - {\bJ^\dag }}}{D}} \right\rangle }&{\frac{{ - \i\bar w}}{N}\left\langle {{\rm{Tr}}\frac{1}{D}} \right\rangle }\\
				{\frac{{ - \i w}}{N}\left\langle {{\rm{Tr}}\frac{1}{D}} \right\rangle }&{\frac{1}{N}\left\langle {{\rm{Tr}}\frac{{z{\id_N} - \bJ}}{D}} \right\rangle }
		\end{array}} \right),
	\end{equation}
where $D = \left( {z{\id_N} - \bJ} \right)\left( {\bar z{\id_N} - {\bJ^\dag }} \right) + {\left| w \right|^2}{\id_N}$. $G(Q)$
	can be simplified in the following form as
	\begin{equation}
		G\left( Q \right) = \left( {\begin{array}{*{20}{c}}
				g&{\i\bar v}\\
				{\i v}&{\bar g}
		\end{array}} \right),\label{matrix-G_in_2times2_form}
	\end{equation}
	\begin{equation}
		g= {\frac{1}{N}{\rm{Tr}}\left\langle {\frac{{\bar z{\id_N} - {\bJ^\dag }}}{D}} \right\rangle }, \quad v = -\frac{{ \bar w}}{N}{\rm{Tr}}\left\langle \frac{1}{D} \right\rangle,
	\end{equation}
	where we find that the Green's function is given by 
	\begin{equation}
		G\left( {z,\bar z} \right) = \mathop {\lim }\limits_{\left| w \right| \to 0} g.\label{reduced_Green}
	\end{equation}
	
        The eigenvector correlator is given by~\cite{Janik-1999}
	\begin{equation}
		O\left( {z,\bar z} \right) = \frac{1}{\pi }\mathop {\lim }\limits_{\left| w \right| \to 0} {\left| v \right|^2}.
	\end{equation}
This shows that the eigenvector correlation is actually the product of the off-diagonal elements of the quaternionic Green's function.
$O\left( {z,\bar z} \right)=0$ provides the condition that determines the boundary in the two-dimension complex space separating holomorphic and nonholomorphic solutions of 
the spectral problem~\cite{Chalker-1998,Janik-1999}. 
	
	Provided that $\|{\cal J}{\cal Q}^{-1}\|<\|\id_{2N}\|$, we can expand the quaternionic Green's function into a geometric series
	\begin{equation}
		{{\cal G}}{\rm{ = }}{{{\cal Q}}^{ - 1}} + \left\langle {{{{\cal Q}}^{ - 1}}{{\cal J}}{{{\cal Q}}^{ - 1}}} \right\rangle  + \left\langle {{{{\cal Q}}^{ - 1}}{{\cal J}}{{{\cal Q}}^{ - 1}}{{\cal J}}{{{\cal Q}}^{ - 1}}} \right\rangle  + ...,\label{series_G}
	\end{equation}
	where ${{{\cal Q}}^{ - 1}}{\rm{ = }}{Q^{ - 1}} \otimes {\id_N}$.
	According to the definition of $\cal J$ in Eq.~\eqref{extend_J}, we can decompose $\cal J$ as follows
	\begin{equation}
		{{\cal J}}{\rm{ = }}\frac{1}{N}{{\cal X}{\cal L}}{{{\cal X}}^\dag },\quad {{\cal X}}{\rm{ = }}\left( {\begin{array}{*{20}{c}}
				\cX&0\\
				0&\cX
		\end{array}} \right),\quad {{\cal L}} = \left( {\begin{array}{*{20}{c}}
				\bm{\Lambda} &0\\
				0&{{\bm{\Lambda} ^\dag}}
		\end{array}} \right).
	\end{equation}
	As a result, we can precisely separate the random parts and the deterministic parts in the series in Eq.~\eqref{series_G} as follows,
	\begin{equation}
		{{\cal G}}{\rm{ = }}{{{\cal Q}}^{ - 1}} + \left\langle {{{{\cal Q}}^{ - 1}}{{\cal X}}\frac{{{\cal L}}}{N}{{{\cal X}}^\dag }{{{\cal Q}}^{ - 1}}} \right\rangle  + \left\langle {{{{\cal Q}}^{ - 1}}{{\cal X}}\frac{{{\cal L}}}{N}{{{\cal X}}^\dag }{{{\cal Q}}^{ - 1}}{{\cal X}}\frac{{{\cal L}}}{N}{{{\cal X}}^\dag }{{{\cal Q}}^{ - 1}}} \right\rangle  + \ldots.\label{series_pre}
	\end{equation}

	\begin{figure}
		\centering
		\includegraphics[bb=5 5 1437 985,width=1.0\linewidth]{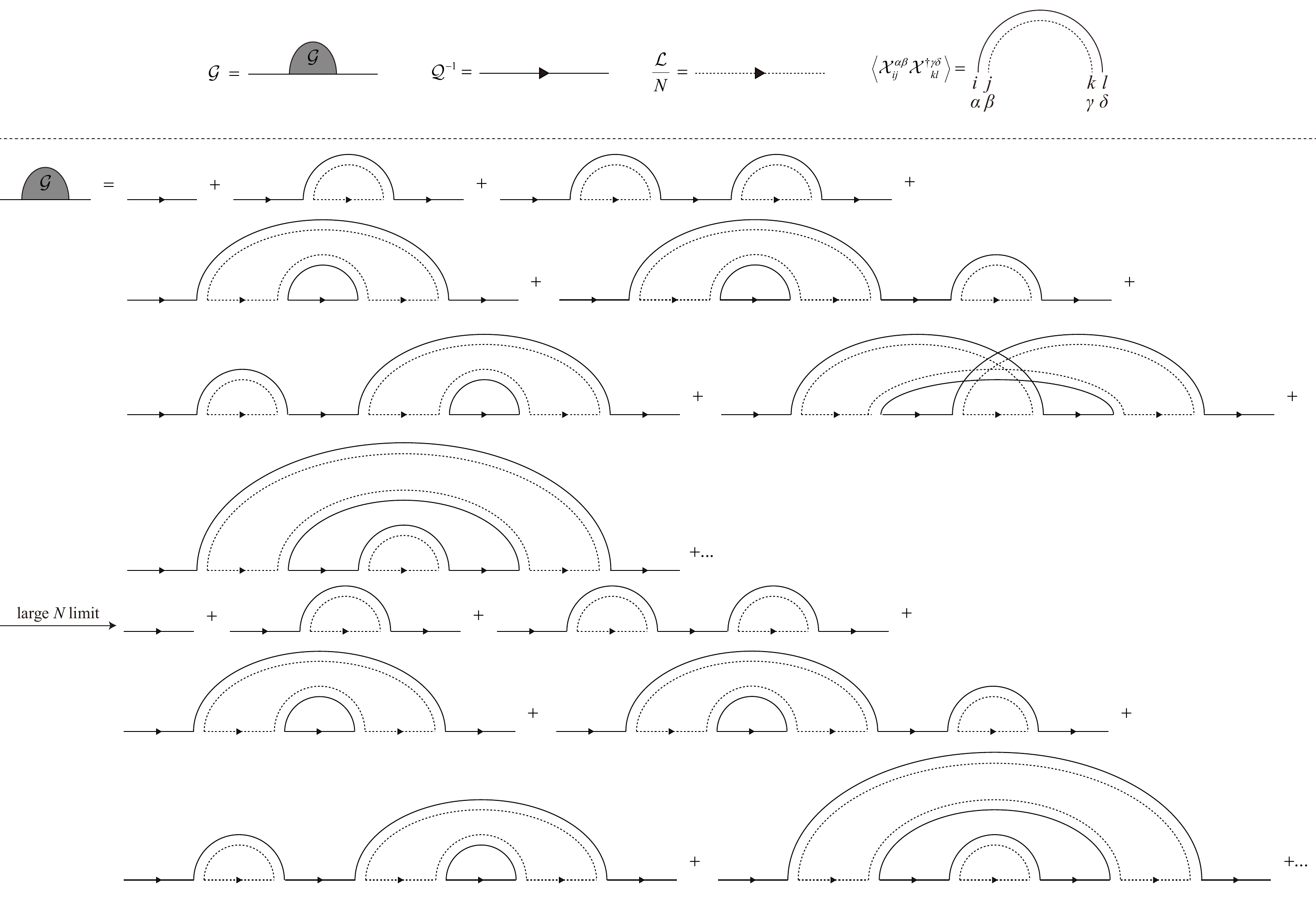}
		\caption{The diagrammatic representation of the Green's function $\cal G$. 
		In the large $N$ limit, only planar diagrams contribute to the final results.}
		\label{fig:Green}
	\end{figure}

	Using the diagrammatic method, we can rewrite the geometric series [Eq.~\eqref{series_pre}] into a graph
	representation (so-called Feynman diagram), which is shown in Fig.~\ref{fig:Green}.
	When the number of neurons $N\to\infty$, the expanded terms in the series contributing to the quaternion Green's function
	have the rainbow-like structure shown in the third part of Fig.~\ref{fig:Green}, 
	which is called planar diagrams since there are non-crossing structures in the diagrams. Non-planar diagrams give rise to vanishing contribution to the 
	expansion in the thermodynamic limit~\cite{Hooft-1974}.
	
\begin{figure}
		\centering
		\includegraphics[bb=5 5 1080 278,width=1.0\linewidth]{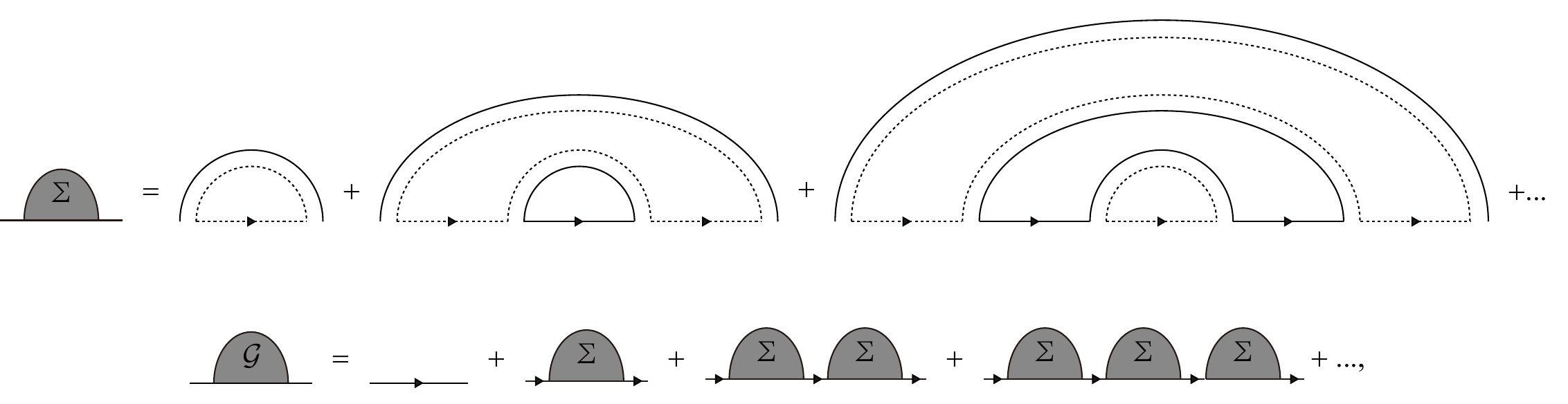}
		\caption{Definition of self-energy $\Sigma$ and the quaternionic Green's function $\cal G$ expressed in terms of the self-energy $\Sigma$.}
		\label{fig:selfenergy}
	\end{figure}

	To proceed, we need to introduce the concept of one-line-irreducible (1LI) diagrams for which
	the diagram terms can not be separated by cutting any horizontal line.
	For example, in the diagrammatic representation shown in Fig.~\ref{fig:Green}, the 1LI diagrams are the rainbow-like graphs.
	Naturally, we define the self-energy $\Sigma$ of the quaternionic Green's function as the sum of all 1LI diagrams. Hence, the quaternionic
	Green's function $\cal G$ can be expressed in terms of the self-energy $\Sigma$, in the form of the Dyson-Schwinger equation as follows~\cite{Nowak-2017},
	\begin{equation}
		\left( {{{\cal Q}}_{\alpha \gamma }^{ac} - \Sigma _{\alpha \gamma }^{ac}} \right){{\cal G}}_{\gamma \beta }^{cb} = {\delta ^{ab}}{\delta _{\alpha \beta }}. \label{Dyson-Schwinger_eq1}
	\end{equation}
	Note that the subscript Greek indices run from 1 to 2 reflecting the quaternion nature, while the Latin indices in the superscript go over all elements (the index takes a value from $1$ to $N$) in the matrix within each block.
	A diagrammatic representation of Eq.~(\ref{Dyson-Schwinger_eq1}) is shown in Fig.~\ref{fig:selfenergy}.
	
	Finally, the problem of computing the quarternion Green's function $\cal G$ is reduced to the problem of computing the self-energy $\Sigma$.
	In order to compute the self-energy, we need to introduce an auxiliary Green's function $\hat{\cal G}$ defined as~\cite{Burda-2004}
	\begin{equation}
		{\hat {\cal G}} = \left\langle {\frac{{\frac{{{\cal L}}}{N}}}{{1 - {{{\cal X}}^\dag }{Q^{ - 1}}\otimes {\id_N}{{\cal X}}\frac{{{\cal L}}}{N}}}} \right\rangle.
	\end{equation}
	The introduction of this auxiliary Green's function leads to the same result if the definition ${\hat {\cal G}} =\left\langle\frac{1}{Q\otimes\id_N-\hat{\cal J}}\right\rangle$
	where $\hat{{\cal J}}=\frac{1}{N}{\cal X}^\dag{\cal X}{\cal L}$ is used for the dual problem~\cite{Nowak-2017}. ${\cal G}$ is the corresponding Green's function.
	Through introducing the dual Green's function, one can then relate $\cal G$ and $\hat{\cal G}$ with self-energies, resulting in a closed-form equation.
	
\begin{figure}[t]
		\centering
		\includegraphics[bb=3 3 1437 986,width=1.0\linewidth]{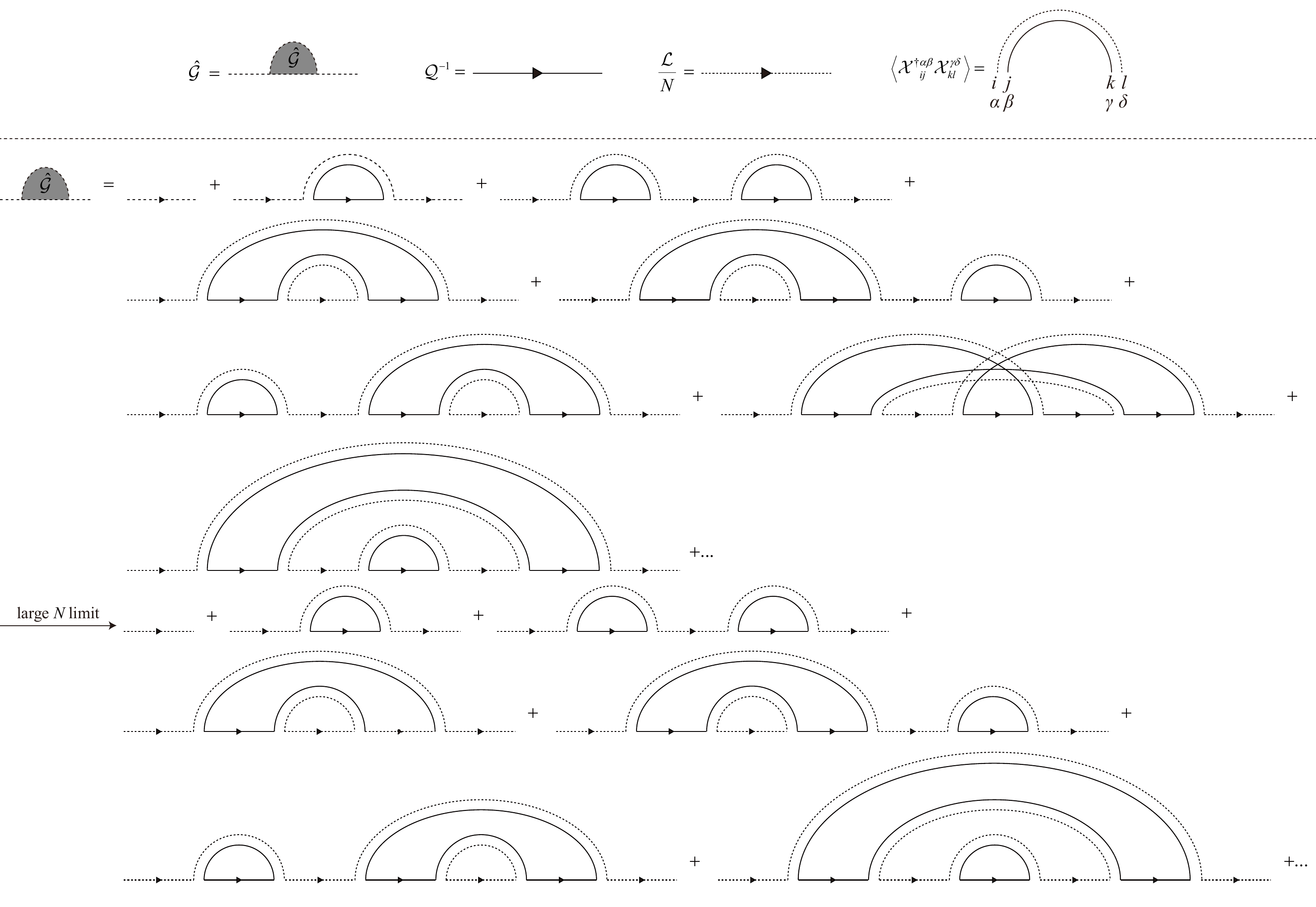}
		\caption{The diagrammatic representation of auxiliary Green's function $\hat{\cal G}$. Similarly, in the large $N$ limit, only planar diagrams contribute to the final results.}
		\label{fig:greenhat}
	\end{figure}

	Similarly, we expand this Green's function as
	\begin{equation}
		{\hat {\cal G}} = \frac{{{\cal L}}}{N} + \left\langle {\frac{{{\cal L}}}{N}{{{\cal X}}^\dag }{{{\cal Q}}^{ - 1}}{{\cal X}}\frac{{{\cal L}}}{N}} \right\rangle  
		+ \left\langle {\frac{{{\cal L}}}{N}{{{\cal X}}^\dag }{{\cal Q}^{ - 1}}{{\cal X}}\frac{{{\cal L}}}{N}{{{\cal X}}^\dag }{{\cal Q}^{ - 1}}{{\cal X}}\frac{{{\cal L}}}{N}} \right\rangle  +\ldots.
	\end{equation}
	Then we apply the diagrammatic methods again to express the precise form of the auxiliary Green's function as shown in Fig.~\ref{fig:greenhat}.
	
\begin{figure}
	\centering
	\includegraphics[bb=5 5 1080 278,width=1.0\linewidth]{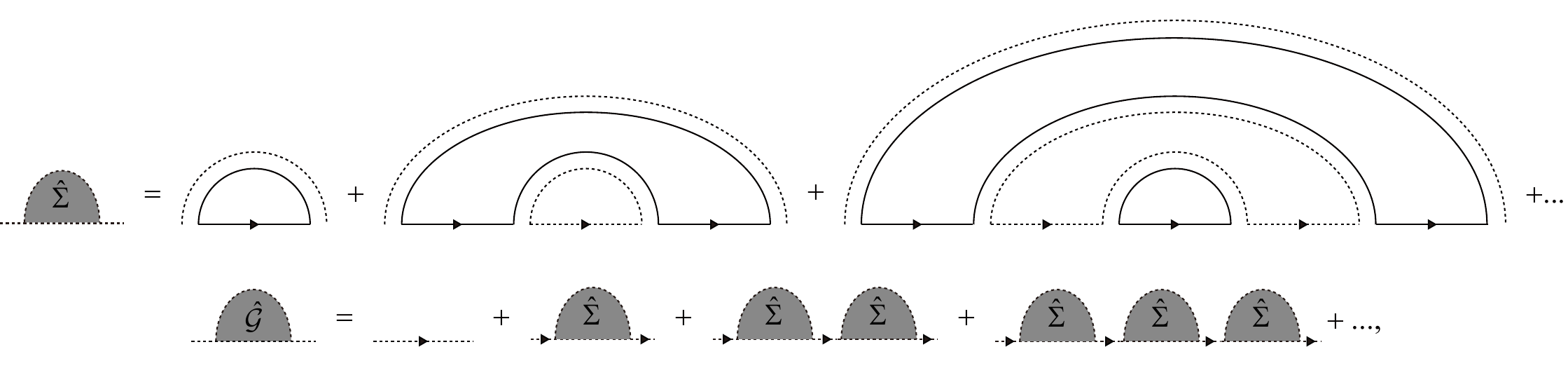}
	\caption{Definition of the self-energy $\hat{\Sigma}$ and the auxiliary Green's function $\hat{\cal G}$ expressed in terms of the 
	self-energy $\hat{\Sigma}$.}
	\label{fig:selfenergyhat}
	\end{figure}

	In an analogous way, we define the self-energy
	$\hat{\Sigma}$ for the auxiliary Green's function $\hat{\cal G}$ and then express the auxiliary Green's function by self-energy $\hat{\Sigma}$,
	which is shown in Fig.~\ref{fig:selfenergyhat}. The associated Dyson-Schwinger equation reads as follows,
	\begin{equation}
		\left( {{\delta _{\alpha \varepsilon }}{\hat{\delta} ^{ad}} - \frac{1}{N}\hat \Sigma _{\alpha \gamma }^{ac}{{\cal L}}_{\gamma \varepsilon }^{cd}} \right){\hat {\cal G}}_{\varepsilon \beta }^{db} = \frac{1}{N}{{\cal L}}_{\alpha \beta }^{ab},\label{Dyson-Schwinger_eq2}
	\end{equation}
	where the Latin index in $\hat{\delta}$ takes a value from $1$ to $P$.
	
Given the self-energy and Green's function shown in Figs.~(\ref{fig:Green}, \ref{fig:selfenergy}, \ref{fig:greenhat} and \ref{fig:selfenergyhat}),
one finds other diagrammatic relations between the Green's function and self-energy apart from Eq.~(\ref{Dyson-Schwinger_eq1}) and Eq.~(\ref{Dyson-Schwinger_eq2}),
	\begin{equation}
		\Sigma _{\alpha \beta }^{ab} = {\rm{Tr}}{{{\hat {\cal G}}}_{\alpha \beta }}{\delta ^{ab}}, \label{self-energy}
	\end{equation}
	and
	\begin{equation}
		\hat \Sigma _{\alpha \beta }^{ab} = {\rm{Tr}}{{{\cal G}}_{\alpha \beta }}{\hat{\delta} ^{ab}}. \label{self-energy_hat}
	\end{equation}
These two additional equations can be proved using the Feynman diagrams,
by adding a double arc joining the external points~\cite{Burda-2004}.
	
	From Eq.~(\ref{Dyson-Schwinger_eq2}) and Eq.~(\ref{self-energy}), we get
	\begin{subequations}
		\begin{align}
			\frac{1}{N}{{\cal L}}_{\alpha \beta }^{ab} &= \left( {{\delta _{\alpha \varepsilon }}{\hat{\delta} ^{ad}} - \frac{1}{N}\hat \Sigma _{\alpha \gamma }^{ac}{{\cal L}}_{\gamma \varepsilon }^{cd}} \right){\hat {\cal G}}_{\varepsilon \beta }^{db} 
			\\ &= \left( {{\delta _{\alpha \varepsilon }}{\hat{\delta} ^{ad}} - \frac{1}{N} {\Tr} {\cal{ G}}_{\alpha \gamma }\hat{\delta}^{ac}{{\cal L}}_{\gamma \varepsilon }^{cd}} \right){\hat {\cal G}}_{\varepsilon \beta }^{db}
			\\ &= \left( {{\delta _{\alpha \varepsilon }}{\hat{\delta} ^{ad}} - G_{\alpha \gamma }\hat{\delta}^{ac}{{\cal L}}_{\gamma \varepsilon }^{cd}} \right){\hat {\cal G}}_{\varepsilon \beta }^{db}.
		\end{align}
	\end{subequations}
	Consequently, we have
	\begin{equation}
		{\hat {\cal G}}_{\varepsilon \beta }^{db} = \frac{1}{N}\left( {{\delta _{\alpha \varepsilon }}{\hat{\delta} ^{ad}} - G_{\alpha \gamma }\hat{\delta}^{ac}{{\cal L}}_{\gamma \varepsilon }^{cd}} \right)^{-1}{{\cal L}}_{\alpha \beta }^{ab},
	\end{equation}
	and then take the block trace, namely, the trace over the index of Latin indices, leading to the following formula
	\begin{equation}
		{{\Tr}}{\hat {\cal G}_{\varepsilon \beta }} = \frac{1}{N}{\Tr}\left[\left( {\delta _{\alpha \varepsilon }}\id_P - G_{\alpha \gamma }\id_P{{\cal L}}_{\gamma \varepsilon } \right)^{-1}{{\cal L}}_{\alpha \beta }\right].
	\end{equation}
	With Eq.~(\ref{Dyson-Schwinger_eq1}) and Eq.~(\ref{self-energy}), we find that
	\begin{subequations}\label{eq40}
		\begin{align}
			{\delta ^{ab}}{\delta _{\alpha \beta }} &= \left( {{{\cal Q}}_{\alpha \gamma }^{ac} - \Sigma _{\alpha \gamma }^{ac}} \right){{\cal G}}_{\gamma \beta }^{cb} 
			\\& = \left( {{{\cal Q}}_{\alpha \gamma }^{ac} - {\rm{Tr}}{{{\hat {\cal G}}}_{\alpha \gamma }}\delta^{ac}} \right){{\cal G}}_{\gamma \beta }^{cb} 
			\\& = \left( {{{ Q}}_{\alpha \gamma }\delta^{ac} - {\rm{Tr}}{{{\hat {\cal G}}}_{\alpha \gamma }}\delta^{ac}} \right){{\cal G}}_{\gamma \beta }^{cb} 
			\\& = \left( {{{ Q}}_{\alpha \gamma } - {\rm{Tr}}{{{\hat {\cal G}}}_{\alpha \gamma }}} \right)\delta^{ac}{{\cal G}}_{\gamma \beta }^{cb}
			\\& = \left( {{{ Q}}_{\alpha \gamma } - \frac{1}{N}{\Tr}\left[\left( {\delta _{\varepsilon \alpha }}\id_P - G_{\varepsilon \zeta }\id_P{{\cal L}}_{\zeta \alpha } \right)^{-1}{{\cal L}}_{\varepsilon \gamma }\right]} \right)\delta^{ac}{{\cal G}}_{\gamma \beta }^{cb}
			\\& = \left( {{{ Q}}_{\alpha \gamma } - \frac{1}{N}{\Tr}\left[\left( {\delta _{\varepsilon \alpha }}\id_P - G_{\varepsilon \zeta }\id_P{{\cal L}}_{\zeta \alpha } \right)^{-1}{{\cal L}}_{\varepsilon \gamma }\right]} \right){{\cal G}}_{\gamma \beta }^{ab}.
		\end{align}
	\end{subequations}
	Finally, taking the block trace of Eq.~\eqref{eq40}, we find that
	\begin{subequations}
		\begin{align}
			{\delta _{\alpha \beta }} & = \left( {{{ Q}}_{\alpha \gamma } - \frac{1}{N}{\Tr}\left[\left( {\delta _{\varepsilon \alpha }}\id_P - G_{\varepsilon \zeta }\id_P{{\cal L}}_{\zeta \alpha } \right)^{-1}{{\cal L}}_{\varepsilon \gamma }\right]} \right)\frac{1}{N}{\Tr}{{\cal G}}_{\gamma \beta }^{ab}
			\\ &= \left( {{{ Q}}_{\alpha \gamma } - \frac{1}{N}{\Tr}\left[\left( {\delta _{\varepsilon \alpha }}\id_P - G_{\varepsilon \zeta }\id_P{{\cal L}}_{\zeta \alpha } \right)^{-1}{{\cal L}}_{\varepsilon \gamma }\right]} \right){ G}_{\gamma \beta }.
		\end{align}	
	\end{subequations} 
Therefore, based on Eqs.~\eqref{Dyson-Schwinger_eq1}, \eqref{Dyson-Schwinger_eq2}, \eqref{self-energy} and \eqref{self-energy_hat}, together with
the definition of the Green's function $G$ shown in Eq.~\eqref{matrix_G}, we obtain finally
	\begin{equation}
		\left[ {Q - \frac{1}{N}{\rm{bTr_2}}\left( {{{\cal L}}{{\left[ {{\id_{2P}} - \left( {G \otimes {\id_P}} \right){{\cal L}}} \right]}^{ - 1}}} \right)} \right]G = {\id_2},\label{self-consistent}
	\end{equation}
	which is the self-consistent equation the Green's function obeys.
	
	In our model setting, $\cal L$ is an invertible matrix. We thus rewrite the above self-consistent equation in a compact form
	\begin{equation}
		\left[ {Q + \alpha {G_{{\bm{\Lambda} ^{ - 1}}}}\left( G \right)} \right]G = {\id_2},
	\end{equation}
	where $\alpha=\frac P N$ is the memory load, and ${G_{{\bm{\Lambda} ^{ - 1}}}}\left( Q \right)$ follows the same definition of the matrix Green's function as above 
	\begin{equation}
		{G_{{\bm{\Lambda} ^{ - 1}}}}\left( Q \right){\rm{ = }}\frac{1}{P}{\rm{bTr}_2}\frac{1}{{Q \otimes {\id_P} - {{{\cal L}}^{ - 1}}}}.
	\end{equation}
	
	Now we can express the explicit form of the matrix Green's function using the block inverse formula [Eq.~\eqref{inverse-matrix}] as 
	\begin{equation}
		{G_{{\bm{\Lambda} ^{ - 1}}}}\left( Q \right) = \frac{1}{P}{\rm{bTr_2}}\left( {\begin{array}{*{20}{c}}
				{\frac{{\bar z\id_P - (\boldsymbol{\Lambda}^{\da}) ^{ -1 }}}{{\left( {z\id_P - {\boldsymbol{\Lambda} ^{ - 1}}} \right)\left( {\bar z\id_P - (\boldsymbol{\Lambda} ^{\dag })^{-1}} \right) + {{\left| w \right|}^2}\id_P}}}&{\frac{{ - \i\bar w\id_P}}{{\left( {z\id_P - {\boldsymbol{\Lambda} ^{ - 1}}} \right)\left( {\bar z\id_P - (\boldsymbol{\Lambda} ^{\dag })^{-1}} \right) + {{\left| w \right|}^2}\id_P}}}\\
				{\frac{{ - \i w\id_P}}{{\left( {z\id_P - {\boldsymbol{\Lambda} ^{ - 1}}} \right)\left( {\bar z\id_P - (\boldsymbol{\Lambda} ^{\dag })^{-1}} \right) + {{\left| w \right|}^2}\id_P}}}&{\frac{{z\id_P - {\boldsymbol{\Lambda} ^{ - 1}}}}{{\left( {z\id_P - {\boldsymbol{\Lambda} ^{ - 1}}} \right)\left( {\bar z\id_P - (\boldsymbol{\Lambda} ^{\dag })^{-1}} \right) + {{\left| w \right|}^2}\id_P}}}
		\end{array}} \right).\label{explicit-matrix-G}
	\end{equation}

	Then we change the argument $Q$ of the matrix Green's function to $G$ defined in Eq.~\eqref{matrix-G_in_2times2_form} and obtain that
	\begin{equation}
		{G_{{\bm{\Lambda} ^{ - 1}}}}\left( G \right) = \left( {\begin{array}{*{20}{c}}
				F&{ - \i\bar wE}\\
				{ - \i wE}&{\bar F}
		\end{array}} \right),
	\end{equation}
	where 
	\begin{equation}
		E\left( {z,w} \right) = \frac{1}{P}\sum\limits_{m = 1}^P {\frac{1}{{{{\left| {z - {{\left( {c + \gamma \sum\limits_{r = 1}^d {{e^{\i2\pi \frac{m}{P}r}}} } \right)}^{ - 1}}} \right|}^2} + {{\left| w \right|}^2}}}} ,
	\end{equation}
	\begin{equation}
		F\left( {z,w} \right) = \frac{1}{P}\sum\limits_{m = 1}^P {\frac{{\bar z - {{\left( {c + \gamma \sum\limits_{r = 1}^d {{e^{ - \i2\pi \frac{m}{P}r}}} } \right)}^{ - 1 }}}}{{{{\left| {z - {{\left( {c + \gamma \sum\limits_{r = 1}^d {{e^{\i2\pi \frac{m}{P}r}}} } \right)}^{ - 1}}} \right|}^2} + {{\left| w \right|}^2}}}} .
	\end{equation}

	From the self-consistent equation [Eq.~(\ref{self-consistent})], we can finally find that
	\begin{equation}
		\left( {\begin{array}{*{20}{c}}
				{z + \alpha F\left( {g,v} \right)}&{\i\bar w - \i\alpha \bar vE\left( {g,v} \right)}\\
				{\i w - \i\alpha vE\left( {g,v} \right)}&{z + \alpha \bar F\left( {g,v} \right)}
		\end{array}} \right)\left( {\begin{array}{*{20}{c}}
				g&{\i\bar v}\\
				{\i v}&{\bar g}
		\end{array}} \right) = \left( {\begin{array}{*{20}{c}}
				1&0\\
				0&1
		\end{array}} \right),
	\end{equation}
	which can be reduced into two independent equations
	\begin{gather}
		 zg - \bar wv + \alpha gF\left( {g,v} \right) + \alpha {\left| v \right|^2}E\left( {g,v} \right) = 1,\\
		 z\bar v + \bar w\bar g + \alpha \bar vF\left( {g,v} \right) - \alpha 	\bar g\bar vE\left( {g,v} \right) = 0.
	\end{gather}
	
	From Eq.~\eqref{reduced_Green}, we need to set $|w| \to 0$ to obtain the Green's function for $\bJ$. Simple algebraic manipulations lead us to the final results:
	\begin{equation}
		z = \frac{\alpha }{P}\sum\limits_{m = 1}^P {\frac{{{{\left( {c + \gamma \sum\limits_{r = 1}^d {{e^{ - \i2\pi \frac{m}{P}r}}} } \right)}^{ - 1 }}}}{{{{\left| {g - {{\left( {c + \gamma \sum\limits_{r = 1}^d {{e^{\i2\pi \frac{m}{P}r}}} } \right)}^{ - 1}}} \right|}^2} + {{\left| v \right|}^2}}}} ,
	\end{equation}
	\begin{equation}
		1 = \frac{\alpha }{P}\sum\limits_{m = 1}^P {\frac{{{{\left| g 	\right|}^2} + {{\left| v \right|}^2}}}{{{{\left| {g - {{\left( {c + \gamma \sum\limits_{r = 1}^d {{e^{\i2\pi \frac{m}{P}r}}} } \right)}^{ - 1}}} \right|}^2} + {{\left| v \right|}^2}}}} .
	\end{equation}
	The spectrum boundary can be obtained by setting $|v|\to0$, i.e.,
	\begin{equation}
		z = \frac{\alpha }{P}\sum\limits_{m = 1}^P {{{\left( {c + \gamma \sum\limits_{r = 1}^d {{e^{ - \i2\pi \frac{m}{P}r}}} } \right)}^{ - 1 }}{{\left| {g - {{\left( {c + \gamma \sum\limits_{r = 1}^d {{e^{\i2\pi \frac{m}{P}r}}} } \right)}^{ - 1}}} \right|}^{ - 2}}} ,
	\end{equation}
	\begin{equation}
		1 = \frac{\alpha }{P}{{\left| g 	\right|}^2}\sum\limits_{m = 1}^P {{{\left| {g - {{\left( {c + \gamma \sum\limits_{r = 1}^d {{e^{\i2\pi \frac{m}{P}r}}} } \right)}^{ - 1}}} \right|}^{ - 2}}} .
	\end{equation}




\end{document}